\documentclass[12pt,twoside,letterpaper,openright]{thesis_msb}
\usepackage{amsmath,amsfonts,amssymb,verbatim,graphics}
\usepackage[amsthm]{ntheorem}

\theoremstyle{plain}
\newtheorem{theorem}{Theorem}
\newtheorem{proposition}[theorem]{Proposition}
\newtheorem{definition}[theorem]{Definition}
\newtheorem{problem}[theorem]{Problem}

\numberwithin{theorem}{chapter}

\let\endproofTEMP\endproof
\renewcommand{\endproof}{\hfill\raisebox{2pt}{\openbox}\endproofTEMP}

\input{thesis_algorithm}

\newcommand{\abs}[1]{\left\vert #1 \right\vert}  
\newcommand{\norm}[1]{\left\Vert #1 \right\Vert}  
\newcommand{\adj}[1]{#1^{\dag}} 
\newcommand{\st}{\,:\,}  
\newcommand{\ip}[2]{\left< \, #1 \, , \, #2 \, \right>} 
\newcommand{\qo}{\mathcal{O}_\Delta}  
\newcommand{\reg}{R_\Delta} 
\newcommand{\cg}{Cl_\Delta}  
\newcommand{\cn}{h_\Delta}  
\newcommand{\iideals}{\mathcal{I}_\Delta}  
\newcommand{\pideals}{\mathcal{P}_\Delta}  

\newcommand{\idealleftchar}{\mathop{\lambda}\!}
\newcommand{\idealleft}[1]{\idealleftchar\paren{#1}}
\newcommand{\ideallefttwo}[2]{\idealleftchar\paren{#1,#2}}

\newcommand{\idealerrorchar}{\mathop{\varepsilon}\!}
\newcommand{\idealerror}[1]{\idealerrorchar\paren{#1}}
\newcommand{\idealerrortwo}[2]{\idealerrorchar\paren{#1,#2}}

\newcommand{\idealdistancechar}{\mathop{\delta}\!}
\newcommand{\idealdistance}[1]{\idealdistancechar\paren{#1}}
\newcommand{\idealdistancetwo}[2]{\idealdistancechar\paren{#1,#2}}

\newcommand{\integers}[1]{\mathcal{O}_{#1}}

\newcommand{\ket}[1]{\left| #1 \right>}

\newcommand{\qgate}[1]{\textup{\textbf{#1}}} 
\newcommand{\Had}{\qgate{H}} 
\newcommand{\Rot}[1]{\qgate{R}_{#1}} 
\newcommand{\CNot}{\qgate{CNOT}} 
\newcommand{\Not}{\qgate{X}} 
\newcommand{\Id}{\qgate{I}} 
\newcommand{\U}{\qgate{U}} 
\newcommand{\Uf}{\qgate{U}_f} 
\newcommand{\QFT}[1]{\qgate{QFT}_{#1}}
\newcommand{\IQFT}[1]{\QFT{#1}^{-1}}
\newcommand{\HilbertInf}{\mathbb{H}}
\newcommand{\Hilbert}[1]{\HilbertInf_{#1}}
\newcommand{\tensor}{\otimes}

\newcommand{\lb}{\left(}  
\newcommand{\rb}{\right)} 
\newcommand{\blb}{\bigl(} 
\newcommand{\brb}{\bigr)} 
\newcommand{\Blb}{\Bigl(} 
\newcommand{\Brb}{\Bigr)} 

\newcommand{\floor}[1]{\left\lfloor #1 \right\rfloor}
\newcommand{\ceil}[1]{\left\lceil #1 \right\rceil}
\newcommand{\round}[1]{\left\lfloor #1 \right\rceil}
\newcommand{\bracket}[1]{\left[ #1 \right]}
\renewcommand{\brace}[1]{\left\{ #1 \right\}}
\newcommand{\paren}[1]{\left( #1 \right)}

\newcommand{\parfrac}[2]{\paren{\frac{#1}{#2}}}

\newcommand{\lcm}{\operatorname{lcm}}
\newcommand{\bigO}[1]{O\!\paren{#1}}

\newcommand{\Z}{\mathbb{Z}}  
\newcommand{\Zn}{\Z^n}
\newcommand{\R}{\mathbb{R}}  
\newcommand{\Rn}{\R^n}
\newcommand{\Q}{\mathbb{Q}}  

\newcommand{\C}{\mathbb{C}}  

\renewcommand{\P}{\textbf{P}}
\newcommand{\NP}{\textbf{NP}}

\newcommand{\tbd}[1]{#1}  

\DeclareMathOperator*{\doublesum}{\sum\sum}

\newcommand{\chapref}[1]{Chapter~\ref{#1}}
\newcommand{\tabref}[1]{Table~\ref{#1}}
\newcommand{\equref}[1]{Equation~\eqref{#1}}
\newcommand{\propref}[1]{Proposition~\ref{#1}}
\newcommand{\secref}[1]{Section~\ref{#1}}
\newcommand{\stepref}[1]{Step~\ref{#1}}
\newcommand{\defref}[1]{Definition~\ref{#1}}
\newcommand{\probref}[1]{Problem~\ref{#1}}
\newcommand{\thmref}[1]{Theorem~\ref{#1}}
\newcommand{\algref}[1]{Algorithm~\ref{#1}}
\newcommand{\figref}[1]{Figure~\ref{#1}}

\title{Classical Cryptosystems In A Quantum Setting}
\author{Michael Stephen Brown}
\discipline{Combinatorics and Optimisation}
\degree{Master of Mathematics}

\begin{document}

\prepages
\maketitle
\sigpages

\begin{acknowledgements}
This thesis would not have been possible without much support and assistance. I would like to thank
my supervisor, Michele Mosca, for sharing wisdom, experience, and guidance.  Thank you to NSERC and
the Department of Combinatorics and Optimisation at the University of Waterloo for their generous
financial support. Thank you also to Phillip Kaye, Edlyn Teske, and Christof Zalka for many helpful
conversations and suggestions.

Finally, thank you very much to my family, who as always, has provided me with so much love and
encouragement.
\end{acknowledgements}

\tableofcontents

\mainbody

\begin{chapter}{Introduction To Public Key Cryptography}
\label{chap:IntroCrypto}\markright{CHAPTER \thechapter. \ INTRODUCTION TO PUBLIC KEY CRYPTOGRAPHY}

Cryptography has been an area of mathematical study for centuries.  Historically, the study of
cryptography focused on the design of systems that provide secret communication over an insecure
channel.  Recently, individuals, corporations, and governments have started to demand privacy,
authenticity, and reliability in all sorts of communication, from online shopping to discussions of
national secrets.  As a result, the goals of cryptography have become more all-encompassing; now,
cryptography might better be defined as the design of systems that need to withstand any malicious
attempts to abuse them.  This thesis will focus on modern algorithms and techniques for
confidentiality, which are also known as encryption schemes.  However, the purposes of cryptography
include not only secret or confidential communication, but also authentication of the entities
involved in the communication, authentication of the data transmitted by those entities, and many
others.

The oldest encryption schemes are known as symmetric key or secret key systems.  Such systems
consist of two main algorithms: an encryption algorithm, which allows one entity to encrypt or
``scramble'' data, and a decryption algorithm, which allows another entity to decrypt or
``unscramble'' data.  Each of these algorithms has an input called a key, which dictates some
aspect of the algorithm's behaviour.  In order for two entities (historically known as Alice and
Bob) to exchange data securely, they must first share a secret key between them.  If Bob wishes to
send Alice a message, he uses the secret key with the encryption algorithm to encrypt the message.
He sends the encrypted message (called the ciphertext) to Alice, and she uses the secret key with
the decryption algorithm to decrypt the ciphertext and recover the original message. Since an
eavesdropper (Eve) does not know the secret key, she should not be able to determine what the
original message was.

A physical analogy of a symmetric key scheme is often given in terms of boxes and padlocks.  Suppose Alice and Bob each have a
copy of a key for a padlock.  If Bob wishes to send Alice a message, he writes the message on a piece of paper and places it in a
box.  He then uses his copy of the key to lock the box with the padlock, and he sends the locked box to Alice. When she receives
it, she uses her copy of the key to unlock the padlock, she opens the box, and she reads the message. If Eve finds the locked box,
however, she cannot open the padlock because she does not have a copy of the key.

Symmetric key encryption schemes are well-suited to many applications.  They tend to be very
efficient in time and space required for their implementation, and they tend to require only a
small amount of key material for a high level of security.  The main drawback of such schemes has
come to be known as the key distribution problem: if Alice and Bob wish to communicate secretly but
have never met, how do they share a secret key?  They cannot send a secret key over an insecure
channel because Eve might be listening and might learn the key; on the other hand, they do not yet
share a secure channel over which to send a secret key. This problem was one of the largest
problems in cryptography for many years.  Some solutions might be for Alice and Bob to meet in
person and agree on a key face-to-face (which is of course impractical if they live far away from
one another) or for them to enlist the services of a third party to courier a secret key between
them (which implies they both must trust the third party not to reveal the key to anyone). Further,
for every pair of parties that wishes to communicate secretly, a unique symmetric key is required;
thus the number of symmetric keys in the system grows rapidly.

In the late 1970\,s, the mathematicians Diffie and Hellman introduced a new idea: public key
cryptography~\cite{DH1976:NewDirections}.  (In fact, a British intelligence researcher had
discovered the same idea earlier~\cite{E1970:NonSecretEncryption}, but his discovery was not made
public until later.)  Like the secret key systems described above, a public key scheme has two main
algorithms for encryption and decryption, each of which has an input called a key. The difference
is that the keys used in the two algorithms are not the same. More specifically, Alice generates
two keys of her own: a public key, which she shares with everyone (even her enemies) and a private
key, which she keeps to herself.  If Bob wishes to send Alice a message, he obtains a copy of
Alice's public key, and uses her public key with the encryption algorithm to encrypt the message.
He sends the ciphertext to Alice, and she uses her private key with the decryption algorithm to
decrypt the ciphertext and recover the original message.  Since Eve does not know Alice's private
key, she should not be able to determine what the original message was. In other words, anyone can
encrypt a message for Alice, since anyone can obtain Alice's public key, but once a message is
encrypted for her, only Alice can decrypt it with her private key.

Again, we can illustrate the idea of a public key encryption scheme with a physical analogy in terms of boxes and padlocks.
Suppose Alice has a number of empty boxes, a number of open padlocks (that can be locked without a key), and a key that opens all
of the padlocks. She freely gives out these boxes and open padlocks to anyone who would like them.  If Bob wishes to send Alice a
message, he writes the message on a piece of paper, gets a box and lock from Alice, and places the message in the box.  He then
locks the box with the padlock, and he sends the locked box to Alice.  When she receives it, she uses her key to unlock the
padlock, she opens the box, and she reads the message.  If Eve finds the locked box, however, she cannot open the padlock because
she does not have a copy of the key.  (After he has locked his message in the box, even Bob cannot get the message back out!)

The major advantage of these public key schemes is that they provide a solution to the key
distribution problem.  Public keys, by design, can be freely distributed to anyone without
compromising the security of the system, so if Alice and Bob wish to communicate secretly but they
have never met before, they need simply obtain one another's public keys.  There are some
disadvantages, in that public key schemes tend to be less efficient and the keys tend to be larger
than in secret key systems, but these disadvantages are small compared to the advantages provided
by such schemes.  There are also ways to use public and secret key schemes together to minimise the
disadvantages.

There are several public key cryptosystems that have been proposed, and many have been studied in great detail.  The study of
these cryptosystems includes studying approaches to breaking them.  Breaking a cryptosystem could have many meanings.  For
example, given only ciphertext, an attacker might try to determine partial or complete information about the corresponding
original message. Given only an entity's public key, an attacker might try to determine partial or complete information about the
corresponding private key.  There are many variations on the same theme.

Whereas the cryptosystems that are currently in use generally have not been broken, attackers are
constantly developing new attacks and improvements in technology are helping to speed up current
attacks.  Especially worrisome to the field of cryptography are developments in the area of quantum
computing, which we will discuss in the next chapter.  Even though a quantum computer of a
sufficient size has not yet been implemented, the theory of quantum computing indicates that many
of the cryptosystems currently in use could easily be broken if this implementation did occur.  If
a quantum computer is successfully built, we will therefore have to change the cryptosystems we use
for encrypted communication so that attackers with quantum computers cannot decrypt it. Further,
encrypted messages captured and stored in the past could also be decrypted by a future quantum
attacker. Since there is a definite possibility that one day quantum computers will become
technologically feasible, we need to prepare for that eventuality by analysing modern cryptosystems
with respect to attacks with a quantum computer.

\end{chapter}

\begin{chapter}{Introduction To Quantum Computing}
\label{chap:IntroQuantumComp}

This chapter provides an overview of some aspects of quantum computing.  For a more complete
treatment of the history of the subject and many more details on the ideas discussed in this
chapter, see for example~\cite{NC2000:Nielsen}.

\begin{section}{Basic Concepts}

The computers that are in widespread use today are sometimes called classical computers.  The behaviour of the elements in these
computers can be described by the laws of classical physics, that is, those laws that were thought to be accurate around the turn
of the twentieth century.  However, early in the twentieth century scientists realised that those laws did not accurately describe
the behaviour of all systems.  For example, objects on an atomic scale behaved differently in experiments than was predicted by
classical physics.   To more accurately describe these systems, scientists developed a new theory of physics called quantum
physics.  This theory includes elements of non-determinism, and it more accurately models the behaviour of all systems.

With this new model of physics, a new type of computer has emerged: the quantum computer.  A
quantum computer is a device that uses the laws of quantum physics to solve problems.  There are
many ways in which quantum computers could be implemented, some of which are summarised
in~\cite{NC2000:Nielsen}. This thesis will not be concerned with specific implementations, but it
is important to note that quantum computers have been implemented successfully, albeit on a small
scale.  However, regardless of the particular implementation, the behaviour of a quantum computer
is governed by a specific set of mathematical rules, namely the laws of quantum physics.  A quantum
computer can therefore be described completely generally and mathematically.

In a classical computer, information is stored and manipulated in the form of ``bits''.  Each bit is represented in the computer
by an object that exists in one of two states, usually referred to as $0$ and $1$.  The computer can manipulate the states of the
bits using various logical operations, and it may examine any bit and determine in which of the two states the bit currently
exists.

In a quantum computer, information is stored and manipulated in the form of quantum bits, or
``qubits''.  (Initially, qubits and classical bits seem to be completely different concepts, but as
we will see, a bit in a classical computer is really a ``restricted'' qubit.)  A qubit can exist in
one of \emph{many} different states. More specifically, we think of the state of a qubit as a unit
vector in a two-dimensional complex vector space. As in any vector space, we could choose any basis
and represent the qubit with respect to that basis. However, for each quantum system we model, we
will choose a convenient orthonormal basis which we will call the ``computational basis''; the
computational basis states are denoted $\ket{0}$ and $\ket{1}$. In other words, the state of a
qubit could be represented as
\[
    \ket{\phi} = \alpha \ket{0} + \beta \ket{1}
\]
where $\alpha$ and $\beta$ are complex numbers.  The condition that $\ket{\phi}$ is a unit vector means that $\abs{\alpha}^2 +
\abs{\beta}^2 = 1$.  Such a linear combination of basis states is often called a superposition.

We cannot examine a qubit directly to determine its exact state (or in other words, the values $\alpha$ and $\beta$). According to the laws of quantum
mechanics, when we measure the qubit, we obtain $\ket{0}$ with probability $\abs{\alpha}^2$ and $\ket{1}$ with probability $\abs{\beta}^2$. Further, when such
a measurement is made, the state of the qubit ``collapses'' from its original superposition to either $\ket{0}$ or $\ket{1}$, depending on the outcome of the
measurement.  Apart from the measurement operation, we will restrict our attention to operations that treat the quantum computer as a closed system; that is,
we will assume that no information about the state of the system is ``leaked'' to the apparatus or to an external system.

As we will see in \secref{sec:IntroQuantumComp:SingleQubitGates}, however, we can manipulate
superposition states without extracting information from them.  This fact allows us to perform
operations that are impossible to implement with a classical computer (even a probabilistic
classical computer).  For example, as a state is manipulated, the amplitudes of each of the basis
states can interfere with each other: two amplitudes of the same sign can combine constructively to
increase the probability associated with a particular measurement outcome, or two amplitudes of
opposite sign can combine destructively to decrease this probability.  The existence of these
quantum interference effects is one of the main differences between quantum and classical
computers.

Despite these apparent differences, bits and qubits both model a physical system with two orthogonal states.  The bits in a classical computer are essentially
restricted qubits in that they do not exist in superposition states for long periods of time: they are continually leaking information about their states to
external systems.  The problems of maintaining coherent superposition states and preventing the computer from coupling with external systems are some of the
main challenges that scientists must overcome when implementing a quantum computer.

\end{section}

\begin{section}{Hilbert Space}\label{sec:IntroQuantumComp:HilbertSpace}

As described above, we can model the state of a qubit as a vector in a two-dimensional complex
vector space.  In fact, the state of any quantum mechanical system can be modeled as a vector
inside a special kind of vector space called a Hilbert space.  For the purposes of this thesis we
will restrict our attention to Hilbert spaces of finite dimension, but to describe general quantum
systems we need to consider infinite-dimensional spaces.  We briefly define a Hilbert space here;
for a more complete description the reader may consult for example~\cite{P1995:QuantumTheory}:
\begin{definition}
A vector space $\HilbertInf$ is called a \tbd{Hilbert space} if it satisfies the following three properties:
\begin{enumerate}
\item For any vectors $u, v \in \HilbertInf$ and any scalars $\alpha, \beta \in \C$, $\alpha u + \beta v \in \HilbertInf$.
\item For any vectors $u, v \in \HilbertInf$ there exists a complex number $\ip{u}{v}$ (known as the inner product of $u$ and
$v$) the value of which is linear in the first component.  Further, $\ip{u}{v}$ and $\ip{v}{u}$ are
complex conjugates of one another, and $\ip{u}{u} \geq 0$ with equality if and only if $u = 0$.
\item Let $\{u_m\}$ be an infinite sequence of vectors in $\HilbertInf$ and define the norm of $u$ by $\norm{u} = \sqrt{\ip{u}{u}}$.  If
$\norm{u_m - u_n} \rightarrow 0$ as $m,n \rightarrow \infty$ then there is a unique $u \in
\HilbertInf$ such that $\norm{u_m - u} \rightarrow 0$ as $m \rightarrow \infty$.  (In other words,
any Cauchy sequence of vectors in the space has a limit which is also a vector in the space.)
\end{enumerate}
\end{definition}
This last property is known as the completeness property, and it is satisfied by every finite
dimensional complex vector space equipped with an inner product. Thus in finite dimensions, every
complex inner product space is a Hilbert space~\cite{NC2000:Nielsen}.  (This fact is not true in
infinite dimensions.)

Suppose we have two quantum systems, the states of which can be modeled by vectors $\ket{\phi} \in
\Hilbert{m}$ and $\ket{\psi} \in \Hilbert{n}$ (where $\Hilbert{m}$ and $\Hilbert{n}$ are Hilbert
spaces of dimension $m$ and $n$, respectively).  To describe the joint state of these systems, we
use the ``tensor product'' of $\ket{\phi}$ and $\ket{\psi}$, denoted $\ket{\phi} \tensor
\ket{\psi}$ or simply $\ket{\phi}\ket{\psi}$. This new vector is an element of a larger Hilbert
space denoted $\Hilbert{m} \tensor \Hilbert{n}$ (which is in fact defined as the set of all linear
combinations of tensor products $\ket{\phi} \tensor \ket{\psi}$ with $\ket{\phi} \in \Hilbert{m}$
and $\ket{\psi} \in \Hilbert{n}$).

By definition, a tensor product over two vector spaces $V$ and $W$ must satisfy the following properties for all $v, v' \in V$,
$w,w' \in W$, and $\alpha \in \C$:
\begin{enumerate}
\item $\alpha ( v \tensor w ) = (\alpha v) \tensor w = v \tensor (\alpha w)$.
\item $( v + v' ) \tensor w = v \tensor w + v' \tensor w$.
\item $ v \tensor ( w + w' ) = v \tensor w + v \tensor w'$.
\end{enumerate}

Also, if $A$ is a linear operator on $V$ and $B$ is a linear operator on $W$, then we can define the linear operator $A \tensor B$
on $V \tensor W$ by
\[
    ( A \tensor B )( v \tensor w ) = Av \tensor Bw
\]
for all $v \in V$ and $w \in W$.

With these definitions, we are ready to describe more of the basic concepts behind quantum computing.

\end{section}

\begin{section}{Single-Qubit Gates}\label{sec:IntroQuantumComp:SingleQubitGates}

In a classical computer, we perform computation using circuits of gates connected by wires that carry the bits between the gates.  An example of a simple gate
in a classical computer is the NOT gate, which maps $0$ to $1$ and $1$ to $0$.  An analogous gate in a quantum computer would map $\ket{0}$ to $\ket{1}$ and
$\ket{1}$ to $\ket{0}$.  The laws of quantum mechanics state that if we are working in a closed system, we should define the gate's behaviour on a
superposition by extending its behaviour on the basis states linearly. In other words, the quantum NOT gate maps
\[
    \alpha\ket{0} + \beta\ket{1} \longmapsto \beta\ket{0} + \alpha\ket{1}.
\]

Because a quantum gate is a linear operator on the space of quantum states, we can express it as a matrix with respect to the
computational basis. We write the state $\alpha\ket{0} + \beta\ket{1}$ in vector form as
\[
    \begin{pmatrix}\alpha \\
                   \beta \end{pmatrix}.
\]
The NOT gate can then be represented by a matrix $\Not$ such that for any $\alpha, \beta \in \C$ with $\abs{\alpha}^2 + \abs{\beta}^2 = 1$,
\[
    \Not \begin{pmatrix} \alpha \\
                      \beta \end{pmatrix}
    = \begin{pmatrix} \beta \\
                      \alpha \end{pmatrix}.
\]
Thus we must have
\[
    \Not = \begin{pmatrix} 0 & 1 \\
                        1 & 0 \end{pmatrix}.
\]

Any quantum gate that acts on a single qubit can similarly be expressed as a $2 \times 2$ matrix.
However, the converse is not true: every $2 \times 2$ matrix does not define a valid quantum gate.
The input and output of a quantum gate are both quantum states, which as mentioned previously are
unit vectors in a two-dimensional complex vector space.  Thus any quantum gate must be an operator
that maps all unit vectors to unit vectors in this vector space; such an operator is called a
unitary operator.  An equivalent definition of a unitary operator says that $\U$ is unitary if and
only if $\adj{\U} \U = \Id$, where $\adj{\U}$ represents the conjugate transpose of $\U$, and $\Id$
represents the identity operator. It is in fact true that any unitary operator does define a
``valid'' quantum gate, although not every unitary operation can be performed efficiently in every
quantum system; so many of these gates cannot be implemented efficiently.

The unitarity condition on quantum gates implies another important aspect of quantum computation: if we assume that our system is
closed, quantum computation is ``reversible''. Since the inverse of any unitary operator is also unitary, the inverse operation of
a quantum gate is also a quantum gate, and hence given the output of a gate $\U$ we can recover the input by applying the valid
gate $\U^\dag$.

We will mention two more important single-qubit gates at this time.  First, the Hadamard gate is defined by the matrix
\[
    \Had = \frac{1}{\sqrt{2}}\begin{pmatrix} 1 & \phantom{-}1 \\
                                          1 & -1 \end{pmatrix},
\]
and maps $\ket{0}$ to $\frac{1}{\sqrt{2}}\blb\ket{0} + \ket{1}\brb$ and $\ket{1}$ to $\frac{1}{\sqrt{2}}\blb\ket{0} - \ket{1}\brb$.  Second, the phase gate is
defined for any angle $\theta$ by the matrix
\[
    \Rot{\theta} = \begin{pmatrix} 1 & 0 \\
                                    0 & e^{i\theta} \end{pmatrix}.
\]
This gate leaves the state $\ket{0}$ unchanged and maps $\ket{1}$ to $e^{i\theta}\ket{1}$.  It is easy to see that these matrices are unitary, since $\Had^{-1}
= \Had^\dag = \Had$ and $\Rot{\theta}^{-1} = \Rot{\theta}^\dag = \Rot{-\theta}$.

\end{section}

\begin{section}{Multiple-Qubit Gates}\label{sec:IntroQuantumComp:MultipleQubitGates}

We can extend the definition of quantum gates to act on $n$ qubits at once.  The state of $n$ qubits can be represented as a unit vector in a $2^n$-dimensional
complex vector space, and again, the only condition on an $n$-qubit gate is that it must be a unitary operator on this vector space.  An example of such a gate
is the controlled-NOT (or $\CNot$) gate, whose two inputs are usually called the control and target qubits.  The gate can be described as follows:
\begin{enumerate}
    \item if the control qubit is $\ket{0}$, the target qubit is not modified, and
    \item if the control qubit is $\ket{1}$, the NOT gate is applied to the target qubit.
\end{enumerate}
There is an interesting result that emphasises the importance of the $\CNot$ gate in quantum computation: any multiple-qubit gate
may be constructed using only $\CNot$ gates and single-qubit gates~\cite{BBCDMSSSW1995:ElementaryGates}.

In a similar fashion we can extend any single-qubit gate to a controlled two-qubit gate.  For example, the controlled-$\Rot{\theta}$ gate works as follows:
\begin{enumerate}
    \item if the control qubit is $\ket{0}$, the target qubit is not modified, and
    \item if the control qubit is $\ket{1}$, the $R_\theta$ gate is applied to the target qubit.
\end{enumerate}
In other words, the gate leaves all of the computational basis states unchanged, except for
$\ket{1}\ket{1}$, which it maps to $e^{i\theta}\ket{1}\ket{1}$.

\end{section}

\end{chapter}

\begin{chapter}{Introduction To Quantum Algorithms}
\label{chap:IntroQuantumAlgs}

An algorithm describes a way to solve a particular problem.  For example, to solve the problem of dividing one number into
another, we could use the algorithm of long division, which consists of many steps that are repeated until we obtain the quotient
and remainder.  In this section we will present several problems, and describe ways to solve them that involve preparing specific
quantum states and applying to them some of the quantum gates defined in \chapref{chap:IntroQuantumComp}.  By examining and
measuring the output of certain sequences of quantum gates, we can solve a variety of problems.

The quantum algorithms that we present in this chapter are the main tools that we will use in later chapters to analyse classical
public key cryptosystems in a quantum setting.  As we will see later, many of the cryptosystems we use today are less secure
against attacks with a quantum computer since the problems on which these systems are based can be solved in polynomial time with
quantum algorithms from this chapter.

\begin{section}{Deutsch's Problem}\label{sec:IntroQuantumAlgs:DP}

Consider the following problem, posed in~\cite{D1985:DeutschQuantumTheory}:
\begin{problem}[Deutsch's Problem (DP)]\label{prob:IntroQuantumAlgs:DP}
Given a function $f \colon \{0,1\} \rightarrow \{0,1\}$ determine $f(0) \oplus f(1)$ using only a
single evaluation of the function $f$, where $\oplus$ is the componentwise XOR operation.
\end{problem}
In other words, we wish to determine with a single evaluation of $f$ whether or not $f$ is a
constant function: if $f(0) \oplus f(1) = 0$ then $f(0) = f(1)$ so $f$ is constant, and if $f(0)
\oplus f(1) = 1$ then $f(0) \neq f(1)$ so $f$ is not constant.

If we consider this problem classically, it is impossible to solve: we can determine $f(0)$ or $f(1)$, but without knowing both we cannot solve the problem.
(In fact, we cannot determine any information whatsoever that would help us to guess the solution correctly with probability greater than $\frac{1}{2}$.)
However, if we consider the problem in a quantum setting and we are given a way to reversibly compute $f$, we can solve it. The solution originally proposed by
Deutsch in~\cite{D1985:DeutschQuantumTheory} was modified and improved slightly in~\cite{CEMM1998:AlgorithmsRevisited} and it is this modified solution that we
present here.

To perform a ``quantum version'' of $f$, we will use an additional qubit (since for a constant function $f$, the mapping $\ket{x} \longmapsto \ket{f(x)}$ is
not reversible).   A typical choice for a reversible implementation of $f$ is the two-qubit unitary operator $\Uf$ which performs the transformation
\[
    \ket{x}\ket{y} \longmapsto \ket{x}\ket{y \oplus f(x)}
\]
for $x,y \in \{0,1\}$.

Suppose that we initialise the second qubit to the state $\frac{1}{\sqrt{2}}\paren{\ket{0} -
\ket{1}}$.
When we apply $\Uf$ to the qubits, by the linearity of quantum operators
as discussed above,
\begin{align*}
    \Uf \Blb \ket{x} \tfrac{1}{\sqrt{2}} \blb \ket{0} - \ket{1} \brb \Brb
        &= \ket{x}\tfrac{1}{\sqrt{2}}\blb \ket{0 \oplus f(x)} - \ket{1 \oplus f(x)}\brb \\
        &= \ket{x}(-1)^{f(x)}\tfrac{1}{\sqrt{2}}\blb\ket{0} - \ket{1}\brb\\
        &= \lb(-1)^{f(x)}\ket{x}\rb\tfrac{1}{\sqrt{2}}\blb\ket{0} - \ket{1}\brb.
\end{align*}

Therefore, if we also initialise the first qubit to the state $\frac{1}{\sqrt{2}}\paren{\ket{0} +
\ket{1}}$ before applying $\Uf$ we obtain
\begin{align*}
    &\Uf \Blb \tfrac{1}{\sqrt{2}}\blb\ket{0} + \ket{1}\brb\tfrac{1}{\sqrt{2}}\blb\ket{0} - \ket{1}\brb \Brb \\
        &= \tfrac{1}{\sqrt{2}}(-1)^{f(0)}\ket{0}\tfrac{1}{\sqrt{2}}\blb\ket{0} - \ket{1} \brb + \tfrac{1}{\sqrt{2}}(-1)^{f(1)}\ket{1}\tfrac{1}{\sqrt{2}}\blb \ket{0} - \ket{1}
        \brb\\
        &= \tfrac{1}{\sqrt{2}}\lb (-1)^{f(0)}\ket{0} + (-1)^{f(1)}\ket{1} \rb\tfrac{1}{\sqrt{2}}\blb \ket{0} - \ket{1} \brb.
\end{align*}

Now we apply the Hadamard gate to the first qubit above:  
\begin{align*}
    &\Had \Blb \tfrac{1}{\sqrt{2}}\paren{(-1)^{f(0)}\ket{0} + (-1)^{f(1)}\ket{1}} \Brb\\
        &= \tfrac{1}{2}(-1)^{f(0)}\blb \ket{0} + \ket{1} \brb + \tfrac{1}{2}(-1)^{f(1)}\blb \ket{0} - \ket{1} \brb \\
        &= \tfrac{1}{2}\lb (-1)^{f(0)} + (-1)^{f(1)} \rb \ket{0} + \tfrac{1}{2}\lb (-1)^{f(0)} - (-1)^{f(1)} \rb \ket{1} \\
        &= (-1)^{f(0)} \ket{ f(0) \oplus f(1) }.
\end{align*}
Apart from the global ``phase'' of $(-1)^{f(0)}$ that precedes it, the qubit's state is the correct
solution to the problem.  Luckily, the laws of quantum physics tell us that the global phase will
not affect the outcome of any measurement we perform on the state, and so we can simply measure
this qubit and recover the solution $f(0) \oplus f(1)$.

We can summarise the quantum algorithm for Deutsch's Problem as follows:
\begin{algorithm}[Solution To DP]\label{alg:IntroQuantumAlgs:DP}
\begin{algone}
    \item Begin with two qubits initialised to the states $\tfrac{1}{\sqrt{2}}\paren{\ket{0} + \ket{1}}$
    and $\tfrac{1}{\sqrt{2}}\paren{\ket{0} - \ket{1}}$.
    \item Apply the two-qubit quantum gate $\Uf$ to the system.
    \item Apply the Hadamard gate $\Had$ to the first qubit.
    \item Measure the first qubit and obtain the integer $y$.
    \item Return $y$.
\end{algone}
\end{algorithm}

\end{section}

\begin{section}{The Hidden Subgroup Problem}\label{sec:IntroQuantumAlgs:HSP}

Deutsch's problem is actually a special case of a more general problem:
\begin{problem}[The Hidden Subgroup Problem (HSP)]\label{prob:IntroQuantumAlgs:HSP}
Let $f$ be a function from a finitely generated group $G$ to a finite set $X$ such that $f$ is
constant on the cosets of a subgroup $K$ of $G$ and distinct on each coset.  Given a quantum
network for evaluating $f$ (namely $\Uf \colon \ket{x}\ket{y} \rightarrow \ket{x}\ket{y \oplus
f(x)}$) find a generating set for $K$.
\end{problem}

In Deutsch's problem we had $G = \Z_2 = \{ 0,1 \}$.  Using the language of HSP,
\begin{enumerate}
    \item if $f$ is a constant function, we have $K = \{0,1\}$ since $f$ is constant on $K$ (and there is only one coset of
$K$, namely $K$ itself); and
    \item if $f$ is not constant, then we have $K = \{0\}$, since $f$ is constant on $K$ and on $K + 1 = \{1\}$, and distinct on these two
cosets.
\end{enumerate}
Thus to solve Deutsch's problem we wish to determine whether $K=\{0,1\}$ or $K=\{0\}$.

There are many other problems that can be thought of as special cases of HSP.  We list two important examples below, and for many more,
see~\cite{M1999:PhDThesis}.

\begin{problem}[The Order Finding Problem (OFP)]\label{prob:IntroQuantumAlgs:OFP}
Given an element $a$ of a finite group $H$, find $r$, the order of $a$.
\end{problem}

Let $f \colon \Z \rightarrow H$ be defined by $f(x) = a^x$.  Then note that
\begin{align*}
  f(x) = f(y) &\Longleftrightarrow a^x = a^y \\
              &\Longleftrightarrow a^{x-y} = 1 \\
              &\Longleftrightarrow x-y \in \{ t \cdot r: t \in \Z \}.
\end{align*}
That is, $f(x) = f(y)$ if and only if $x$ and $y$ are in the same coset of the hidden subgroup $K = r\Z$ of $\Z$.  By finding a generator for $K$ we can
determine $r$.  Thus OFP is a special case of HSP.

\begin{problem}[The Discrete Logarithm Problem (DLP)]\label{prob:IntroQuantumAlgs:DLP}
Given an element $a$ of a finite group $H$ and $b = a^k$, find $k$.  (This $k$ is called the discrete logarithm of $b$ to the base $a$.)
\end{problem}

Suppose the order of $a$ is $r$.  Let $f \colon \Z_r \times \Z_r \rightarrow H$ be defined by
$f(x_1,x_2) = a^{x_1}b^{x_2}$.  Then note that
\begin{align*}
    f(x_1,x_2) = f(y_1,y_2) &\Longleftrightarrow a^{x_1}b^{x_2} = a^{y_1}b^{y_2} \\
                            &\Longleftrightarrow a^{x_1-y_1}b^{x_2-y_2} = 1 \\
                            &\Longleftrightarrow a^{x_1-y_1}a^{k(x_2-y_2)} = 1 \\
                            &\Longleftrightarrow (x_1-y_1) + k(x_2-y_2) = 0 \quad\text{(in $\Z_r$)} \\
                            &\Longleftrightarrow (x_1,x_2) - (y_1,y_2) \in \{ (-tk,t) \st t \in \Z_r \}.
\end{align*}
That is, $f(x_1,x_2) = f(y_1,y_2)$ if and only if $(x_1,x_2)$ and $(y_1,y_2)$ are in the same coset
of the hidden subgroup $K = \langle( -k, 1 )\rangle$ of $\Z_r \times \Z_r$. By finding a generator
for $K$ we can determine $k$.  Thus DLP is a special case of HSP.

We mention these two problems in particular because as we will see in the remainder of this thesis,
if we have algorithms to solve these problems efficiently, we can break many of the classical
cryptosystems that are in widespread use today.  There do exist polynomial-time quantum algorithms
that solve these problems, and we will discuss these algorithms later.  In fact, there exist
efficient quantum algorithms that solve the general HSP when the group $G$ is Abelian, as described
in~\cite{M1999:PhDThesis}.  Some work has been done to design algorithms for HSP in non-Abelian
groups, although success has been limited.  For example, an efficient algorithm was presented
in~\cite{EH2000:NoncommutativeHSP} that is able to determine some information about the generator
of a hidden subgroup in a dihedral group, but there is no known way to recover the subgroup in
polynomial time from this information.  In~\cite{IMS2001:NonAbelianHSP} some special cases of the
problem were solved in non-Abelian groups, but the general case still remains open.

We now introduce one of the most important ingredients in many quantum algorithms: the Quantum
Fourier Transform.

\end{section}

\begin{section}{The Quantum Fourier Transform}\label{sec:IntroQuantumAlgs:QFT}

The Quantum Fourier Transform (QFT) provides a way to estimate parameters that are encoded in a
specific way in the phases and amplitudes of quantum states.  We will begin with a small
three-qubit example, and then define the general QFT.

Assume $a$ is an integer, $0 \leq a < 8$.  Now suppose that we are given the three-qubit state
\[
    \blb \ket{0} + e^{2\pi i \frac{a}{2} }\ket{1} \brb \blb \ket{0} + e^{2\pi i \frac{a}{4} }\ket{1} \brb \blb \ket{0} + e^{2\pi i\frac{a}{8}}\ket{1} \brb
\]
(ignoring the normalisation factors) and we wish to find $a$.

We can write $a = 4a_2 + 2a_1 + a_0$ where each $a_j \in \{0, 1 \}$ and then rewrite the state as
\[
    \blb \ket{0} + e^{2\pi i(\frac{a_0}{2})}\ket{1} \brb \blb \ket{0} + e^{2\pi i(\frac{2a_1 + a_0}{4})}\ket{1} \brb \blb \ket{0} + e^{2\pi i(\frac{4a_2 + 2a_1 + a_0}{8})}\ket{1} \brb.
\]

Recall the Hadamard gate $\Had$ from \secref{sec:IntroQuantumComp:SingleQubitGates} and note that
ignoring normalisation factors we could equivalently define it by the map
\[
    \ket{x} \longrightarrow \ket{0} + e^{2\pi i \frac{x}{2} } \ket{1}
\]
for $x \in \{ 0, 1 \}$. So if we apply $\Had^{-1} = \Had$ to the first qubit, we obtain $\ket{a_0}$.

Next, we will try to determine $a_1$.  Consider the following two cases:
\begin{enumerate}
\item If $a_0 = 0$, the second qubit is actually in the state $\ket{0} + e^{2\pi i(\frac{a_1}{2})}\ket{1}$.
\item If $a_0 = 1$, the second qubit is in the state $\ket{0} + e^{2\pi i(\frac{2a_1 + 1}{4})}\ket{1}$. In this case,
if we apply a $\Rot{-\frac{\pi}{2}}$ gate to the qubit, we get
\begin{align*}
      &\ket{0} + e^{2\pi i(\frac{2a_1 + 1}{4})}e^{-i\frac{\pi}{2}}\ket{1}\\
    &= \ket{0} + e^{2\pi i(\frac{a_1}{2})}\ket{1}.
\end{align*}
\end{enumerate}
Thus we will obtain a common state if we can decide, based on the state of the first qubit, whether or not to apply a $\Rot{-\frac{\pi}{2}}$ gate to the second
qubit. In other words, we wish to apply a controlled-$\Rot{-\frac{\pi}{2}}$ gate to the first and second qubits.  After this gate has been applied, our second
qubit will be in the state $\ket{0} + e^{2\pi i(\frac{a_1}{2})}\ket{1}$, and we can apply an $\Had$ gate to the second qubit to obtain the state $\ket{a_1}$.

Similarly, if we now apply a controlled-$\Rot{-\frac{\pi}{4}}$ to the first and third qubits, the
third qubit will be in the state $\ket{0} + e^{2\pi i(\frac{2a_2 + a_1}{4})}\ket{1}$.  Then,
applying a controlled-$\Rot{-\frac{\pi}{2}}$ to the second and third qubits will put the third
qubit into the state $\ket{0} + e^{2\pi i(\frac{a_2}{2})}\ket{1}$.  Finally, we can apply an $\Had$
gate to the third qubit to obtain $\ket{a_2}$.

The sequence of gates we have described, illustrated in \figref{fig:IntroQuantumAlgs:3qubitQFT},
implements the transformation
\[
    \blb \ket{0} + e^{2\pi i \frac{a}{2} }\ket{1} \brb \blb \ket{0} + e^{2\pi i \frac{a}{4} }\ket{1} \brb \blb \ket{0} + e^{2\pi i\frac{a}{8}}\ket{1} \brb
    \longrightarrow \ket{a_0}\ket{a_1}\ket{a_2}
\]
and by reversing the order of the qubits, we obtain the $3$-qubit state
\[
    \ket{a_2}\ket{a_1}\ket{a_0}\;=\;\ket{a}.
\]

\begin{figure}[ht]
\begin{center}
\fbox{\includegraphics{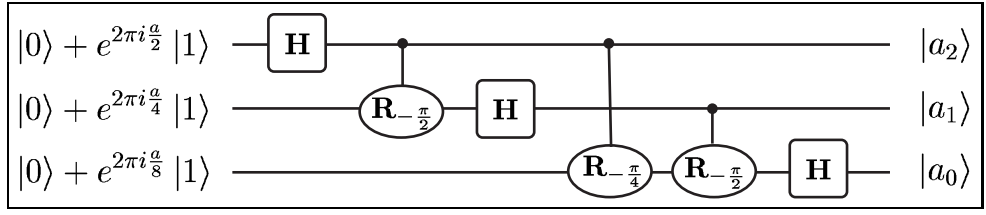}}
\caption{A 3-qubit Quantum Fourier Transform}
\label{fig:IntroQuantumAlgs:3qubitQFT}
\end{center}
\end{figure}

We can generalise this quantum circuit so that if $a$ is an integer with $0 \leq a < 2^n$ and $a =
2^{n-1}a_{n-1} + 2^{n-2}a_{n-2} + \cdots + 2 a_1 + a_0$ for $a_j \in \{0,1\}$, we can start with
the $n$-qubit state
\begin{equation}\label{equ:IntroQuantumAlgs:QFTStartState}
    \blb \ket{0} + e^{2\pi i \frac{a}{2}}\ket{1} \brb \blb \ket{0} + e^{2\pi i \frac{a}{4}}\ket{1} \brb \cdots \blb \ket{0} + e^{2\pi i \frac{a}{2^n}}\ket{1} \brb
\end{equation}
and transform it into the state
\[
    \ket{a_{n-1}}\ket{a_{n-2}} \cdots \ket{a_1}\ket{a_0}\;=\;\ket{a}.
\]

As pointed out in~\cite{M1999:PhDThesis}, the start state~\eqref{equ:IntroQuantumAlgs:QFTStartState} can be rewritten as
\[
    \sum_{x=0}^{2^n-1} e^{2\pi i x \frac{a}{2^n}} \ket{x}.
\]
The states of this form, for $a=0,1, \ldots, 2^n-1$ are called the Fourier basis states.  The transformation we have discussed in this section therefore maps a
state in the Fourier basis to its corresponding state in the computational basis.  We call this transformation the \emph{inverse} QFT; the QFT therefore maps
states from the computational basis to the Fourier basis.

We can define the QFT more generally as follows:
\begin{definition}\label{def:IntroQuantumAlgs:QFT}
For any integer $m > 1$, the \tbd{$m$-bit Quantum Fourier Transform} $\QFT{m}$ acts on the vector
space generated by the states
\[
    \ket{0}, \ket{1}, \ldots, \ket{m-1}
\]
and maps
\[
    \ket{a} \longmapsto \frac{1}{\sqrt{m}}\sum_{x=0}^{m-1} e^{2\pi i x \frac{a}{m}} \ket{x}.
\]
\end{definition}

We have described an efficient implementation of $\QFT{m}$ in the case where $m$ is a power of $2$,
as originally presented in~\cite{C1994:QFTDefinition}. There are also efficient exact
implementations of $\QFT{m}$ in the case where the prime factors of $m$ are distinct and in
$\bigO{\log m}$~\cite{S1994:DLFactoring}, or more generally when the prime factors are not
necessarily distinct but still in $\bigO{\log m}$~\cite{C1994:FourierTransforms}.  It was shown
in~\cite{K1995:MeasurementAbelianStabilizer} that for arbitrary values of $m$ we can approximate
$\QFT{m}$ efficiently; very recently, it was shown that we can in fact implement $\QFT{m}$ exactly
for arbitrary values of $m$~\cite{MZ2003:ExactQFT}.

We now make some important observations about the QFT.  First, if we start in the state $\ket{0}$
and apply $\QFT{m}$ we obtain the state
\begin{align*}
    &\frac{1}{\sqrt{m}}\sum_{x=0}^{m-1} e^{2\pi i x \frac{0}{m}} \ket{x}\\
    &= \frac{1}{\sqrt{m}}\sum_{x=0}^{m-1} \ket{x}
\end{align*}
which is an equally weighted superposition of the $m$ computational basis states.

Also, given the state $\ket{\phi} = \frac{1}{\sqrt{m}}\sum_{x=0}^{m-1} e^{2\pi i x \omega} \ket{x}$
where $\omega = \frac{a}{m}$ for some integer $a$, then by definition, if we apply the inverse QFT
to $\ket{\phi}$ we will obtain the state $\ket{a}$ and we can recover $\omega$ exactly.  If, on the
other hand, $\omega$ is any real number, we can still use the inverse QFT to obtain an estimate of
$\omega$, and we can bound the distance of this estimate from the true value of $\omega$. More
precisely:

\begin{theorem}\label{thm:IntroQuantumAlgs:QFTEstimateBound}
Given an integer $m > 0 $ and the state $\ket{\phi} = \frac{1}{\sqrt{m}}\sum_{x=0}^{m-1} e^{2\pi i
x \omega} \ket{x}$, where $\omega$ is an any real number, applying $\IQFT{m}$ to $\ket{\phi}$ and
measuring the result yields an integer $y$ satisfying the following conditions:
\begin{itemize}
\item If $\omega = \frac{a}{m}$ for some integer $a$, then with probability $1$, $y = a$.
\item Otherwise, with probability at least $\frac{8}{\pi^2}$, $\abs{\frac{y}{m}-\omega} \leq \frac{1}{m}$.
\end{itemize}
\end{theorem}

For a proof of this theorem, see~\cite{C2003:GeneralizedQFTsForQPE}.

\end{section}

\begin{section}{Solving A Special Case Of The Hidden Subgroup
Problem}\label{sec:IntroQuantumAlgs:IHSP}\markright{\thesection. \ SOLVING A SPECIAL CASE OF HSP}

We first consider the task of solving HSP where $G = \Z$; that is, $f$ is a function from $\Z$ to
some finite set $X$, and $f(x) = f(y)$ if and only if $x-y \in r\Z$ for some fixed (unknown)
integer $r$.  We will call this special case of HSP the Integer Hidden Subgroup Problem (IHSP).

We choose an integer $n \geq \log\abs{X}$ and an integer $m$ which is a power of $2$, and we are
given the unitary operator $\Uf$ which acts on $\Hilbert{m} \tensor \Hilbert{n}$ and maps
$\ket{x}\ket{y} \longmapsto \ket{x}\ket{y \oplus f(x)}$.  We can then implement the following
algorithm, which will form the core of an algorithm to solve IHSP:
\begin{algorithm}[Core Of Solution To IHSP]\label{alg:IntroQuantumAlgs:IHSP}
\begin{algone}
\item Start in the state $\ket{0}\ket{0} \in \Hilbert{m} \tensor \Hilbert{n}$.
\item Apply $\QFT{m}$ to the first register.\label{step:IntroQuantumAlgs:IHSP:QFT}
\item Apply $\Uf$ to the system.\label{step:IntroQuantumAlgs:IHSP:Uf}
\item Apply $\IQFT{m}$ to the first register.
\item Measure the first register to obtain the integer $y$.
\item Return $y$.
\end{algone}
\end{algorithm}

We now have the following well-known result (see for example~\cite{NC2000:Nielsen}):

\begin{proposition}\label{prop:IntroQuantumAlgs:IHSPInterference}
After \stepref{step:IntroQuantumAlgs:IHSP:Uf} of \algref{alg:IntroQuantumAlgs:IHSP}, our system is
in the state
\[
    \ket{\phi} = \frac{1}{\sqrt{rm}}\sum_{k=0}^{r-1} \lb \sum_{x=0}^{m-1}e^{2\pi i x \frac{k}{r}} \ket{x} \rb \ket{\psi_k}
\]
where $\smash{\ket{\psi_k} = \frac{1}{\sqrt{r}}\sum\limits_{j=0}^{r-1} e^{-2\pi i j \frac{k}{r}}
\ket{f(j)}}$.
\end{proposition}

\begin{proof}
After \stepref{step:IntroQuantumAlgs:IHSP:QFT} of \algref{alg:IntroQuantumAlgs:IHSP}, our system is
in the state $\frac{1}{\sqrt{m}}\sum_{x=0}^{m-1}\ket{x}\ket{0}$, and applying $\Uf$ in
\stepref{step:IntroQuantumAlgs:IHSP:Uf} produces the state
$\frac{1}{\sqrt{m}}\sum_{x=0}^{m-1}\ket{x}\ket{f(x)}$. We will show that this state is in fact
equal to $\ket{\phi}$.
 Note that
\begin{align}
    \ket{\phi} &= \frac{1}{r\sqrt{m}}\sum_{k=0}^{r-1} \lb \sum_{x=0}^{m-1} e^{2\pi i x \frac{k}{r}} \ket{x} \rb \sum_{j=0}^{r-1} e^{-2\pi i j \frac{k}{r}} \ket{f(j)}\notag \\
               &= \frac{1}{r\sqrt{m}}\sum_{x=0}^{m-1} \sum_{j=0}^{r-1} \lb \sum_{k=0}^{r-1} e^{2\pi i \frac{k}{r}(x-j)} \rb \ket{x} \ket{f(j)}. \label{equ:IntroQuantumAlgs:IHSPInterference}
\end{align}

Now fix $x$ and $j$, and consider the coefficients $c_{k} = e^{2\pi i \frac{k}{r}(x-j)}$ for $0
\leq k < r$.  There are two cases:
\begin{enumerate}
\item If $x \equiv j \pmod{r}$ then $\frac{k}{r}(x-j)$ is an integer for all $k$, so each of the $c_{k}$ is $1$,
and the sum of the $c_{k}$ is $r$.  In this case, we say that there is \emph{constructive
interference} between the coefficients.
\item If $x \not\equiv j \pmod{r}$, consider $c_{1} = e^{2 \pi i \frac{1}{r}(x-j)}$.  Note that $c_{1}^{\;r} - 1 = 0$, and since
$c_{1}^{\;r} - 1 = (c_{1} - 1)(1 + c_{1} + \cdots + c_{1}^{\;r-1})$ and $c_{1} - 1 \neq 0$, it must
be true that $1 + c_{1} + \cdots + c_{1}^{\;r-1} = 0$. Furthermore, $c_{1}^{\;k} = c_{k}$ for $0
\leq k < r$, so the sum of the $c_{k}$ is $0$.  In this case we say that there is \emph{destructive
interference} between the coefficients.
\end{enumerate}

Thus from~\eqref{equ:IntroQuantumAlgs:IHSPInterference}, we see that
\begin{align*}
  \ket{\phi} &= \frac{1}{r\sqrt{m}}\sum_{x=0}^{m-1} \sum_{j=0}^{r-1} \lb \sum_{k=0}^{r-1} c_{k} \rb \ket{x} \ket{f(j)} \\
             &= \frac{1}{r\sqrt{m}}\sum_{x=0}^{m-1} \ket{x} r \ket{f(x \bmod{r})}\\
             &= \frac{1}{\sqrt{m}}\sum_{x=0}^{m-1} \ket{x} \ket{f(x)}
\end{align*}
since by the periodicity of $f$, $f(x) = f(x \bmod{r})$.  Therefore the result is proven.
\end{proof}

So after \stepref{step:IntroQuantumAlgs:IHSP:Uf} our system is in the state
$\frac{1}{\sqrt{m}}\sum_{k=0}^{r-1} \blb \sum_{x=0}^{m-1} e^{2\pi i x \frac{k}{r}} \ket{x} \brb
\ket{\psi_k}$. Letting $\omega = \frac{k}{r}$, by \thmref{thm:IntroQuantumAlgs:QFTEstimateBound}
and by linearity we can see that with probability at least $\frac{8}{\pi^2}$, applying $\IQFT{m}$
to the first register and measuring the result yields an integer $y_k$ such that
$\abs{\frac{y_k}{m} - \frac{k}{r}} \leq \frac{1}{m}$, where $k$ is chosen at random from $\{0, 1,
\ldots, r-1 \}$.

We will now make use of a theorem from the theory of continued fractions. Given any real number
$\lambda$, we can use the theory of continued fractions to compute a sequence of rational numbers
called ``convergents'' that approximate $\lambda$ with increasing precision. If $\lambda$ is
positive and rational (say $\lambda = \frac{x}{m}$ for positive integers $x$ and $m$) we have the
following result (see for example~\cite{R1993:ElementaryNumberTheory}):
\begin{theorem}\label{thm:IntroQuantumAlgs:ContinuedFraction}
Let $x$, $m$, $k$, and $r$ be positive integers, with
\[
    \abs{ \frac{x}{m} - \frac{k}{r} } < \frac{1}{2r^2}
\]
Then $\frac{k}{r}$ appears as a convergent in the continued fraction expansion of $\frac{x}{m}$.
\end{theorem}

There exist efficient algorithms to compute the continued fraction expansion of $\frac{x}{m}$, as
described for example in~\cite{K1994:NumberTheoryAndCryptography}.  Clearly there exists at most
one fraction $\frac{a}{b}$ with $b \leq r$ such that $\abs{ \frac{x}{m} - \frac{a}{b} } \leq
\frac{1}{2r^2}$, so when we find such a convergent $\frac{a}{b}$, we know that $\frac{a}{b} =
\frac{k}{r}$ and we can stop computing convergents.  The continued fractions algorithms guarantee
that we will have to compute at most $\bigO{\log{m}}$ convergents before we can stop.

 So by setting $x$ equal to our measurement output $y_k$ and running
these algorithms, provided we have chosen $m > 2r^2$, we can efficiently find a fraction
$\frac{a}{b} = \frac{k}{r}$.

Combining \algref{alg:IntroQuantumAlgs:IHSP} and \thmref{thm:IntroQuantumAlgs:ContinuedFraction},
we obtain an efficient probabilistic quantum algorithm to solve IHSP if we have a bound on the size
of $r$:
\begin{algorithm}[Solution To IHSP When $r$ Is Bounded]\label{alg:IntroQuantumAlgs:IHSPBounded}
\begin{algone}
\item Choose an integer $m > 2r^2$.
\item Repeat \algref{alg:IntroQuantumAlgs:IHSP} two times to obtain two values $y_{k_1}, y_{k_2}$.
\item Use the continued fractions algorithm to obtain fractions $\frac{a_1}{b_1}, \frac{a_2}{b_2}$ such
that\linebreak $b_1, b_2 \leq \sqrt{\frac{m}{2}}$ and
\begin{align*}
    \abs{\frac{y_{k_1}}{m} - \frac{a_1}{b_1}}, \abs{\frac{y_{k_2}}{m} - \frac{a_2}{b_2}} &\leq \frac{1}{m}.
\end{align*}
If two such fractions cannot be found, return
FAIL.\label{step:IntroQuantumAlgs:IHSPBounded:ContinuedFraction}
\item Let $t = \lcm(b_1,b_2)$.  If $t > \sqrt{\frac{m}{2}}$, return FAIL.
\item If $f(0) \neq f(t)$, return FAIL.\label{step:IntroQuantumAlgs:IHSPBounded:TestMultipleOfR}
\item Return $t$.
\end{algone}
\end{algorithm}

\begin{theorem}\label{thm:IntroQuantumAlgs:IHSPBoundedSuccess}
\algref{alg:IntroQuantumAlgs:IHSPBounded} finds the correct value of $r$ with probability at least
$\frac{32}{\pi^4}$.  If it does not return FAIL, it returns a multiple of $r$.
\end{theorem}

\begin{proof}
We run \algref{alg:IntroQuantumAlgs:IHSP} twice independently, obtaining results $y_{k_1}$ and
$y_{k_2}$.  The values $\frac{y_{k_1}}{m}$ and $\frac{y_{k_2}}{m}$ are estimates of $\frac{k_1}{r}$
and $\frac{k_2}{r}$, respectively.

By \thmref{thm:IntroQuantumAlgs:QFTEstimateBound}, $\abs{\frac{y_{k_i}}{m} - \frac{k_i}{r}} \leq
\frac{1}{m}$ with probability at least $\frac{8}{\pi^2}$ for $i=1$, and independently for $i=2$.
The probability that the inequality is satisfied for both $i=1$ and $i=2$ is therefore at least
$\frac{64}{\pi^4}$. If this is the case, then since $m>2r^2$, by
\thmref{thm:IntroQuantumAlgs:ContinuedFraction} the continued fractions algorithm will successfully
find fractions $\frac{a_1}{b_1}$ and $\frac{a_2}{b_2}$ that satisfy the conditions in
\stepref{step:IntroQuantumAlgs:IHSPBounded:ContinuedFraction}. Thus with probability at least
$\frac{64}{\pi^4}$ we will have found $\frac{a_i}{b_i} = \frac{k_i}{r}$ for $i=1,2$.

Now note that $\gcd(k_i,r)$ is not necessarily $1$ because $\frac{a_i}{b_i}$ could be the reduced
form of $\frac{k_i}{r}$.  It is true however that $b_i = \frac{r}{\gcd(k_i,r)}$, so $\lcm(b_1,b_2)
= \frac{r}{\gcd(k_1,k_2,r)}$.  If, whenever we have measured a $0$ we replace it by $r$ (for
mathematical convenience) we can treat $k_1$ and $k_2$ as having been selected uniformly at random
from the integers between $1$ and $r$; so $k_1$ and $k_2$ are coprime with probability at least
$\frac{1}{2}$~\cite{CEMM1998:AlgorithmsRevisited}. In this case, $\lcm(b_1,b_2) = r$ as desired.
Thus the algorithm finds the correct value of $r$ with probability at least
$\parfrac{64}{\pi^4}\parfrac{1}{2} = \frac{32}{\pi^4}$.

The final test in \stepref{step:IntroQuantumAlgs:IHSPBounded:TestMultipleOfR} checks to make sure
that $t$ is a multiple of $r$.  Thus the algorithm either returns FAIL or a multiple of $r$.
\end{proof}

If we do not have a bound on $r$ to begin with, we can guess at an initial value of $m$, and repeat
\algref{alg:IntroQuantumAlgs:IHSPBounded} three times, say. If all three repetitions return FAIL,
we can assume that our $m$ is not large enough, double it, and try again.  Eventually, we will
obtain an $m > 2r^2$, and with high probability, one of the repetitions of the algorithm for that
value of $m$ will succeed.  The number of iterations of this process that are required to ensure $m
> 2r^2$ is polynomial in $\log{r}$.

Thus we have an efficient quantum algorithm to solve IHSP.  It should be noted that if $f$ is given
to us as a ``black box'', there is no efficient classical algorithm to solve IHSP: it is a problem
for which an efficient quantum algorithm exists but for which no known efficient classical
algorithm exists.

\end{section}

\begin{section}{Solving The Order Finding Problem}\label{sec:IntroQuantumAlgs:OFP}

Given an element $a$ of a group $H$, to solve OFP we must compute $r$, the order of $a$. As
illustrated in \probref{prob:IntroQuantumAlgs:OFP}, OFP is a special case of IHSP where $f(x) =
a^x$.  Thus, the algorithm we have described in \secref{sec:IntroQuantumAlgs:IHSP} allows us to
solve OFP in polynomial time. A polynomial-time quantum algorithm to solve OFP was first proposed
in~\cite{S1994:DLFactoring}.

\end{section}

\begin{section}{Solving The Factoring Problem}\label{sec:IntroQuantumAlgs:Factoring}

In this section, we describe how to find a non-trivial factor of an integer in polynomial time
using a quantum computer. Given a polynomial-time algorithm to solve OFP we can use a classical
reduction to develop an algorithm that allows us to find a non-trivial factor of an integer in
polynomial time. This reduction was first described by Miller in~\cite{M1976:RiemannPrimality}. The
idea of solving the factoring problem using the polynomial-time quantum algorithm for OFP and
Miller's reduction was proposed in~\cite{S1994:DLFactoring}.  The resulting quantum factoring
algorithm has become the most famous quantum algorithm.

We present a sketch of Miller's reduction. Suppose we wish to factor a positive integer $n$.  First, we assume that $n$ is odd, since factors of $2$ are easy
to detect.  We also assume that $n$ is not a prime power, since there are efficient classical algorithms to determine the factors of $n$ in this case.
Consider the following algorithm:

\begin{algorithm}[Finding A Non-Trivial Factor]\label{alg:IntroQuantumAlgs:Factoring}
\begin{algone}
\item Choose an integer $a$ at random from $\{ 0, 1, \ldots, n-1 \}$.
\item Let $s = \gcd(a,n)$.  If $s > 1$ then return $s$.  Otherwise, $a \in \Z_n^*$.  (Recall that $\Z_n^*$ is the multiplicative group of
all integers (modulo $n$) that are coprime with $n$.)
\item Apply \algref{alg:IntroQuantumAlgs:IHSPBounded} three times with $m > 2n^2$ to attempt to determine the order of $a$ in $\Z_n^*$.
If all three results are FAIL, return FAIL.
Otherwise, take $r$ to be the minimum non-FAIL output.\label{step:IntroQuantumAlgs:Factoring:FindOrder}
\item If $r$ is odd, return FAIL.
\item Let $t = \gcd( a^{r/2}-1, n )$.  If $t = 1$, return FAIL.
\item Return $t$.
\end{algone}
\end{algorithm}

\begin{theorem}
\algref{alg:IntroQuantumAlgs:Factoring} correctly returns a non-trivial factor of $n$ with probability at least $\frac{1}{3}$.
\end{theorem}

\begin{proof}
For any integer $a \in \{ 0, 1, \ldots, n-1 \}$ which is coprime with $n$ and whose order in $\Z_n^*$ is $r$, we know that
\begin{align*}
    a^r &\equiv 1 \pmod{n} \\
    a^r - 1 &\equiv 0 \pmod{n} \\
    (a^{r/2} - 1)(a^{r/2} + 1) &\equiv 0 \pmod{n} \quad\mbox{(if $r$ is even)}.
\end{align*}
Since $r$ is the order of $a$, we know that $a^{r/2} - 1 \not\equiv 0 \pmod{n}$.  So if
\begin{enumerate}
\item $r$ is even, and
\item $a^{r/2} + 1 \not\equiv 0 \pmod{n}$,
\end{enumerate}
then $t = \gcd( a^{r/2}-1, n )$ must be a non-trivial factor of $n$.

We now show that a randomly selected $a$ satisfies both of these conditions with probability at
least $1-(\frac{1}{2})^{k-1}$, where $k$ is the number of distinct odd prime factors of $n$. Let $n
= \prod_{i=1}^k p_i^{e_i}$ where the $p_i$ are distinct odd primes, and let $r_i$ be the order of
$a$ in $\Z_{p_i^{e_i}}^*$. Then $r$ is the least common multiple of the $r_i$.  Consider the
multiplicity of $2$ in the prime factorisation of each $r_i$.
\begin{enumerate}
\item If each of these multiplicities is $0$ (ie. each $r_i$ is odd) then $r$ is odd.
\item If each of these multiplicities is larger than $0$ but they are all equal, then $r_i$ does not divide $\frac{r}{2}$
for any $i$, and thus it must be true that $a^{r/2} \equiv -1 \pmod{p_i^{e_i}}$ for each $i$.  Then
by the Chinese Remainder Theorem, $a^{r/2} \equiv -1 \pmod{n}$.
\item Otherwise, there is some $i$ for which $a^{r/2} \equiv 1 \pmod{p_i^{a_i}}$ and thus $a^{r/2} \not\equiv -1 \pmod{n}$.
\end{enumerate}

Thus a randomly selected $a$ will fail to satisfy both required conditions if and only if the
multiplicities of $2$ in the prime factorisations of the $r_i$ are all the same. By the Chinese
Remainder Theorem, there is a one-to-one correspondence between $\Z_n^*$ and the set $\{ (x_1,
\ldots, x_k): x_i \in \Z_{p_i^{e_i}}^*, 1 \leq i \leq k \}$. Thus selecting an $a$ at random from
$\Z_n^*$ is the same as selecting a $k$-tuple at random from the above set.  For each $i$,
$\Z_{p_i^{e_i}}^*$ is cyclic since $p_i$ is odd; so if we choose a random element $x_i$ with order
$r_i$, the probability of obtaining a particular multiplicity of $2$ in the prime factorisation of
$r_i$ is at most $\frac{1}{2}$.  Thus the probability of obtaining the same multiplicity for each
$i$ is at most $(\frac{1}{2})^{k-1}$. In other words, the probability of choosing an appropriate
$a$ is at least
\begin{equation}\label{equ:IntroQuantumAlgs:FactoringGoodAProb}
    1-\paren{\tfrac{1}{2}}^{k-1}.
\end{equation}

If such an $a$ is chosen, the algorithm will succeed in finding a non-trivial factor of $n$
provided that at least one of the applications of \algref{alg:IntroQuantumAlgs:IHSPBounded} is
successful in correctly determining $r$, the order of $a$ in $\Z_n^*$.  By
\thmref{thm:IntroQuantumAlgs:IHSPBoundedSuccess} each individual application of
\algref{alg:IntroQuantumAlgs:IHSPBounded} succeeds with probability at least $\frac{32}{\pi^4}$.
Thus the probability that at least one of them succeeds is
\begin{equation}\label{equ:IntroQuantumAlgs:FactoringSuccessfulOrderProb}
    1 - \left( 1 - \paren{\tfrac{32}{\pi^4}} \right)^3 > \tfrac{2}{3}.
\end{equation}

Combining~\eqref{equ:IntroQuantumAlgs:FactoringGoodAProb}
and~\eqref{equ:IntroQuantumAlgs:FactoringSuccessfulOrderProb}, we see that the probability that the
entire algorithm succeeds is at least
\[
    \left( 1 - \paren{\tfrac{1}{2}}^{k-1} \right) \paren{\tfrac{2}{3}}.
\]
Since we have assumed $n$ is not a prime power, $k \geq 2$.  Thus the probability of success is at least $\parfrac{1}{2} \parfrac{2}{3} > \frac{1}{3}.$
\end{proof}

By applying this algorithm recursively, we can split $n$ into its prime factors.  We therefore have a polynomial time probabilistic quantum algorithm to solve
the factoring problem.

The core of this quantum factoring algorithm has been successfully implemented on a small quantum
computer to attempt to factor the integer $15$ into its prime factors ($3$ and $5$).  Scientists
are still only able to tackle problems with small parameter sizes using the current implementations
of quantum computers, but the successful implementation of this and other quantum algorithms
indicates that the theory currently being developed can actually be applied to a physical
realisation of a quantum computer.

\end{section}

\begin{section}{Solving Another Special Case Of The Hidden Subgroup
Problem}\label{sec:IntroQuantumAlgs:PHSP}\markright{\thesection. SOLVING ANOTHER SPECIAL CASE OF
HSP}

Next consider the task of solving HSP where $G = \Z_{p} \times \Z_{p}$; that is, $f$ is a function
from $\Z_{p} \times \Z_{p}$ to some finite set $X$, and $f(x_1,x_2) = f(y_1,y_2)$ if and only if
$(x_1,x_2)$ and $(y_1,y_2)$ are in the same coset of some hidden subgroup $K$ (which is of size
$p$). We will call this special case of HSP the Prime Hidden Subgroup Problem (PHSP).  In this
section we will present a sketch of a well-known algorithm to solve this special case; the
algorithm is a generalisation of \algref{alg:IntroQuantumAlgs:IHSP} and can be found also
in~\cite{NC2000:Nielsen}, for example.

We choose an integer $n \geq \log{\abs{X}}$.  We use the natural generalisation of the definition
of $\Uf$; that is, $\Uf$ implements the unitary transformation
\[
    \ket{x}\ket{y}\ket{z} \longrightarrow \ket{x}\ket{y}\ket{z \oplus f(x,y)}.
\]
We also assume that we can implement $\QFT{p}$ exactly.  In practice we would likely use an
approximation of $\QFT{p}$, for example by performing $\QFT{2^l}$ where $2^l \approx p$; in this
case the algorithm still succeeds with high probability as formalised
in~\cite{HH1999:QuantumFourierSampling}.  However, by using the methods described
in~\cite{MZ2003:ExactQFT} we could instead implement $\QFT{p}$ exactly and subsequently obtain an
exact algorithm for PHSP.

Consider the following algorithm, which will form the core of an algorithm to solve PHSP:
\begin{algorithm}[Core Of Solution To PHSP]\label{alg:IntroQuantumAlgs:PHSP}
\begin{algone}
\item Start in the state $\ket{0}\ket{0}\ket{0}$ in $\Hilbert{p} \tensor \Hilbert{p} \tensor \Hilbert{n}$.
\item Apply $\QFT{p}$ to each of the first two registers.\label{step:IntroQuantumAlgs:PHSP:QFT}
\item Apply $\Uf$ to the system.\label{step:IntroQuantumAlgs:PHSP:Uf}
\item Apply $\IQFT{p}$ to each of the first two registers.
\item Measure the first two registers and output the ordered pair $(s,t)$.
\end{algone}
\end{algorithm}

Define the set $T = \left\{ (s,t) \st su + tv \equiv 0 \pmod{p} \mbox{ for every } (u,v) \in K
\right\}.$ Note that $\abs{T} = p$.  For each $(s,t) \in T$ define the state
\[
    \ket{\psi_{(s,t)}} = \frac{1}{\sqrt{p}}\sum_{(u,v) \in G / K} e^{-2\pi i \frac{su + tv}{p}} \ket{f(u,v)}.
\]
(Each $(u,v)$ in the above sum is a representative of one of the cosets of $K$ in $G$.) We now
prove a result similar to \propref{prop:IntroQuantumAlgs:IHSPInterference}:
\begin{proposition}\label{prop:IntroQuantumAlgs:PHSPInterference}
After \stepref{step:IntroQuantumAlgs:PHSP:Uf} of \algref{alg:IntroQuantumAlgs:PHSP}, our system is
in the state
\[
    \ket{\phi} = \frac{1}{p\sqrt{p}}\sum_{(s,t) \in T} \lb \sum_{x=0}^{p-1}e^{2\pi i x \frac{s}{p}} \ket{x} \rb \lb \sum_{y=0}^{p-1}e^{2\pi i y \frac{t}{p}} \ket{y} \rb \ket{\psi_{(s,t)}}.
\]
\end{proposition}

\begin{proof}
After \stepref{step:IntroQuantumAlgs:PHSP:QFT} of \algref{alg:IntroQuantumAlgs:PHSP}, our system is
in the state
\[
    \lb \frac{1}{\sqrt{p}}\sum_{x=0}^{p-1}\ket{x} \rb \lb \frac{1}{\sqrt{p}}\sum_{y=0}^{p-1}\ket{y} \rb \ket{0}
\]
and applying $\Uf$ in \stepref{step:IntroQuantumAlgs:PHSP:Uf} produces the state
\[
    \frac{1}{p}\sum_{(x,y) \in G} \ket{x}\ket{y}\ket{f(x,y)}.
\]
We will show that this state is in fact equal to $\ket{\phi}$.

Note that
\begin{align}
    \ket{\phi} 
               &= \frac{1}{p^2}\sum_{(s,t) \in T} \lb \sum_{(x,y) \in G}e^{2\pi i \frac{sx + ty}{p}} \ket{x} \ket{y} \rb \sum_{(u,v) \in G / K} e^{-2\pi i \frac{su + tv}{p}} \ket{f(u,v)}\notag \\
               &= \frac{1}{p^2}\sum_{(x,y) \in G} \sum_{(u,v) \in G / K} \lb \sum_{(s,t) \in T} e^{2\pi i \frac{s(x-u)+t(y-v)}{p}(x-j)} \rb \ket{x}\ket{y} \ket{f(u,v)}
                        \label{equ:IntroQuantumAlgs:PHSPInterference}
\end{align}

Now fix $x$, $y$, $u$, and $v$, and consider the coefficients $c_{(s,t)} = e^{2\pi i
\frac{s(x-u)+t(y-v)}{p}}$ for\linebreak
$(s,t) \in T$.  There are two cases:
\begin{enumerate}
\item If $(x,y)$ and $(u,v)$ are in the same coset of $K$, then $(x-u,y-v) \in K$. By the definition of $T$, $s(x-u) + t(y-v) \equiv 0 \pmod{p}$.  Thus $\frac{s(x-u)+t(y-v)}{p}$ is an
integer, each of the $c_{(s,t)}$ is $1$, and the sum of the $c_{(s,t)}$ is $p$.  In this case, we
say that there is \emph{constructive interference} between the coefficients.
\item If $(x,y)$ and $(u,v)$ are in different cosets of $K$, then using a method similar to that in the proof of \propref{prop:IntroQuantumAlgs:IHSPInterference} we can show that the sum of
the $c_{(s,t)}$ is $0$.  In this case we say that there is \emph{destructive interference} between the coefficients.
\end{enumerate}

From~\eqref{equ:IntroQuantumAlgs:PHSPInterference} we see that
\begin{align*}
  \ket{\phi} &= \frac{1}{p^2}\sum_{(x,y) \in G} \sum_{(u,v) \in G / K} \lb \sum_{(s,t) \in T} c_{(s,t)} \rb \ket{x}\ket{y} \ket{f(u,v)} \\
             &= \frac{1}{p^2}\sum_{(x,y) \in G} \ket{x}\ket{y} p \ket{f(\hat{u},\hat{v})}
\end{align*}
where $(\hat{u},\hat{v})$ is the representative for the coset containing $(x,y)$.  Therefore,
\[
    \ket{\phi} = \frac{1}{p}\sum_{(x,y) \in G} \ket{x} \ket{y}\ket{f(x,y)}
\]
since by definition $f(\hat{u},\hat{v}) = f(x,y)$. Thus the result is proven.
\end{proof}

So after \stepref{step:IntroQuantumAlgs:PHSP:Uf} our system is in the state
\[
    \frac{1}{p\sqrt{p}}\sum_{(s,t) \in T} \lb \sum_{x=0}^{p-1}e^{2\pi i x \frac{s}{p}} \ket{x} \rb \lb \sum_{y=0}^{p-1}e^{2\pi i y \frac{t}{p}} \ket{y} \rb \ket{\psi_{(s,t)}}.
\]
Assuming we can implement $\QFT{p}$ exactly, then we can see intuitively that by applying
$\IQFT{p}$ to each of the first two registers and measuring the results we will obtain,
respectively, random values $s$ and $t$ such that $(s,t) \in T$.

By running \algref{alg:IntroQuantumAlgs:PHSP} several times to obtain several random elements of
$T$, we can use methods from linear algebra to determine a generating set for $K$.  (In fact, in
some special cases, such as the solution to the Discrete Logarithm Problem discussed below, it is
sufficient to run \algref{alg:IntroQuantumAlgs:PHSP} only once.)

It follows, therefore, that \algref{alg:IntroQuantumAlgs:PHSP} forms the core of an efficient
quantum algorithm to solve PHSP.

\end{section}

\begin{section}{Computing Discrete Logarithms}\label{sec:IntroQuantumAlgs:DLP}

Given a generator $a$ of a group $H$ of order $n$, and $b = a^k$, to solve DLP we must compute $k$.
As illustrated in \probref{prob:IntroQuantumAlgs:DLP}, DLP in a group of prime order is a special
case of PHSP. Thus, the algorithm from the previous section allows us to solve DLP in a group of
prime order in polynomial time.

If we wish to solve DLP in a group of general order, we can use a slight modification of the
classical Pohlig-Hellman algorithm, which was proposed in~\cite{PH1978:PohligHellmanDLP}.  In
short, given the prime factorisation $n = p_1^{e_1} \cdots p_w^{e_w}$ where the $p_i$ are distinct
primes, the algorithm computes $k_i = k \bmod{p_i^{e_i}}$ for each $i$, and then uses the Chinese
Remainder Theorem to recombine these values into the discrete logarithm $k$.  We modify the
original algorithm in a natural way by using some quantum algorithms as subroutines.
\begin{algorithm}[Solution to DLP]\label{alg:IntroQuantumAlgs:DLP}
\begin{algone}
\item Apply \algref{alg:IntroQuantumAlgs:Factoring} recursively to split $n$ into its prime factorisation, say \linebreak
$n = p_1^{e_1} \cdots p_w^{e_w}$ where the $p_i$ are
distinct primes.\label{step:IntroQuantumAlgs:DLP:Factor}
\item For $i$ from $1$ to $w$ do the following:
  \begin{algtwo}
  \item Set $p = p_i$ and $e = e_i$.
  \item Set $\gamma = 1$ and $l_{-1} = 0$.
  \item Compute $\alpha = a^{n/p}$.
  \item For $j$ from $0$ to $e-1$ do the following:
    \begin{algthree}
    \item Compute $\gamma = \gamma a^{l_{j-1}p^{j-1}}$ and $\beta = (b\gamma^{-1})^{n/p^{j+1}}$.
    \item Compute $l_j = \log_{\alpha}\beta$ using the quantum algorithm from \secref{sec:IntroQuantumAlgs:PHSP}.
    \end{algthree}\label{step:IntroQuantumAlgs:DLP:InnerDLP}
  \item Set $k_i = l_0 + l_1 p + \cdots + l_{e-1}p^{e-1}$.
  \end{algtwo}
\item Use the Chinese Remainder Theorem to combine the $k_i$ to determine the discrete logarithm $k$.
\item Return $k$.
\end{algone}
\end{algorithm}

The element $\alpha$ computed in each iteration is an element of order $p$, since $a^n = 1$ and
$\alpha = a^{n/p}$.  Thus the instance of DLP in \stepref{step:IntroQuantumAlgs:DLP:InnerDLP} is an
instance of DLP in the group $\langle\alpha\rangle$, which is a group of order $p$.  We can
therefore indeed apply the quantum algorithm from \secref{sec:IntroQuantumAlgs:PHSP} to compute
$l_j$.  For a proof that the remainder of the algorithm is correct,
see~\cite{PH1978:PohligHellmanDLP}.

To find a factor of $n$ using \algref{alg:IntroQuantumAlgs:Factoring} requires time polynomial in
$\log n$, and there are $\bigO{\log n}$ factors, so the factoring in
\stepref{step:IntroQuantumAlgs:DLP:Factor} requires polynomial time.  In total, the number of
iterations of the inner loop is $\sum_{i=1}^{w}e_i$ (which is in $\bigO{\log n}$) and each
iteration uses the efficient quantum algorithm to compute a discrete logarithm.
\algref{alg:IntroQuantumAlgs:DLP} is therefore an efficient quantum algorithm to solve DLP.

This quantum algorithm will succeed for any group $H$, provided that we can efficiently perform the
group operation, and that each group element can be represented by a unique quantum state. (If a
single group element can be represented by multiple quantum states, these quantum states will not
interfere with one another as required.)  Of considerable interest is the group of points on an
elliptic curve over a finite field, which is fast becoming an important group in cryptographic
applications (see \chapref{chap:ElGamal}).  For a detailed discussion of quantum circuits for
solving DLP in the group of points on an elliptic curve over $GF(p)$
see~\cite{PZ2003:GFpEllipticCurveDLP}. A polynomial-time quantum algorithm to solve DLP was first
proposed in~\cite{S1994:DLFactoring}.

\end{section}
\bigskip This chapter has introduced several quantum algorithms, most importantly algorithms to solve OFP, the factoring problem, and DLP.  These algorithms are
the main tools that we will use in the subsequent chapters as we analyse various public key cryptosystems in a quantum setting.  Further quantum algorithms
that depend on more specialised concepts will be described as the required definitions and results are introduced.

\end{chapter}

\begin{chapter}{The RSA Cryptosystem}
\label{chap:RSA}

The RSA cryptosystem, the first published realisation of a public key cryptosystem, was proposed in
1977 by Rivest, Shamir, and Adleman~\cite{RSA1978:Cryptosystem}.  It is similar to the system
proposed in~\cite{C1973:NonSecretEncryption}, although that system was not made public until later.
Since the late 1970\,s, the RSA cryptosystem has become the most widely used public key encryption
scheme in many applications from electronic commerce to national security.

\begin{section}{The Cryptosystem}\label{sec:RSA:Cryptosystem}

To generate an RSA key, Alice performs the following steps:
\begin{algorithm}[RSA Key Generation]\label{alg:RSA:KeyGen}
\begin{algone}
  \item Alice selects at random two distinct primes $p$ and $q$.
  \item She calculates $n = pq$ and $\phi(n) = (p-1)(q-1)$.
  \item She selects some integer $e$, $1 < e < \phi(n)$, such that $\gcd (e,\phi(n)) = 1$.
  \item She computes the unique integer $d$, $1 < d < \phi(n)$, such that $ed \equiv 1 \pmod{\phi(n)}$.
  \item Alice's public key is $(n,e)$, and her private key is $d$.
\end{algone}
\end{algorithm}

To encrypt a message for Alice using the RSA cryptosystem, Bob performs the following steps:
\begin{algorithm}[RSA Encryption]\label{alg:RSA:Encryption}
\begin{algone}
  \item Bob obtains Alice's public key $(n,e)$.
  \item He converts the message to an integer $m$, such that $0 \leq m \leq n-1$.
  \item Bob computes the encrypted message $c = m^e \bmod n$.
\end{algone}
\end{algorithm}

To recover the original message, Alice does the following:
\begin{algorithm}[RSA Decryption]\label{alg:RSA:Decryption}
\begin{algone}
    \item She uses her private key $d$ and computes $m = c^d \bmod n$.
\end{algone}
\end{algorithm}

\begin{theorem}
RSA decryption works properly.
\end{theorem}

\begin{proof}
First note that since $ed \equiv 1 \bmod{\phi(n)}$ there exists some integer $t$ such that $ed = 1 + t\phi(n)$.

We now have two cases.
\begin{enumerate}
  \item If $\gcd(m,p) = 1$,
    \begin{align*}
      m^{p-1} &\equiv 1 \pmod{p} \quad \text{(by Fermat's Little Theorem)}\\
      (m^{p-1})^{t(q-1)} &\equiv 1 \pmod{p} \\
      m^{1 + t\phi(n)} &\equiv m \pmod{p}.
    \end{align*}
  \item If $\gcd(m,p) = p$,
    \begin{align*}
      m &\equiv 0 \pmod{p} \\
      m^{1+ t\phi(n)} &\equiv 0 \pmod{p}.
    \end{align*}
\end{enumerate}

Thus in both cases, $m^{1 + t\phi(n)} \equiv m \pmod{p}$.  In a similar way, we can prove that
$m^{1 + t\phi(n)} \equiv m \pmod{q}$.  Combining these two congruences, since $p$ and $q$ are
distinct primes,
\begin{align*}
    m^{1 + t\phi(n)} &\equiv m \pmod{n} \\
    m^{ed} &\equiv m \pmod {n} \\
    c^d &\equiv m \pmod {n}
\end{align*}
so decryption indeed works properly.
\end{proof}

\end{section}

\begin{section}{Security Of The System}\label{sec:RSA:Security}

The security of the RSA cryptosystem is based on the hardness of the RSA problem, which is the problem of finding $e^{th}$ roots
in the ring $\Z_n = \Z/n\Z$.
\begin{problem}[The RSA Problem (RSAP)]\label{prob:RSA:RSA}
Given $n$, $e$, and $m^e \bmod{n}$ for some $m \in \Z_n$, find $m$.
\end{problem}

Determining the plaintext from an RSA ciphertext is equivalent to solving RSAP, which is thought to
be hard for a classical computer.  Alternatively, if Eve can successfully factor $n$ to recover $p$
and $q$, she can compute $\phi(n)$ and $d$ just as Alice did when she generated the keys, and
obtain complete knowledge of Alice's private key.  It has been conjectured that these two attacks
are computationally equivalent (that is, solving RSAP is equivalent to factoring $n$) but this
conjecture has not been proven.  However, it can be shown that determining the private key $d$ from
the public key $(n,e)$ is indeed equivalent to factoring $n$~\cite{MvOV1996:HAC}, and most current
attacks on the RSA cryptosystem attempt to factor $n$.

The problem of factoring integers has been studied in detail for many years; some of the current
known classical factoring algorithms are listed in \tabref{tab:RSA:ClassicalFactoringAlgs}.
\begin{table}[ht]
  \begin{center}
    \begin{tabular}{ll}
      \hline
      Algorithm & Expected running time\\
                & \qquad(neglecting logarithmic factors) \\ \hline
      Trial division & $\bigO{n^{1/2}}$ \\
      Pollard rho & $\bigO{n^{1/4}}$ \\
      Quadratic sieve & $\exp\left[\; \bigO{ ( \log n )^{1/2} ( \log \log n )^{1/2} } \;\right]$ \\
      Number field sieve & $\exp\left[\; \bigO{ ( \log n )^{1/3} ( \log \log n )^{2/3} } \;\right]$ \\
      \hline
    \end{tabular}
    \caption{Some classical factoring algorithms}\label{tab:RSA:ClassicalFactoringAlgs}
  \end{center}
\end{table}
Other than the trial division algorithm, the algorithms in \tabref{tab:RSA:ClassicalFactoringAlgs}
are probabilistic algorithms.  The running times presented in the table are upper bounds on the
expected running times of the algorithms, taken over the random bits used as input.  In general,
rigorous analysis of an algorithm leads to an expected running time that is valid for any input. In
the case of the last two algorithms above, however, some additional assumptions on the input are
required in order for the expected running times to be valid; so the estimates are heuristic ones.
(These assumptions are conjectured to hold true for all inputs, but the conjectures are unproven.)
There are also algorithms that have rigorously proven expected running times of $\exp\left[\;
\bigO{ ( \log n )^{1/2} ( \log \log n )^{1/2} } \;\right]$, such as the algorithm
in~\cite{P1987:RigorousFactoringAndDLP}.

The last two algorithms in \tabref{tab:RSA:ClassicalFactoringAlgs} are both ``sieving'' algorithms,
and they operate on the same basic premise: each of them tries to find positive integers $x$ and
$y$ less than $n$ such that
\begin{align*}
    x^2 &\equiv y^2 \pmod{n}, \;\;\mbox{and}\\
    x   &\not\equiv \pm y \pmod{n}.
\end{align*}
Once two such integers have been found, we know that
\[
    (x-y)(x+y) \equiv 0 \pmod{n}
\]
and $n$ does not divide either $x-y$ or $x+y$.  Thus $\gcd(x-y,n)$ is a non-trivial factor of $n$.
As mentioned earlier, these algorithms are randomised: they find congruences of the desired form by
choosing random integers and performing specific series of operations on them.

Another popular factoring algorithm is the elliptic curve method proposed
in~\cite{L1987:EllipticCurveFactoring}.  This algorithm works especially well when the smallest
prime factor of $n$ is much smaller than $\sqrt{n}$: the algorithm's expected running time is $\exp
\left[\;(2+\varepsilon)( \log p )^{1/2} ( \log \log p )^{1/2}\;\right]$ where $p$ is the smallest
prime factor of $n$ and $\varepsilon \rightarrow 0$ as $p \rightarrow \infty$.  (This is a
heuristic estimate.) In general, the two prime factors of an RSA modulus are chosen to be
approximately equal in size, and so the elliptic curve algorithm may not run significantly faster
than the other algorithms in \tabref{tab:RSA:ClassicalFactoringAlgs}; on the other hand, it
requires considerably less storage space~\cite{K1994:NumberTheoryAndCryptography}.

For more details on these classical factoring algorithms, see for
example~\cite{K1994:NumberTheoryAndCryptography} or~\cite{C1993:CourseInComputationalANT}.

Because all of these algorithms require superpolynomial time, the RSA cryptosystem is still
considered secure against classical factoring attacks for sufficiently large $n$. Generally, a
modulus of $1024$ bits or more is thought to be secure against today's computers. Recent
developments in specialised hardware indicate that this modulus length may not be sufficient for
much longer, however: the device proposed in~\cite{ST2003:FactoringTWIRL} would reportedly cost
\$10 million and would be capable of factoring a $1024$-bit modulus in less than a year.

We do not know of an efficient classical algorithm for factoring; so the RSA cryptosystem may be
hard to break with any classical algorithm.  However, as we have seen in
\secref{sec:IntroQuantumAlgs:Factoring}, \algref{alg:IntroQuantumAlgs:Factoring} is a probabilistic
polynomial-time quantum algorithm that solves the factoring problem.  Thus the RSA cryptosystem is
insecure in a quantum setting.

It is also interesting to note that given a particular RSA ciphertext $c$, we can use a quantum
computer to solve RSAP directly; that is, to determine the corresponding plaintext $m$ without
having to factor $n$~\cite{CEMM1998:AlgorithmsRevisited}.  Since $e$ is relatively prime to
$\phi(n)$, we know that $m$ and $m^e = c$ have the same order, say $r$. To determine $m$ from $c$,
we first give $c$ as input to the quantum order-finding algorithm described in
\secref{sec:IntroQuantumAlgs:OFP}, and obtain $r$ as output.  Next we compute the unique $a$ such
that $ea \equiv 1 \pmod{r}$.  Finally, we compute
\begin{align*}
      &c^a \bmod{n}\\ &= m^{ea} \bmod{n}\\ &= m
\end{align*}
and recover the plaintext $m$.

\end{section}

\end{chapter}

\begin{chapter}{The Rabin Cryptosystem}
\label{chap:Rabin}

The Rabin cryptosystem was proposed in~\cite{R1979:RabinCryptosystem}.  Like the RSA cryptosystem,
an adversary can attack the scheme by factoring a product of two large primes. However, unlike the
RSA cryptosystem, it has been proven that performing this factorisation is computationally
equivalent to determining the plaintext corresponding to a given ciphertext. If we assume that the
factoring problem is intractable, then the Rabin cryptosystem is provably secure against a passive
adversary.

\begin{section}{The Cryptosystem}\label{sec:Rabin:Cryptosystem}

To generate Rabin keys, Alice does the following:
\begin{algorithm}[Rabin Key Generation]\label{alg:Rabin:KeyGen}
\begin{algone}
  \item Alice selects two distinct primes $p$ and $q$.
  \item She calculates $n = pq$.
  \item Alice's public key is $n$, and her private key is $(p,q)$.
\end{algone}
\end{algorithm}

To encrypt a message for Alice using the Rabin cryptosystem, Bob does the following:
\begin{algorithm}[Rabin Encryption]\label{alg:Rabin:Encryption}
\begin{algone}
  \item Bob obtains Alice's public key $n$.
  \item He converts the message to an integer $m$, such that $0 \leq m \leq n-1$.
  \item Bob computes the encrypted message $c = m^2 \bmod n$.
\end{algone}
\end{algorithm}

To recover the original message, Alice does the following:
\begin{algorithm}[Rabin Decryption]\label{alg:Rabin:Decryption}
\begin{algone}
  \item She computes the four square roots of $c \bmod n$.
  \item Somehow, she decides which of the four square roots corresponds to the original message sent by Bob.
\end{algone}
\end{algorithm}

One problem with the Rabin cryptosystem is that in order to recover the original message, Alice
must somehow choose between the four square roots of the ciphertext.  One way to avoid this problem
is to include some redundancy in the message before encrypting it, so that with high probability
only one of the four square roots will have this redundancy.

If $p$ and $q$ are chosen to be congruent to $3 \bmod 4$, there is a simple algorithm to calculate the four square roots of $c$:
\begin{algorithm}[Computing Square Roots]\label{alg:Rabin:SquareRoots}
\begin{algone}
  \item Alice computes $r = c^{(p+1)/4} \bmod p$ and $s = c^{(q+1)/4} \bmod q$.
  \item She uses the Extended Euclidean Algorithm to find integers $a$ and $b$ such that $ap + bq = 1$.
  \item She calculates $x = ( aps + bqr ) \bmod n$ and $y = ( aps - bqr ) \bmod n$.
  \item The four square roots of $c$ are $x \bmod n$, $-x \bmod n$, $y \bmod n$, and $-y \bmod n$.
\end{algone}
\end{algorithm}

The steps of the algorithm correspond to finding the square roots of $c$ modulo $p$ and $q$, and
then combining them using the Chinese Remainder Theorem. It is easily verified that the resulting
integers are indeed the four square roots of $c$. Note that if one or both of $p$ or $q$ is
congruent to $1 \bmod 4$, the square roots can still be efficiently computed, but the algorithm is
more complicated~\cite{MvOV1996:HAC}. For this reason, during the key generation procedure it makes
sense to choose the primes $p$ and $q$ to be congruent to $3 \bmod 4$.

It is also interesting to note as in~\cite{MvOV1996:HAC} that Rabin encryption is more efficient
than RSA encryption, since it requires a single modular squaring operation.  (RSA encryption will
require more squaring and multiplication operations since the encryption exponent is always greater
than $2$.)  The efficiencies of the RSA and Rabin decryption algorithms are comparable.

\end{section}

\begin{section}{Security Of The System}\label{sec:Rabin:Security}

If an adversary can factor Alice's modulus $n$ and recover the primes $p$ and $q$, then the
adversary has complete knowledge of Alice's private key and hence the scheme is broken.  As
mentioned previously, in fact it is easy to see that decrypting Rabin ciphertexts is
computationally equivalent to the factoring problem. If we can decrypt a Rabin ciphertext, $c$, we
can find the four square roots of $c \bmod{m}$, say $\pm x, \pm y$, where $y \not\equiv \pm x
\pmod{n}$. Then we know that $c \equiv x^2 \pmod{n}$ and $c \equiv y^2 \pmod{n}$, or in other
words, $x^2 \equiv y^2 \pmod{n}$.  (This congruence is one of the type that the sieving algorithms
in \secref{sec:RSA:Security} attempt to find.)  Then we know that $(x-y)(x+y) \equiv 0 \pmod{n}$,
so $\gcd( x-y, n )$is a non-trivial factor of $n$.

Because of this equivalence, factoring algorithms like the ones mentioned in \chapref{chap:RSA} are
the only available tools for a passive adversary.  Since the best known classical factoring
algorithms require superpolynomial time, the scheme is thought to be secure against a passive
classical adversary. However, an adversary with a quantum computer can factor in polynomial time
using \algref{alg:IntroQuantumAlgs:Factoring}.  Hence, the Rabin cryptosystem is not secure in a
quantum setting.

\end{section}

\end{chapter}

\begin{chapter}{The ElGamal Cryptosystem}\label{chap:ElGamal}

The ElGamal cryptosystem was proposed in~\cite{E1985:ElGamalCryptosystem}, and is based on the
hardness of the Discrete Logarithm Problem (DLP).  It may be used with any finite cyclic group,
although as stated in~\cite{MvOV1996:HAC} in order for the group to be a good choice, it should
satisfy two main criteria:
\begin{enumerate}
  \item the group operation should be easy to apply so that the cryptosystem is efficient, and
  \item DLP in the group should be computationally infeasible so that the cryptosystem is secure.
\end{enumerate}
Some examples of groups for which these criteria seem to be met are the multiplicative group $\Z_p^*$ of the integers modulo a
prime $p$, and the group of points on an elliptic curve over a finite field.  For more examples of groups where the ElGamal
cryptosystem is thought to be secure, see~\cite{MvOV1996:HAC}.

\begin{section}{The Cryptosystem}\label{sec:ElGamal:Cryptosystem}

To generate ElGamal keys, Alice does the following:
\begin{algorithm}[ElGamal Key Generation]\label{alg:ElGamal:KeyGen}
\begin{algone}
  \item She selects a cyclic group $G$ that meets the above criteria.  Let $n$ denote the order of $G$.
  \item She finds a generator $\alpha$ of $G$.
  \item Alice selects a random integer $a$ such that $ 1 \leq a \leq n-1$, and computes $\alpha^a$.
  \item Her public key is $( G, n, \alpha, \alpha^a )$ and her private key is $a$.
\end{algone}
\end{algorithm}

To encrypt a message for Alice using the ElGamal cryptosystem, Bob does the following:
\begin{algorithm}[ElGamal Encryption]\label{alg:ElGamal:Encryption}
\begin{algone}
  \item Bob obtains Alice's public key $( G, n, \alpha, \alpha^a )$.
  \item He converts his message to an element $m \in G$.
  \item He selects a random integer $k$ such that $1 \leq k \leq n-1$.
  \item Bob computes $\gamma = \alpha^k$ and $\delta = m ( \alpha^a )^k$.
  \item The ciphertext is $c = (\gamma, \delta)$.
\end{algone}
\end{algorithm}

To decrypt the ciphertext, Alice performs the following steps:
\begin{algorithm}[ElGamal Decryption]\label{alg:ElGamal:Decryption}
\begin{algone}
  \item Alice uses her private key $a$ to compute $\gamma^a$ and then $\gamma^{-a}$.
  \item She computes $m = \gamma^{-a} \delta$.
\end{algone}
\end{algorithm}

\begin{theorem}
ElGamal decryption works properly.
\end{theorem}

\begin{proof}
Note that
\begin{align*}
    \gamma^{-a} \delta &= \gamma^{-a} m ( \alpha^a )^k \\
                       &= m \gamma^{-a} (\alpha^k)^a \\
                       &= m \gamma^{-a} \gamma^a \\
                       &= m
\end{align*}
so decryption indeed works properly.
\end{proof}

\end{section}

\begin{section}{Security Of The System}

It is clear that an attacker can compute Alice's private key $a$ by finding the discrete logarithm
of $\alpha^a$ to the base $\alpha$, both of which are public quantities.  Also, if the attacker can
find the particular value of $k$ that was used to encrypt a message, she can decrypt the message,
but to determine $k$ she must find the discrete logarithm of $\gamma$ to the base $\alpha$.  These
facts imply that the security of the scheme depends on the hardness of DLP in the group $G$.

There are many classical algorithms to solve DLP.  These algorithms can be divided into two main
categories: algorithms that work in any group $G$, and algorithms that depend on a particular group
$G$.

In the first category are algorithms like Shanks's baby-step giant-step algorithm, which runs in
$\bigO{n^{1/2}}$ time (ignoring logarithmic factors)~\cite{C1993:CourseInComputationalANT}.  The
Pohlig-Hellman algorithm mentioned in \secref{sec:IntroQuantumAlgs:DLP} also works in any group,
and it performs especially well if the factors of $n$ are known and are all small; however, in the
worst case it also requires $\bigO{n^{1/2}}$ time. In fact, in a ``generic'' group $G$ of prime
order $p$ (that is, a group where the elements have unique encodings but where the encodings do not
reveal any group structure that algorithms can take advantage of) a lower bound on the complexity
of any classical algorithm to solve DLP is $\Omega(p^{1/2})$ steps~\cite{S1997:DLPLowerBound}. Some
algorithms may take advantage of the structure of a specific group, however, which gives rise to
the second main category of algorithms for DLP.

In this second category are algorithms like the index calculus algorithms, which work in the
multiplicative group of $GF(p^k)$, where $p$ is a prime and $k$ is a positive integer. The index
calculus algorithms are similar in structure to the sieving algorithms for factoring discussed in
\secref{sec:RSA:Security}, and there are methods with a rigorous expected running time of
$\exp\left[\; \bigO{ ( \log p^k )^{1/2} ( \log \log p^k )^{1/2} } \;\right]$.  There is also an
analogue of the Number Field Sieve which is slightly more efficient, with an expected running time
of $\exp\left[\; \bigO{ ( \log p^k )^{1/3} ( \log \log p^k )^{2/3} } \;\right]$ (although this is a
heuristic estimate).  Even though they run in subexponential time, like the best known factoring
algorithms, these algorithms for DLP still require superpolynomial time.

However, \algref{alg:IntroQuantumAlgs:DLP} requires time polynomial in $\log n$, and thus is an
efficient quantum algorithm to solve DLP in any group $G$.  The existence of this efficient quantum
algorithm implies that the ElGamal cryptosystem is not secure in a quantum setting: an attacker
could use \algref{alg:IntroQuantumAlgs:DLP} to compute Alice's private key $a$ from her public key
$\alpha^a$.  Alternatively, the attacker could use \algref{alg:IntroQuantumAlgs:DLP} to determine
$k$ from $\gamma$ for a particular message and hence decrypt the message by computing $\delta(
\alpha^a )^{-k} = m$.

\end{section}

\end{chapter}

\begin{chapter}{The McEliece Cryptosystem}\label{chap:McEliece}

The McEliece cryptosystem was proposed in~\cite{M1978:McElieceCryptosystem}, and is based on
problems in algebraic coding theory. To generate a key pair, Alice constructs a linear
error-correcting code that has an efficient decoding algorithm, and then uses some secret
parameters to transform it into a different linear code with no apparent efficient decoding
algorithm.  The secret parameters that Alice has chosen allow her to perform the inverse
transformation and then use an efficient algorithm to decrypt the ciphertext she receives.

\begin{section}{The Cryptosystem}\label{sec:McEliece:Cryptosystem}

Before describing the McEliece cryptosystem, we give a brief introduction to the theory of
error-correcting codes.  For more information about error-correcting codes and information theory
see, for example,~\cite{M1977:CodingTheory}.  Originally, coding theory was developed to allow data
to be reliably transmitted through a channel that may distort the data during transmission. The
idea of an error-correcting code is to introduce a certain amount of redundancy into the message
being transmitted, so that even if errors do occur, they can be detected and possibly corrected. It
should be noted that these codes are not ``encryption schemes'', in that they are not designed to
protect the confidentiality of the data, and they have publicly known encoding and decoding
procedures.

Codes can be defined over any set of messages, but for our purposes, we will only consider binary codes, which use messages
constructed from the alphabet $\mathbb{Z}_2 = \{0,1\}$.  If we wish to send messages of $k$ bits in length, we consider our
message space to be the set of $k$-tuples with entries from $\mathbb{Z}_2$.  The idea of a code is to choose some $n > k$ and
define a one-to-one mapping between the message space and a subset of size $2^k$ of the set of binary $n$-tuples.  This subset is
called a code, and the elements of the subset are the code words.   Since there are more bits in each code word than there are in
each message, the code words can carry more information than the messages: namely the redundancy that we need to achieve the goals
stated above.

We begin with some basic definitions.

\begin{definition}
Let $x$ be an $n$-tuple with entries in $\Z_2$.  The \tbd{Hamming weight} of $x$, denoted $w(x)$, is the number of components of $x$ that are equal to $1$.
\end{definition}

\begin{definition}
Let $x$ and $y$ be $n$-tuples with entries in $\Z_2$.  The \tbd{distance} between $x$ and $y$,
denoted $d(x,y)$, is the number of components in which $x$ and $y$ differ. Equivalently, $d(x,y) =
w(x \oplus y)$.
\end{definition}

We now introduce the concept of a linear code, which is one of the most common types of error-correcting codes.

\begin{definition}
Let $V_n(\Z_2)$ be the $n$-dimensional vector space consisting of the $n$-tuples with entries from $\Z_2$.  Let $k < n$.  A \tbd{$(k,n)$ linear binary code} is
a $k$-dimensional subspace of $V_n(\Z_2)$.
\end{definition}

\begin{definition}
Let $C$ be a linear binary code.  A \tbd{generator matrix} for $C$ is a $k \times n$ matrix with entries from $\Z_2$ whose rows form a basis for $C$.
\end{definition}

Let $m$ be a $k$-bit (row) vector. The code word corresponding to the message $m$ is the $n$-bit
(row) vector $mG$, where $G$ is the generator matrix for the code.  When the code word is
transmitted, errors may be introduced by the communication channel or by a malicious third party,
and ideally the receiver will be able to detect and correct these errors.  The decoding procedure
(that is, the process of recovering the original message from the received binary string) may be
complicated, and varies depending on the type of code being used. The codes that are of interest in
coding theory are those with which the receiver can detect and correct a large number of errors
relative to the size of the code words, and for which the decoding procedure is efficient.  When
describing a code, we often state its error-correcting capability, which is an integer representing
the number of errors that may be introduced in a transmitted code word without affecting the
receiver's ability to properly decode the received binary string.  We also often state the distance
of the code, which is the minimum distance between any two codewords.

However, there are many linear codes for which there apparently exists no efficient decoding procedure.  In fact, given a generator matrix for a random
subspace of $V_n(\Z_2)$ the problem of decoding a received binary string can be shown to be \NP-complete~\cite{BMvT1978:Coding}.  The security of the McEliece
cryptosystem is based on the hardness of this general decoding problem. The cryptosystem uses a specific type of code called a binary Goppa code.

\begin{definition}
Let $GF(2^l)$ denote the finite field with $2^l$ elements.  Let $G(x)$ be a polynomial of degree
$s$ with coefficients in $GF(2^l)$ and let $\alpha_1, \ldots, \alpha_n \in GF(2^l)$ be chosen such
that $G(\alpha_i) \ne 0$ for $i = 1, \ldots, n$.  These parameters define a \tbd{binary Goppa code}
in which $c = (c_1, \ldots, c_n) \in V_n(\Z_2)$ is a codeword if and only if
\[
    \sum_{i=1}^n c_i ( x - \alpha_i )^{-1} \equiv 0 \pmod{G(x)}.
\]
\end{definition}

\begin{proposition}
A Goppa code defined as above is a $(k,n)$ linear binary code with $k \geq n - ls$ and distance at
least $s+1$.
\end{proposition}

The bounds given in this proposition are not necessarily tight bounds, and many choices of
parameters may result in codes with larger $k$ and larger distances. Goppa codes are among the
classes of codes that are of interest in coding theory because they have an efficient decoding
procedure, which is described in~\cite{M1977:CodingTheory}.  The idea of the McEliece cryptosystem
is to transform a randomly selected binary Goppa code into a general linear code using some secret
parameters.  Without knowledge of these secret parameters, the best decoding procedures for the
resulting general code are thought to require superpolynomial time; with knowledge of the secret
parameters, the general code can be transformed back to a Goppa code, where an efficient decoding
procedure does exist. The cryptosystem is described below.

To generate a McEliece key, Alice performs the following steps:
\begin{algorithm}[McEliece Key Generation]
\begin{algone}
  \item Alice constructs a linear $t$ error-correcting Goppa code $C$ with a $k \times n$ generator matrix $G$.
  \item Alice selects a $k \times k$ invertible matrix $S$ (called a ``scrambling matrix'') and an $n \times n$ permutation matrix
  $P$.
  \item She computes $\bar{G} = SGP$.
  \item Alice's public key is $\bar{G}$, and her private key is $( S, G, P )$.
\end{algone}
\end{algorithm}

Note that the matrix $\bar{G}$ is a generator matrix for a general linear code that is related to $C$, but for which there is no apparent efficient decoding
algorithm.

To encrypt a message for Alice using the McEliece cryptosystem, Bob performs the following steps:
\begin{algorithm}[McEliece Encryption]
\begin{algone}
  \item Bob obtains Alice's public key $\bar{G}$.
  \item He converts the message to a $k$-bit binary vector $m$.
  \item He selects a random $n$-bit vector $e$ of weight $t$.
  \item Bob computes the encrypted message $c = m \bar{G} \oplus e$.
\end{algone}
\end{algorithm}

In other words, to encrypt a message, Bob starts with the message vector, computes the corresponding codeword in the general linear code, and adds a random
``error'' vector to the message.  Since the new code has no apparent efficient decoding algorithm, it is hard for an attacker to correct the error and recover
the original message.  However, since Alice knows the matrices $S$ and $P$ she can use them to transform the codeword from this new code back to the original
code, and then use the efficient decoding algorithm for that code.

In other words, to recover the original message, Alice does the following:
\begin{algorithm}[McEliece Decryption]
\begin{algone}
  \item She computes $cP^{-1} = (mS)G \oplus eP^{-1}$.
  \item Since $P$ is a permutation matrix, $eP^{-1}$ also has weight $t$.  Alice can therefore use the efficient decoding algorithm for the
  original Goppa code $C$ to remove the error $eP^{-1}$ and recover the codeword $mS$.
  \item She applies $S^{-1}$ to recover $m$.
\end{algone}
\end{algorithm}

\end{section}

\begin{section}{Security Of The System}\label{sec:McEliece:Security}

The best known classical attack on the McEliece cryptosystem is described
in~\cite{AM1987:McElieceSecurity}; minor improvements to the algorithm have been suggested
in~\cite{LB1988:McElieceSecurity} and others, however the general idea of the attack remains the
same.

Suppose the attacker obtains a ciphertext $c = m \bar{G} \oplus e$.  She chooses $k$ components of
$c$, and uses them to form the shorter vector $\hat{c}$.  Let the positions of the chosen
components be $i_1, i_2, \ldots, i_k$; so $\hat{c} = ( c_{i_1}, c_{i_2}, \ldots, c_{i_k} )$. Let
$\hat{e} = ( e_{i_1}, e_{i_2}, \ldots, e_{i_k} )$ denote the corresponding $k$ components of $e$,
and let $\hat{\bar{G}}$ denote the square matrix formed by taking the corresponding $k$ columns of
$\bar{G}$. Then
\[
    \hat{c} = m \hat{\bar{G}} \oplus \hat{e}
\]
and provided $\hat{\bar{G}}$ is invertible,
\[
    ( \hat{c} \oplus \hat{e} ) \hat{\bar{G}}^{-1} = m.
\]

Thus if the $k$ components of $\hat{e}$ all happen to be $0$, the attacker is able to recover $m$
by computing $\hat{c} \hat{\bar{G}}^{-1} = m$.  The idea of the attack is to choose various sets of
components until a set is found for which those $k$ components of $e$ are all $0$, at which point
the message will be recovered.

It is important for the attacker to have a method to recognise that the correct message $m$ has been obtained, especially in cases where the message does not
contain any redundancy; such a method was proposed in~\cite{LB1988:McElieceSecurity}.  The method can be summarised in the following proposition:
\begin{proposition}\label{prop:McEliece:AttackSuccess}
The attack has succeeded (that is, $\hat{c} \hat{\bar{G}}^{-1} = m$) if and only if \linebreak $w(
c \oplus \hat{c} \hat{\bar{G}}^{-1} \bar{G} ) \leq t$, where $t$ is the error-correcting capability
of $C$.
\end{proposition}

\begin{proof}
First consider the case where $\hat{c} \hat{\bar{G}}^{-1}$ is indeed the true message $m$.  Recall
that $c = m \bar{G} \oplus e$, so
\begin{align*}
    w( c \oplus \hat{c} \hat{\bar{G}}^{-1} \bar{G} ) &= w( m \bar{G} \oplus e \oplus \hat{c} \hat{\bar{G}}^{-1} \bar{G} ) \\
                                             &= w( m \bar{G} \oplus e \oplus m \bar{G} ) \\
                                             &= w( e ) \\
                                             &\leq t.
\end{align*}

Next consider the case where $\hat{c} \hat{\bar{G}}^{-1}$ is not the true message $m$, but instead
some other binary string $m'$.  Since $C$ can correct $t$ errors, by definition it must have
distance greater than $2t$, meaning that the number of components in which any two codewords differ
is greater than $2t$.  So the codewords corresponding to $m$ and $m'$ must differ in more than $2t$
components.  That is,
\[
    w( m \bar{G} \oplus m' \bar{G} ) > 2t.
\]
Note that $m \bar{G} = c \oplus e$ so we have
\[
    w( c \oplus e \oplus \hat{c} \hat{\bar{G}}^{-1} \bar{G} ) > 2t
\]
and since $w(e) \leq t$,
\[
    w( c \oplus \hat{c} \hat{\bar{G}}^{-1} \bar{G} ) > t
\]
and the proposition is proven.
\end{proof}

Thus the attacker has a method to easily determine when the attack has succeeded: she can
exhaustively search all sets of components until the correct one is found and the message is
recovered. The attack clearly requires time exponential in $k$. In~\cite{LB1988:McElieceSecurity},
improvements and generalisations are suggested that improve the running time of the attack,
although only by a polynomial factor.

It is interesting to note that this approach will decode any code, not just one of the special form
used in the McEliece cryptosystem; in other words, this attack solves the general decoding problem.
However, as mentioned above, the general decoding problem is known to be
\NP-complete~\cite{BMvT1978:Coding}. It is possible that if a polynomial-time attack is desired,
the special form of a McEliece code will have to be exploited by the attacker.

In~\cite{H1987:SecurityOfErrorCorrectingCodeCrypto} it is shown that determining the plaintext from
the ciphertext is polynomially equivalent to determining the Hamming weight of the plaintext. This
and other facts are then used to develop and propose partial attacks on the McEliece cryptosystem
and its variants.  The attacks are shown to be generally unsuccessful against the original
cryptosystem, but empirical evidence mentioned in~\cite{H1987:SecurityOfErrorCorrectingCodeCrypto}
suggests these attacks may be successful against some variants of the cryptosystem, such as schemes
that use a different class of error correcting codes in the same general way.

Unlike the first cryptosystems we discussed, it is not clear that there is a way to efficiently
break the McEliece cryptosystem using a quantum computer.  None of the algorithms presented in
\chapref{chap:IntroQuantumAlgs} seem to give a quantum attacker any advantage over a classical one.
The hard problem on which the idea for the cryptosystem is based, namely the decoding of an
arbitrary linear code, does not seem to fit well into the Hidden Subgroup Problem framework, and so
it is unlikely that any of the algorithms we have discussed will be helpful to the attacker in
developing a polynomial-time attack on the scheme.

However, we could use a different quantum algorithm to speed up the best known classical attack (by
a polynomial factor).  The quantum algorithm proposed in~\cite{G1996:QuantumSearching} (sometimes
called Grover's algorithm) allows us to improve the performance of searching algorithms.
Specifically, given a number of possible solutions to a problem, only some of which are correct,
the algorithm allows us to find a correct solution more efficiently with a quantum computer than we
can classically.  This type of search is often referred to as a ``needle in a haystack'' problem
since typically there are many incorrect solutions and only a few correct ones.  For a description
of Grover's algorithm, see for example~\cite{G1996:QuantumSearching} or~\cite{NC2000:Nielsen}.

Suppose we have a set of $n$ elements indexed by the set of integers $S = \{ 0, 1, \ldots, n - 1
\}$. Further suppose that we are given a function $f \colon S \rightarrow \{ 0, 1 \}$ such that:
\begin{enumerate}
    \item $f(x) = 1$ if $x$ is the index of an element which is a solution to the problem, and
    \item $f(x) = 0$ otherwise.
\end{enumerate}
The search succeeds when it finds an $x$ such that $f(x) = 1$.  If there is a constant number of
such $x$, then the best classical searching algorithm (brute force search) requires $\bigO{n}$
time.  However, Grover's quantum search algorithm  requires only $\bigO{\sqrt{n}}$ time, which is a
considerable improvement (although just a polynomial one).

In the case of the attack on the McEliece cryptosystem mentioned above, our set of elements is the
set of all $k$-subsets of the components of the codeword.  For any particular subset $A$, according
to \propref{prop:McEliece:AttackSuccess}, $f(A)$ should be $1$ if and only if $w( c \oplus \hat{c}
\hat{\bar{G}}^{-1} \bar{G} ) \leq t$.  This condition provides us with an efficient way of
evaluating $f$.

Using Grover's algorithm, then, we can achieve a square-root speedup over the classical version of
the search algorithm.  While this still represents only a polynomial improvement in the running
time, such an improvement could pose a significant security threat for many of the McEliece
parameter sizes that are currently thought of as secure.

\end{section}

\end{chapter}

\begin{chapter}{The Ajtai-Dwork Cryptosystem}\label{chap:AjtaiDwork}

The Ajtai-Dwork cryptosystem~\cite{AD1997:Cryptosystem} was one of the first proposed cryptosystems whose security was based on
the hardness of problems involving lattices.  This cryptosystem is currently not of practical interest since messages are
encrypted bit-by-bit, and the ciphertext is very long compared to the plaintext; also recent attacks by Nguyen and
Stern~\cite{NS1998:ADCryptanalysis} have shown that the scheme with small parameters is insecure. Nonetheless, the cryptosystem
has received a good deal of theoretical interest, especially since a proof of its security in~\cite{AD1997:Cryptosystem} was based
on worst-case instead of average-case analysis.

Since the proposal of the Ajtai-Dwork scheme, there have been other proposals for cryptosystems based on lattices, some of which we will discuss later. There
has been much recent interest in these ``lattice-based'' cryptosystems, perhaps because their security is based on problems that are fundamentally different
from integer factorisation and computing discrete logarithms, and because the encryption and decryption rates for several of the schemes are asymptotically
faster than those for the more widely-used cryptosystems.  For an excellent overview of the many uses of lattices in cryptography and cryptanalysis, including
simple descriptions of many of these schemes, we refer the reader to~\cite{NS2000:LatticesCryptology}.

\begin{section}{The Cryptosystem}\label{sec:AjtaiDwork:Cryptosystem}

The details of the cryptosystem are quite complicated; so we present the general idea of the scheme
and refer the reader to~\cite{AD1997:Cryptosystem} for a more rigorous presentation.  We begin with
a few definitions.

\begin{definition}
Let $w_1, \ldots, w_n \in \Rn$.  The \tbd{parallelepiped} $W$ spanned by the $w_i$ is defined as $\left\{ \sum_{i=1}^n \lambda_i
w_i \st 0 \leq \lambda_i < 1 \right\}$.  In other words, $W$ is the set of all points that are a linear combination of the $w_i$
with coefficients between $0$ and $1$.
\end{definition}

\begin{definition}
Let $w_1, \ldots, w_n \in \Rn$ and let $W$ be the parallelepiped spanned by the $w_i$.  Let $H_i$
be the $(n-1)$-dimensional hyperplane spanned by the set $\left\{ w_j \st 1 \leq j \leq n, j \neq i
\right\}$.  The \tbd{width} of $W$ is defined as the maximum of the perpendicular distances between
$w_i$ and $H_i$ for $1 \leq i \leq n$.
\end{definition}

To generate Ajtai-Dwork keys, Alice performs the following steps:
\begin{algorithm}[Ajtai-Dwork Key Generation (sketch)]\label{alg:AjtaiDwork:KeyGen}
\begin{algone}
  \item Alice selects a vector $u$ ``uniformly at random'' from the $n$-dimensional unit ball.
  (Note that she actually selects $u$ from a large discrete set of vectors in the unit ball,
  as described in~\cite{AD1997:Cryptosystem}.)
  \item According to the procedure in~\cite{AD1997:Cryptosystem} she defines a distribution $\mathcal{H}_u$ of points in the $n$-dimensional
  ball of radius $2^{ n \log n }$ such that for each point $h$ in the distribution, the inner product $\ip{h}{u}$ is very close to an integer.
  \item She sets $m = n^3$ and selects the $m+n$ points $v_1, \ldots, v_m, w_1, \ldots w_n$ uniformly at random from the distribution
  $\mathcal{H}_u$ defined above.
  \item She verifies that the width of the parallelepiped spanned by $w_1, \ldots, w_n$ is at least $2^{n \log n} /
  n^2$.\label{step:AjtaiDwork:KeyGen:ParallelpipedWidth}
  (With high probability this is true; otherwise, she begins the procedure again.)
  \item Alice's public key is $( v_1, \ldots, v_m, w_1, \ldots, w_n )$ and her private key is $u$.
\end{algone}
\end{algorithm}

To encrypt a message for Alice, Bob does the following:
\begin{algorithm}[Ajtai-Dwork Encryption]
\begin{algone}
  \item Bob obtains Alice's public key $( v_1, \ldots, v_m, w_1, \ldots, w_n )$.  Let $W$ be the parallelepiped spanned by the $w_i$.
  \item He encrypts each bit $z$ of the message as follows:
  \begin{algtwo}
    \item If $z = 0$ then
    \begin{algthree}
      \item Bob chooses $m$ values $a_1, \ldots, a_m$ uniformly at random from $\{0,1\}$ and computes the linear combination
      $x = \sum_{i=1}^m a_i v_i$.
      \item He reduces $x$ ``modulo $W$'', meaning he computes the unique vector $c$
      in $W$ such that $x - c$ is an integer linear combination of the $w_i$.
    \end{algthree}
    \item If $z = 1$ then
    \begin{algthree}
      \item Letting $2^{-n}\Zn = \{ 2^{-n}v \st v \in \Zn \}$, Bob selects a vector $c$ ``uniformly at random'' from $W \cap
      2^{-n}\Zn$.
    \end{algthree}
  \end{algtwo}
  \item In either case, the encrypted message is the vector $c$.
\end{algone}
\end{algorithm}

Each bit of the plaintext is essentially encoded as a decision problem: to decrypt the ciphertext
$c$, Alice (or an attacker) must decide whether $c$ is a linear combination of the $v_i$ (suitably
reduced) or a random vector.  More specifically, Alice can do the following:
\begin{algorithm}[Ajtai-Dwork Decryption]
\begin{algone}
  \item Alice computes $\ip{c}{u} = z + \delta$, where $z \in \Z$ and $-1/2 < \delta \leq 1/2$.
  \item If $\abs{\delta} < 1/n$ then $c$ is decrypted as $0$; otherwise it is decrypted as $1$.
\end{algone}
\end{algorithm}
In other words, if the inner product of the ciphertext and the private key is very close to an
integer, Alice decrypts the ciphertext as $0$.

\begin{theorem}
Ajtai-Dwork decryption works properly (with high probability).
\end{theorem}

For a complete proof of this result, refer to~\cite{AD1997:Cryptosystem}.

\begin{proof}[Sketch of Proof]
First, we justify that if the original message bit was $0$, it is always decrypted correctly. Using
the encryption procedure, Bob selects random $a_i$ and computes $x = \sum_{i=1}^m a_i v_i$. When he
reduces $x$ modulo $W$, he obtains the unique vector $c$ such that $x-c = w$, where $w =
\sum_{j=1}^n b_j w_j$ and the $b_j$ are integers.

Note that
\begin{align*}
  \ip{c}{u} &= \ip{x-w}{u} \\
            &= \ip{x}{u} - \ip{w}{u} \\
            &= \ip{\sum_{i=1}^m a_i v_i}{u} - \ip{\sum_{j=1}^n b_j w_j}{u} \\
            &= \sum_{i=1}^m a_i \ip{v_i}{u} - \sum_{j=1}^n b_i \ip{w_j}{u}.
\end{align*}

Since Alice chose the $v_i$ from the distribution $\mathcal{H}_u$ specifically so that their inner
product with $u$ was very close to an integer, we note that $\ip{v_i}{u}$ is ``close to'' an
integer for $1 \leq i \leq m$.  The $w_j$ are also chosen from $\mathcal{H}_u$, so $\ip{w_j}{u}$ is
``close to'' an integer for $1 \leq j \leq n$.  Finally, since the coefficients $a_i$ and $b_j$ are
all integers, the inner product $\ip{c}{u}$ is ``close to'' an integer.

With sufficient restrictions on the distribution $\mathcal{H}_u$, the authors of~\cite{AD1997:Cryptosystem} formalise this
intuitive argument and show that indeed this inner product is always within $1/n$ of an integer.  Hence if a $0$ is sent, Alice
always correctly recovers the plaintext.

We now consider the case where the original message bit was $1$.  In this case, Bob randomly
selects a vector $c$ from $W \cap 2^{-n}\Z^n$, so it is possible that the inner product of $c$ with
$u$ could be close to an integer.  As a result, there is a small probability that a $1$ could be
decrypted incorrectly as a $0$.  However, in \stepref{step:AjtaiDwork:KeyGen:ParallelpipedWidth} of
\algref{alg:AjtaiDwork:KeyGen} the parallelepiped $W$ was chosen to be wide enough that this event
occurs with probability at most $1/n$~\cite{AD1997:Cryptosystem}.  Thus, when a $1$ is sent,
decryption works properly with probability at least $1-1/n$.
\end{proof}

\end{section}

\begin{section}{Security Of The System}\label{sec:AjtaiDwork:Security}

It is interesting to note that despite its classification as a lattice-based cryptosystem, the
Ajtai-Dwork scheme does not explicitly use lattices to encrypt or decrypt data.  However, it is
usually considered to be a lattice-based cryptosystem because in~\cite{AD1997:Cryptosystem} its
security was shown to be based on the worst-case hardness of a problem in lattice reduction. Before
introducing this problem, we first present some basic definitions relating to lattices.

\begin{definition}
Let $B = \{ b_1, b_2, \ldots, b_d \}$ be a set of $d$ linearly independent vectors in $\Rn$.  The
\tbd{lattice} spanned by $B$ is the set of all possible integer linear combinations of the vectors
in $B$, denoted
\[
    L(B) = \left\{ \sum_{i=1}^d a_i b_i : a_i \in \Z, 1 \leq i \leq d \right\}.
\]
$B$ is called a basis for the lattice $L(B)$.  We say $L(B)$ has dimension $d$.
\end{definition}

Note that there are several possible bases for any given lattice.  For example, given a basis for a lattice $L$, if we take any
basis vector and add to it an integer linear combination of the other basis vectors, we obtain a different basis for the same
lattice.

\begin{definition}
Let $L$ be a lattice in $\Rn$.  The length of the shortest non-zero vector in $L$ (with respect to the Euclidean norm) is called
the \tbd{first minimum} of the lattice, and denoted $\lambda_1(L)$.
\end{definition}

\begin{definition}
Let $L$ be a $d$-dimensional lattice in $\Rn$.  For $1 \leq i \leq d$, the \tbd{$i^{th}$ successive minimum} of the lattice,
denoted $\lambda_i(L)$, is the smallest real number $a$ such that there exist $i$ linearly independent vectors in $L$ whose norms
are at most $a$.  In other words,
\[
    \lambda_i(L) = \min_{\substack{x_1, \ldots, x_i \in L \\
                                   \textup{\tiny and lin.indep.}}} \left\{ \max_{1 \leq j \leq i} \left\{ \norm{x_j}
    \right\} \right\}.
\]
\end{definition}

\begin{definition}
Let $L$ be a lattice in $\Rn$.  The \tbd{lattice gap} of $L$ is defined as the ratio between the second and first successive
minima, $\lambda_2(L) / \lambda_1(L)$.
\end{definition}

There are several problems that are thought to be hard problems in lattice theory; we mention two of the best-known such problems.

\begin{problem}[The Shortest Vector Problem (SVP)]
Given a lattice $L$ of dimension $d$ in $\Rn$, find a vector $v \in L$ such that $\norm{v} =
\lambda_1(L)$.
\end{problem}

There is no known polynomial time algorithm to solve SVP, or to approximate it to within a
polynomial factor.  The problem has been shown to be \NP-hard under randomised reductions; that is,
there is a probabilistic polynomial-time algorithm to reduce an instance of any problem in \NP\;to
an instance of SVP~\cite{A1998:SVPRandomisedReductions}.  In fact, approximating the problem to
within a factor of $\sqrt{2}$ is \NP-hard under randomised
reductions~\cite{NS2000:LatticesCryptology}. An important contrasting complexity result has been
proven as well: approximating SVP to within a factor of $\sqrt{d/\bigO{\log d}}$ is not \NP-hard
unless $\P=\NP$~\cite{GG1998:LatticeComplexity}. Despite all of these results, it has not been
proved or disproved that SVP is \NP-hard under deterministic reductions, and its \NP-hardness is an
important open question in lattice theory. The best known classical algorithms to approximate SVP
are based on the LLL algorithm~\cite{LLL1982:LLL} and its variants, which can approximate the
solution to within a factor of $2^{(d-1)/2}$.  In practice, the algorithm tends to outperform this
theoretical bound.

\begin{problem}[The Closest Vector Problem (CVP)]\label{prob:AjtaiDwork:CVP}
Given a lattice $L$ of dimension $d$ in $\Rn$ and a vector $u \in \Rn$, find a vector $v \in L$
such that $\norm{u-v}$ is minimised.
\end{problem}

As mentioned in~\cite{NS2000:LatticesCryptology}, this problem is known to be \NP-hard; in fact,
approximating the problem to within any constant factor is \NP-hard, and there is no known
polynomial-time algorithm that can approximate CVP to within a polynomial factor of $d$.  It is at
least as hard as SVP (since SVP is a special case of CVP where the given vector is $u = 0$) and
similarly approximating CVP to within a factor of $\sqrt{d/\bigO{\log d}}$ is not \NP-hard unless
$\P=\NP$~\cite{GG1998:LatticeComplexity}. There are algorithms that can approximate CVP in
$\mathbb{R}^n$ to within a factor of $2^{d/2}$ in the worst case; these algorithms are also based
on the LLL algorithm and its variants, and as mentioned previously they tend to outperform this
theoretical bound.

The hardness of the Ajtai-Dwork scheme is not based on either of these problems exactly, but rather on a variant of SVP:
\begin{problem}[The Unique Shortest Vector Problem (USVP)]
Given a lattice $L$ of dimension $d$ in $\Rn$ with lattice gap $\lambda_2(L)/\lambda_1(L)
> n^8$, find a vector $v \in L$ such that $\norm{v} = \lambda_1(L)$.
\end{problem}

The name for this new problem comes from the fact that the shortest vector in a lattice with such a
gap is ``unique'', in that it is polynomially shorter than any other non-parallel vector in the
lattice. In~\cite{AD1997:Cryptosystem} the following equivalence between USVP and the Ajtai-Dwork
cryptosystem is established: if for random instances of the cryptosystem there exists a
probabilistic polynomial-time algorithm that is capable of distinguishing an encryption of $0$ from
an encryption of $1$, then there exists a probabilistic polynomial-time algorithm to solve a
worst-case instance of USVP.

Despite this promising result, Nguyen and Stern have proven that one can construct a probabilistic
decryption algorithm for the Ajtai-Dwork cryptosystem, given an oracle capable of approximating CVP
to within a factor of $n^{1.33}$ (or equivalently an oracle capable of approximating SVP to within
a factor of $n^{0.5-\epsilon}$)~\cite{NS1998:ADCryptanalysis}.  Thus as pointed out
in~\cite{NS2000:LatticesCryptology}, since approximating CVP to within such a factor is not
\NP-hard~\cite{GG1998:LatticeComplexity} it is likely that breaking the Ajtai-Dwork cryptosystem is
not \NP-hard.

The result of Nguyen and Stern suggests that attacks on the system should be possible, and
indeed~\cite{NS1998:ADCryptanalysis} presents a heuristic attack on the scheme based on lattice
reduction algorithms.  The attack is based on the fact that for each $v_i$, $\ip{v_i}{u}$ is close
to some (unknown) integer, and by finding short linear combinations of the $v_i$, one can obtain
information about these unknown integers, which in turn reveals information about the private key
$u$. The implementation of the attack uses lattice reduction algorithms like the LLL algorithm.

For $n=8$, their experiments were able to recover the private key in under three hours, and for
$n=32$, the authors predicted that the attack would succeed in several days, if computations were
done on several machines in parallel. Further, since for $n=32$ storing the public key requires
approximately $20$ Megabytes and the ciphertext for each message bit is $768$ bytes long, the
scheme is impractical.

With a classical attack like this one already known, the question of the scheme's quantum
vulnerabilities becomes a question of purely theoretical interest.  There has been little work done
in applying quantum algorithms to these well-known lattice problems.  Some preliminary results were
proven in~\cite{ME1997:ShortLatticeVectors}, but they do not seem to provide much advantage in this
case.  More recently, in~\cite{R2002:QuantumComputationAndLattices} it was shown that USVP can be
reduced to the Hidden Subgroup Problem (HSP) in the dihedral group.  However, while some work has
been done on solving HSP in the dihedral group~\cite{EH2000:NoncommutativeHSP} there is still no
known efficient quantum algorithm to solve it completely.  The quantum tools from
\chapref{chap:IntroQuantumAlgs} which solve the Abelian HSP do not seem to be of much use to a
quantum attacker.

Grover's algorithm, introduced briefly in \chapref{chap:McEliece}, does not have a clear
application in this context, either.  The best known attacks on the system all rely on the LLL
algorithm, which does not seem well-suited to being sped up with Grover's algorithm.

If quantum algorithms can provide a quantum attacker with an advantage over a classical one, it is
likely that new algorithms will have to be developed.  The problems in lattice theory form one
class of hard problems for which the known quantum algorithms cannot significantly outperform the
known classical algorithms.  It is possible that lattice-based cryptosystems (that resist classical
attacks more successfully than the Ajtai-Dwork system) may be cryptosystems that are also resistant
to quantum attacks.
\end{section}

\end{chapter}

\begin{chapter}{The Goldreich-Goldwasser-Halevi Cryptosystem}\label{chap:GGH}

Like the Ajtai-Dwork cryptosystem, the Goldreich-Goldwasser-Halevi (GGH)
cryptosystem~\cite{GGH1997:Cryptosystem} is based on the hardness of problems in lattice reduction.
We can choose many different bases to represent the same lattice, and using a different basis can
make it much more difficult to solve particular instances of these problems.  It is this fact on
which the GGH cryptosystem is based.

The GGH algorithms for encryption and decryption are more efficient than the corresponding algorithms in the more popular RSA and ElGamal schemes; however, the
increased efficiency of encryption and decryption is offset by the fact that GGH public keys are considerably longer.

Recently, an attack has been discovered that successfully breaks the cryptosystem for most
practical parameter sizes~\cite{N1999:GGHCryptanalysis}. Despite the resulting impracticality of
the scheme, it is still an important cryptosystem from a theoretical point of view, since along
with the Ajtai-Dwork scheme, it was one of the first lattice-based cryptosystems.

\begin{section}{The Cryptosystem}\label{sec:GGH:Cryptosystem}

We begin by presenting some more definitions that are important to the study of lattice theory.

\begin{definition}
Let $\mathcal{B}$ be a $n \times n$ non-singular matrix with real entries, and let $L$ be the
$n$-dimensional lattice in $\Rn$ with the rows of $\mathcal{B}$ as a basis.  The determinant of the
lattice $L$ is defined to be the determinant of the matrix $\mathcal{B}$.
\end{definition}

Note that the determinant of the lattice is independent of the choice of basis.  We also define the orthogonality defect of a lattice basis, which is a
quantity that represents how ``non-orthogonal'' the basis vectors are.

\begin{definition}
Let $b_1, b_2, \ldots, b_n$ be a basis for an $n$-dimensional lattice $L$ in $\Rn$, and let $\mathcal{B}$ be the $n \times n$ non-singular matrix with the
$b_i$ as its rows. The orthogonality defect of the basis (or equivalently of the matrix $\mathcal{B}$) is defined as
\[
 \frac{ \prod_{i=1}^n \norm{b_i} }{ \abs{\det{\mathcal{B}}} }
\]
where $\norm{\cdot}$ represents the Euclidean norm.
\end{definition}

By Hadamard's Inequality~\cite{C1993:CourseInComputationalANT} we know that
$\abs{\det{\mathcal{B}}} \leq \prod_{i=1}^n \norm{b_i}$ with equality if and only if the $b_i$ are
orthogonal.  Thus a matrix $\mathcal{B}$ has orthogonality defect $1$ if and only if its rows are
orthogonal to one another, and otherwise, its orthogonality defect is greater than $1$. In other
words, the smaller the orthogonality defect, the more orthogonal the rows of $\mathcal{B}$.

In general, it is easier to solve most lattice problems (like SVP and CVP) if we have a basis with
vectors that are more orthogonal.  The idea of the GGH cryptosystem is that the public key consists
of a basis with \emph{high} orthogonality defect that Bob uses to encode a message in an instance
of CVP, and the private key consists of a basis with \emph{low} orthogonality defect that Alice
uses to solve the instance of CVP and recover the message.




We now present the cryptosystem; for more details, refer to~\cite{GGH1997:Cryptosystem}.

To generate a GGH key, Alice does the following:
\begin{algorithm}[GGH Key Generation]
\begin{algone}
\item Using a procedure described in~\cite{GGH1997:Cryptosystem}, Alice generates a full-rank lattice $L$, and two matrices $\mathcal{R}$ and $\mathcal{B}$ whose rows
form bases for $L$.  The generation procedure ensures that $\mathcal{B}$ has high orthogonality defect, and $\mathcal{R}$ has low orthogonality defect.
\item She also selects a positive integer $\sigma$, as described in~\cite{GGH1997:Cryptosystem}, which acts as a security parameter.
\item Her public key is $(\mathcal{B},\sigma)$ and her private key is $\mathcal{R}$.
\end{algone}
\end{algorithm}

To encrypt a message for Alice using the GGH cryptosystem, Bob does the following:
\begin{algorithm}[GGH Encryption]
\begin{algone}
\item Bob obtains Alice's public key $(\mathcal{B},\sigma)$.
\item Bob represents his message as a vector $m \in \Zn$.
\item He generates an error vector $e$ by setting each entry in $e$ to either $\sigma$ or $-\sigma$, each with probability $1/2$.
\item He computes the ciphertext $c=m\mathcal{B}+e$.
\end{algone}
\end{algorithm}

The error vector $e$ disguises the message $m$ from an attacker; however, it is designed to be
small enough that $m$ is still the closest vector in the lattice to $c$.  Ideally, the public basis
is not ``orthogonal enough'' to allow an attacker to find that closest vector, but Alice can use
her private, more orthogonal basis to find it. More specifically, to decrypt the ciphertext, Alice
does the following:
\begin{algorithm}[GGH Decryption]
\begin{algone}
\item Alice represents $c$ as a linear combination of the vectors in $\mathcal{R}$, where the coefficients are not
necessarily integers.
\item She rounds off each coefficient in the linear combination to the nearest integer and obtains a lattice vector $v$.
\item She represents $v$ as a linear combination of the columns of $\mathcal{B}$.
\item With high probability, the coefficients of this linear combination are the entries in the message vector $m$.
\end{algone}
\end{algorithm}

We note that it is possible for the decryption procedure to fail, since the rounding off technique
(which was proposed by Babai in~\cite{B1986:NearestPoint}) may not result in the correct lattice
point.  However, the authors of~\cite{GGH1997:Cryptosystem} show that by selecting the parameter
$\sigma$ properly, Alice can ensure that decryption works with high probability.  It is also true
that when decryption fails, Alice can detect that it has failed. For more details about these
facts, see~\cite{GGH1997:Cryptosystem}.

Note that despite the fact that they are based on the hardness of different problems, the GGH cryptosystem and the McEliece cryptosystem (described in
\chapref{chap:McEliece}) are quite similar. In the GGH scheme, the public and private keys are different representations of the same lattice, and in the
McEliece scheme, the public and private keys are different representations of the same linear code.  In both cases, encrypting a message corresponds to
performing a transformation involving the public key representation and adding a random error vector in such a way that it can easily be removed only with
knowledge of the private key representation.

\end{section}

\begin{section}{Security Of The System}

First consider the task of determining the plaintext given only the ciphertext.  Correctly
performing this task amounts to solving an instance of CVP: the eavesdropper, given only
$c=m\mathcal{B}+e$ needs to first find $m\mathcal{B}$, which (assuming $\sigma$ is not too large)
is the vector in the lattice closest to $c$.  As mentioned previously, there is no known polynomial
time algorithm to solve CVP exactly, or to approximate it to within a polynomial factor.

Next note that by construction the public basis $\mathcal{B}$ has high orthogonality defect and the
private basis $\mathcal{R}$ has low orthogonality defect.  Thus in order to determine the private
key given only the public key, an eavesdropper would need to solve (or at least approximate a
solution to) a different problem that is also thought to be hard:
\begin{problem}[The Smallest Basis Problem (SBP)]\label{prob:GGH:SBP}
Given a basis $B$ for a lattice $L$ in $\mathbb{R}^n$, find the ``smallest'' basis $B'$ for $L$.
\end{problem}
There are many ways that the ``smallest'' basis of a lattice could be defined (although in this
case we consider the basis with the smallest orthogonality defect). As with the other lattice
problems defined previously, there are no known polynomial-time algorithms to solve SBP or to
approximate it to within a polynomial factor, although there are algorithms based on the LLL
algorithm and its variants that can approximate SBP in $\Rn$ to within a factor of $2^{\bigO{n^2}}$
in the worst case.

Originally the authors of the cryptosystem suggested three classical attacks on the system, each of
which is shown to require an infeasible amount of work in sufficiently high dimension. We briefly
mention these attacks, and refer the reader to~\cite{GGH1997:Cryptosystem} for more details.  All
of the attacks assume that the public basis $\mathcal{B}$ has been reduced to a new basis
$\mathcal{B}'$ (with smaller orthogonality defect) using the LLL algorithm or one of its variants,
since this is a logical first step for any solution to the problems on which the cryptosystem is
based.

In the first attack, Eve uses the reduced basis $\mathcal{B}'$ to perform the same rounding off
technique as Alice uses in the decryption procedure with the private basis $\mathcal{R}$. The
vector Eve obtains will be an approximation to the correct message vector, and can be used as a
starting point for an exhaustive search for the message. According to experiments cited
in~\cite{GGH1997:Cryptosystem}, in dimensions up to 80 this attack works well since the LLL
algorithm tends to perform very well in practice, but in higher dimensions the attack quickly
becomes infeasible since a measure of the work required grows exponentially with the dimension.

The second attack also involves using the reduced basis $\mathcal{B}'$ to approximate CVP, but
using Babai's ``nearest plane'' algorithm (a better approximation algorithm also proposed
in~\cite{B1986:NearestPoint}).  Essentially, whereas the rounding off algorithm rounds off all of
the coefficients of the resulting vector at the same time, the nearest plane algorithm rounds them
off one by one in a more adaptive way.  Again according to the experiments performed by the authors
of~\cite{GGH1997:Cryptosystem} the attack is much more successful than the rounding off attack and
is generally successful in dimensions up to 120, but again in higher dimensions the work required
grows exponentially.

To perform the third attack, for a particular ciphertext $c = m\mathcal{B} + e$ Eve creates a new
lattice $L'$ of dimension $n+1$ as specified by the rows of the matrix
\[
    \mathcal{B'} = \begin{pmatrix}b_1 & 0 \\ \vdots & \vdots \\ b_n & 0 \\ c & 1 \end{pmatrix}.
\]
The vector $v = e \Vert (1)$ (where the operator $\Vert$ indicates vector concatenation) is a short
vector in $L'$, and in fact, as explained in~\cite{N1999:GGHCryptanalysis} it is likely that it
will be the shortest vector in $L'$.  Thus if we attempt to solve SVP in $L'$ using the LLL
algorithm or one of its variants, we hope that we will find the vector $v$, from which we can
recover $m$. (Note that this attack is a general approach to solving CVP by ``embedding'' an
instance of CVP in an instance of SVP.)  Unlike in the first two attacks, if this heuristic fails
to recover the correct message, it is not clear whether the incorrect message can be used as a
starting point for an exhaustive search. Nonetheless, the attack seems to be fast and successful in
dimensions up to about 120.

Based on their experiments, the authors of~\cite{GGH1997:Cryptosystem} conjectured that the problem
of breaking the cryptosystem was intractable in dimension $300$ or higher.  However,
in~\cite{N1999:GGHCryptanalysis} the author presents a different attack that exploits some
weaknesses in the encryption scheme and does break it in higher dimensions. Recall that $c =
m\mathcal{B} + e$ where $e$ is a vector with each entry equal to $\pm \sigma$. Defining $s = (
\sigma, \ldots, \sigma ) \in \Zn$ we see that $e + s \equiv 0 \pmod{2\sigma}$ and hence $c + s
\equiv m\mathcal{B} \pmod{2\sigma}$.  It is further shown in~\cite{N1999:GGHCryptanalysis} that
this modular equation has very few solutions with high probability, and it is not hard to compute
all of them. So we can easily determine $m \bmod{2\sigma}$. With this partial information, we can
simplify the CVP instance defined by a ciphertext and obtain a new CVP instance where the error
vector is much shorter than $e$. Then by applying the embedding technique (or some other algorithm
for CVP) we are more likely to be able to recover the original message.  As predicted, experiments
cited in~\cite{N1999:GGHCryptanalysis} indicate that this attack can break the scheme in dimensions
up to about 400.  In dimensions higher than 400, the parameters become so large as to make the
scheme practically infeasible.

As with the Ajtai-Dwork cryptosystem presented in \chapref{chap:AjtaiDwork}, quantum attacks on the
GGH cryptosystem are only theoretically interesting, since there are effective classical attacks
against the scheme.  Again, however, there seem to be very few known quantum algorithms that could
assist an attacker further.  The Abelian Hidden Subgroup Problem (HSP) framework does not seem
useful, nor does Grover's searching algorithm.  The reduction
in~\cite{R2002:QuantumComputationAndLattices} is interesting in this context, although it is not
currently useful to a quantum attacker since we know of no efficient algorithm to solve HSP in the
dihedral group. Again, the current evidence suggests that lattice-based cryptosystems could perhaps
be systems that resist quantum attacks as well as they do classical ones.
\end{section}

\end{chapter}

\begin{chapter}{The NTRU Cryptosystem}\label{chap:NTRU}

The NTRU cryptosystem~\cite{HPS1998:NTRUCryptosystem} is a relatively new cryptosystem that uses
polynomial arithmetic for encryption and decryption.  One of the most efficient known classical
attacks on the cryptosystem is based on a problem in lattice reduction.  Because of this attack,
the scheme is often referred to as a ``lattice-based'' cryptosystem, even though the description of
the system does not rely on lattices.

The cryptosystem has the interesting property that there exist valid ciphertexts that cannot be
decrypted properly using the private key.  For this reason, many of the security properties that
can be proven for traditional public key encryption schemes do not hold for the NTRU cryptosystem.
In~\cite{P2003:NTRUImperfectDecryption} a new class of encryption schemes is defined called
imperfect public key encryption schemes, which allow for the possibility of such ``indecipherable''
ciphertexts.  The paper also presents a new attack on the scheme that attempts to recover the
private key by searching for indecipherable ciphertexts.  In experiments, this attack has been
successful against the system parameter sets originally suggested
in~\cite{HPS1998:NTRUCryptosystem}.

\begin{section}{The Cryptosystem}\label{sec:NTRU:Cryptosystem}

We work in the ring $\Z[x]/(x^N-1)$ for some integer $N$.
We first define a notation for classes of polynomials in this ring:
\begin{definition}
The set $\mathcal{L}(d_1, d_2)$ is the set of polynomials in $\Z[x]/(x^N-1)$ with $d_1$
coefficients equal to 1, $d_2$ coefficients equal to $-1$, and the remaining coefficients equal
to~$0$.
\end{definition}

To generate an NTRU key, Alice performs the following steps:
\begin{algorithm}[NTRU Key Generation]
\begin{algone}
  \item Alice selects two coprime integers $p$ and $q$ with $q$ considerably larger than
  $p$, and an integer $N$.  She also selects integers $d_f$, $d_g$, and $d_r$ considerably smaller than $N$.
  (These parameters may be chosen to provide the desired level of security for the cryptosystem as described in~\cite{HPS1998:NTRUCryptosystem}.)
  \item She randomly selects two polynomials $F \in \mathcal{L}(d_f, d_f-1)$ and $G \in \mathcal{L}(d_g, d_g)$.
  \item She computes the polynomials $F_p^{-1}$ and $F_q^{-1}$, the inverses of $F$ modulo $p$ and
  modulo $q$, respectively. That is, $F F_p^{-1} = 1$ (when the coefficients are taken modulo $p$) and $F F_q^{-1} = 1$
  (when the coefficients are taken modulo $q$). (Such inverses
  will exist with high probability; if they do not, she begins the procedure again.)
  \item Alice calculates $H = F_q^{-1} G \bmod q$.
  \item Alice's public key is $( p, q, N, d_r, H )$, and her private key is $F$.
\end{algone}
\end{algorithm}

To encrypt a message for Alice using the NTRU cryptosystem, Bob performs the following steps:
\begin{algorithm}[NTRU Encryption]
\begin{algone}
  \item Bob obtains Alice's public key.
  \item He converts the message to a polynomial $M \in \Z[x]/(x^N-1)$ with coefficients in the range
  $\left[\, -\frac{p-1}{2}, \frac{p-1}{2} \,\right]$.
  \item He selects a random polynomial $R \in \mathcal{L}(d_r, d_r)$.
  \item Bob computes the encrypted message $C = \paren{ p \, R H + M } \bmod q$.
\end{algone}
\end{algorithm}

To decrypt the ciphertext and recover the original message, Alice does the following:
\begin{algorithm}[NTRU Decryption]
\begin{algone}
  \item She computes the polynomial $A = F C \bmod{q}$, choosing the coefficients of $A$ to be integers in the interval
  $\left[\, -\frac{q}{2}, \frac{q}{2} \,\right]$.
  \item She computes $M' = F_p^{-1} A \bmod p$.
  \item With high probability, $M'$ is the original message.
\end{algone}
\end{algorithm}

\begin{theorem}
NTRU decryption works properly (with high probability).
\end{theorem}

\begin{proof}[Justification]
As mentioned above, there exist certain ciphertexts that cannot be properly decrypted using the
private key.  Although not proven rigorously, there is heuristic evidence that such indecipherable
ciphertexts occur rarely, as demonstrated in~\cite{S2001:NTRUWrapGapFailure}
and~\cite{S2002:SilvermanPrivateCommunication}.

Note that
\begin{align*}
    A &= F C \bmod{q}\\
      &= ( F p \, R H + F M ) \bmod{q} \\
      &= ( F p \, R F_q^{-1} G + F M ) \bmod{q} \\
      &= ( p \, R G + F M ) \bmod{q}.
\end{align*}

Suppose the coefficients of the polynomial $A = ( p \, R G + F M ) \bmod{q}$ computed by Alice
happen to be exactly the same as those of the unreduced polynomial $B = p \, R G + F M$. In that
case, the decryption algorithm will work properly, since reducing $A$ modulo $p$ produces the
polynomial $F M$, and multiplying by $F_p^{-1}$ correctly recovers $M$. In other words, we wish to
show that we can choose parameters for the system so that the polynomial $A$ computed by Alice is
exactly equal to the polynomial $B$ in $\Z[x]/(x^N-1)$. Since Alice computes $A$ choosing its
coefficients to lie in the interval $\left[\, -\frac{q}{2}, \frac{q}{2} \,\right]$, it is
sufficient to ensure that with high probability the coefficients of $B$ lie in the same interval.

In~\cite{S2001:NTRUWrapGapFailure} the ways in which this sufficient condition may not be met are
classified into two categories:

\begin{enumerate}
\item ``Wrapping failure'' is said to occur if the maximum coefficient of $B$ is greater than or equal to
$q/2$, or if the minimum coefficient of $B$ is less than or equal to $-q/2$.  In this case, when
Alice chooses the coefficients of $A$ to be in the interval $\left[\, -\frac{q}{2}, \frac{q}{2}
\,\right]$, she will not obtain $B$, and hence she will obtain the incorrect decrypted message. (To
allow her to detect such a failure, the authors of~\cite{HPS1998:NTRUCryptosystem} suggest
including some kind of redundancy in the message so that its proper decryption can be verified.)

\item ``Gap failure'' is said to occur if the difference between the maximum and minimum coefficients of
$B$ (called the ``spread'' of $B$~\cite{S2001:NTRUWrapGapFailure}) is greater than $q$.  In this
case, when Alice chooses the coefficients of $A$ to be in \emph{any} interval of width $q$, she
will not obtain $B$, and hence she will obtain the incorrect decrypted message.
\end{enumerate}

For a particular parameter set, the probability of wrapping or gap failure can be estimated by
performing many encryptions of random messages and calculating the proportion of them that exhibit
each type of failure when decryption is attempted.  This is the strategy employed
in~\cite{S2001:NTRUWrapGapFailure} and for the parameter sets suggested in that paper the estimates
for the failure probabilities are indeed low (on the order of $10^{-5}$ to $10^{-6}$ for wrapping
failure and $10^{-9}$ to $10^{-13}$ for gap failure). It should be noted that these are empirical
estimates, and that the gap failure probabilities in~\cite{S2001:NTRUWrapGapFailure} were
calculated using an approximation formula (whose correctness is justified further
in~\cite{S2002:SilvermanPrivateCommunication}) since the chance of actually observing an instance
of gap failure is so small.

This evidence indicates that indeed NTRU decryption tends to work properly in practice.
\end{proof}

It should be noted, however, that while the probability of obtaining one of these indecipherable
ciphertexts may indeed be small, an attacker can use one of them to obtain information about the
corresponding private key~\cite{P2003:NTRUImperfectDecryption}.

\end{section}

\begin{section}{Security Of The System}\label{sec:NTRU:Security}

NTRU is usually considered to be a lattice-based cryptosystem; despite the fact that lattices are
not used in the encryption or decryption algorithms, one of the most efficient known classical
attacks on the cryptosystem is based on finding short vectors in a lattice. We present the attack
briefly below, as it is presented in~\cite{HPS1998:NTRUCryptosystem}.

Recall that Alice's public key is $( p, q, N, d_r, H )$.  Let the coefficients of $H$ be given by
$h_0, h_1, \ldots, h_{N-1}$ so that $H = \sum_{i=0}^{N-1} h_i x^i$.  We define the following $2N
\times 2N$ matrix, where $\alpha$ is a parameter chosen by the attacker:
\[
    \mathcal{B} = \left( \begin{array}{cccc|cccc}
                    \alpha & 0 & \cdots & 0 & h_0 & h_1 & \cdots & h_{N-1} \\
                    0 & \alpha & \cdots & 0 & h_{N-1} & h_0 & \cdots & h_{N-2} \\
                    \vdots & \vdots & \ddots & \vdots & \vdots & \vdots & \ddots & \vdots \\
                    0 & 0 & \cdots & \alpha & h_1 & h_2 & \cdots & h_0 \\ \hline
                    0 & 0 & \cdots & 0 & q & 0 & \cdots & 0 \\
                    0 & 0 & \cdots & 0 & 0 & q & \cdots & 0 \\
                    \vdots & \vdots & \ddots & \vdots & \vdots & \vdots & \ddots & \vdots \\
                    0 & 0 & \cdots & 0 & 0 & 0 & \cdots & q
                  \end{array} \right).
\]
Let the rows of $\mathcal{B}$ be $b_0, b_1, \ldots b_{2N-1}$.  Recall that $H = F_q^{-1} G \bmod
q$, so $G = H F \bmod q$.  In other words, there exists some polynomial $K \in \Z[x]/(x^N-1)$ such
that $G = H F + q K$. Let the coefficients of $F$ be $f_0, f_1, \ldots, f_{N-1}$ and the
coefficients of $K$ be $k_0, k_1, \ldots, k_{N-1}$; these coefficients are all integers. Then note
that
\begin{align*}
    &\phantom{=} f_0 b_0 + f_1 b_1 + \cdots + f_{N-1} b_{N-1} + k_0 b_N + k_1 b_{N+1} + \cdots + k_{N-1} b_{2N-1} \\
    &= ( \alpha F ) || ( H F + q K )\\
    &= ( \alpha F ) || \,G
\end{align*}
where the operator $||$ indicates vector concatenation. In other words, if we let $L$ be the lattice spanned by the rows of
$\mathcal{B}$, we see that the vector $\tau = (\alpha F) || \, G$ is in $L$. The goal of the attacker will be to choose $\alpha$
so that $\tau$ is a short vector in the lattice $L$ and to attempt to find it using lattice reduction techniques like the LLL
algorithm~\cite{LLL1982:LLL} and its variants.

In~\cite{HPS1998:NTRUCryptosystem} the authors next make use of the Gaussian heuristic, which does
not seem to be well-known, but which bounds the expected length of the shortest vector in a
``random'' lattice of dimension $d$.  The heuristic says that a sphere that contains a lattice
point at its centre and exactly one other lattice point is expected to have a volume equal to the
determinant $D$ of the lattice; the radius of such a sphere clearly provides an upper bound on the
shortest vector in the lattice~\cite{W2003:WhytePrivateCommunication}.  Specifically, the heuristic
says that the expected length of the shortest vector in a random lattice of dimension $d$ and
determinant $D$ is between $D^{1/d} \sqrt{\frac{d}{2 \pi e}}$ and $D^{1/d} \sqrt{\frac{d}{\pi e}}$.

In our case,  the determinant of $L$ is equal to $\det \mathcal{B} = \alpha^N q^N$; we also have $d
= 2N$. Thus the expected length of the shortest vector in $L$ should be close to
\begin{align*}
    s &= ( \alpha^N q^N )^{1/2N} \sqrt{\frac{2N}{2 \pi e}} \\
      &= ( \alpha q )^{1/2} \sqrt{\frac{N}{\pi e}} \\
      &= \sqrt{\frac{N \alpha q}{\pi e}}.
\end{align*}

In order for the lattice reduction algorithms to have the greatest chance of finding the vector
$\tau$, the attacker would like to maximise the probability that it is one of the shortest vectors
in the lattice.  This will be likely if $\tau$ is considerably shorter than this expected length of
the shortest vector.  In other words, the attacker should choose $\alpha$ to maximise the ratio
$s/\norm{\tau}$.  Note that $\norm{\tau} = \sqrt{\alpha^2 \norm{F}^2 + \norm{G}^2}$, where the norm
of a polynomial is taken to mean the norm of the vector of its coefficients.  Thus,
\begin{equation}\label{equ:NTRU:ExpectedToTargetRatio}
    \frac{s}{\norm{\tau}} = \sqrt{ \frac{ N \alpha q }{ \pi e ( \alpha^2 \norm{F}^2 + \norm{G}^2 ) } }.
\end{equation}

Since $N$, $q$, $\pi$, and $e$ are all fixed, the attacker should attempt to maximise
\[
    \frac{\alpha}{\alpha^2 \norm{F}^2 + \norm{G}^2} = (\alpha \norm{F}^2 + \alpha^{-1} \norm{G}^2)^{-1}.
\]
Differentiating the expression with respect to $\alpha$ and setting it equal to zero, we see that
it is maximised when $\alpha = \norm{G}/\norm{F}$. We assume that the attacker has knowledge of
$\norm{F}$ and $\norm{G}$ (or equivalently of $d_f$ and $d_g$) which is not an unrealistic
assumption since the values of $d_f$ and $d_g$ are specified in the sets of suggested system
parameters listed in~\cite{HPS1998:NTRUCryptosystem}.  The attacker can therefore compute this
optimal value for $\alpha$, and proceed to use the LLL algorithm to find short vectors in $L$ (as
described for example in~\cite{C1993:CourseInComputationalANT}).

In~\cite{HPS1998:NTRUCryptosystem} this optimal value for $\alpha$ is substituted back into
\equref{equ:NTRU:ExpectedToTargetRatio} to obtain the constant
\[
    c = \sqrt{\frac{Nq}{2 \pi e \norm{F} \norm{G}}}.
\]
It is noted that this $c$ can be used as a measure of the ``randomness'' of the lattice defined by $\mathcal{B}$.  If $c$ is close
to $1$, the vector $\tau$ is not considerably larger than the expected length of the shortest vector in a random lattice, and so
in that sense, $L$ is fairly ``random'' and typical reduction algorithms should work less effectively than when $c$ is larger.

Based on limited evidence, it would appear as though the time required for this attack is still exponential in $N$, with a constant in the exponent
proportional to $1/c$~\cite{HPS1998:NTRUCryptosystem}.  In~\cite{M1999:NTRUCryptanalysis} a modification of this attack was proposed that requires a lattice of
smaller dimension, and as a result the attack runs more quickly.  The new attack is especially successful against certain classes of keys, even when using
parameters of a size that were originally thought to provide high security.  These classes of keys should therefore be avoided.

The imperfection of the decryption algorithm has recently been shown to be a serious weakness of
the scheme.  The attack proposed in~\cite{P2003:NTRUImperfectDecryption} is effective against the
parameter sets proposed in~\cite{HPS1998:NTRUCryptosystem} provided that the attacker has access to
an oracle that given a ciphertext returns only whether the ciphertext could be properly decrypted
using the corresponding private key.  It is therefore desirable to choose parameter sets that
minimise the probability of obtaining such an indecipherable ciphertext, or in other words, to
minimise the probability of wrapping failure and gap failure.

Another strategy to avoid such an attack is to perform further processing on indecipherable
ciphertexts in an attempt to recover the correct plaintext.  Examples of this further processing
are suggested in~\cite{S2001:NTRUWrapGapFailure}.  If wrapping failure occurs, Alice may
re-calculate the polynomial $A$ with coefficients in the interval $\left[\, -\frac{q}{2}+x,
\frac{q}{2}+x \,\right]$ for various (positive and negative) values of $x$ and try again. Provided
gap failure has not occurred, this correction mechanism will likely succeed for some small value of
$x$, and Alice will be able to recover the correct plaintext.  Correction for gap failure is more
difficult than for wrapping failure, since in order to obtain $B$ Alice would have to move
\emph{some} of the coefficients of $A$ outside the interval $\left[\, -\frac{q}{2}, \frac{q}{2}
\,\right]$ and try again.  Since the set of coefficients that need to be moved is unknown, this
correction method is much less feasible.

However, in both of these cases, as noted in~\cite{P2003:NTRUImperfectDecryption}, the attacker may
still be able to use timing and power analysis to determine when further processing is required,
and hence when a ciphertext was not decipherable using the standard decryption algorithm.  In that
case, the attack could still be ultimately successful.

It is unclear whether the system is more vulnerable in a quantum setting; the algorithms from
\chapref{chap:IntroQuantumAlgs} do not seem to provide the quantum attacker with any useful tools.
As discussed with respect to some of the previous schemes, it may be possible to use Grover's
algorithm to speed up the known classical attacks (or parts of them).  In the case of the first
attack, such an improvement is not immediately obvious since the majority of the running time is
spent in the LLL algorithm (which as we have mentioned previously is not easily improved upon with
quantum resources).  In the case of the second attack, such an improvement might be more feasible
since the initial search for an indecipherable ciphertext could possibly be sped up by a square
root factor. Other steps of the second attack, such as modifying the first indecipherable
ciphertext to find another one which is ``nearly decipherable'', might also run faster using
Grover's algorithm.

\end{section}
\end{chapter}

\begin{chapter}{A Quantum Public Key Cryptosystem}\label{chap:Okamoto}

All of the cryptosystems presented so far have been classical cryptosystems, in that they use only classical operations in all of the key generation,
encryption, and decryption algorithms.  If we wish to find cryptosystems that resist attacks with a quantum computer, however, it seems natural to also allow
the use of quantum operations in any of the three algorithms.  The cryptosystem presented in~\cite{OTU2000:QuantumPKC} and summarised below uses some quantum
operations to generate keys, and then uses purely classical algorithms to encrypt and decrypt messages.  We will refer to this scheme as the Quantum Public Key
cryptosystem (QPKC).

\begin{section}{The Cryptosystem}\label{sec:Okamoto:Cryptosystem}

Before presenting the cryptosystem, we first present some definitions and results from algebraic number theory that are important to understanding the
cryptosystem.  First we introduce a few concepts related to algebraic numbers and algebraic integers.

\begin{definition}
Let $\alpha \in \C$.  Then we say $\alpha$ is an \tbd{algebraic number} if there exists a non-zero
$p \in \Z[x]$ such that $p(\alpha) = 0$. Further, if $p$ can be chosen to be monic (that is, with a
leading coefficient of $1$) then we say $\alpha$ is an \tbd{algebraic integer}.
\end{definition}

\begin{definition}
Let $\alpha$ be an algebraic number.  Let $m \in \Z[x]$ be chosen such that $m(\alpha) = 0$, the leading coefficient of $m$ is positive, and the coefficients
of $m$ are coprime.  If we further choose $m$ to be of minimal degree, then $m$ is unique and irreducible and called the \tbd{minimal polynomial} of $\alpha$.
\end{definition}

\begin{definition}
Let $R \subseteq \C$.  The \tbd{set of integers} of $R$, denoted $\integers{R}$, is the
intersection of $R$ with the set of all algebraic integers.  If $R$ is a ring, then $\integers{R}$
is also a ring.
\end{definition}

We also introduce the concept of a number field, and the embedding of a number field in $\C$.

\begin{definition}
A \tbd{number field} $K$ is a subfield of $\C$ which is finite-dimensional when considered as a vector space over $\Q$.  The \tbd{degree} of $K$ is the
dimension of this vector space.
\end{definition}

\begin{proposition}
Let $K$ be a number field of degree $n$.  There exists $\theta \in K$ such that $K = \Q[\theta]$,
and the minimal polynomial of $\theta$ has degree $n$. There exist exactly $n$ embeddings of $K$ in
$\C$, which are maps in which $\theta \mapsto \theta_i$ for $i = 1, \ldots, n$, where the
$\theta_i$ are the distinct roots in $\C$ of the minimal polynomial of $\theta$.
\end{proposition}

\begin{definition}
Let $K$ be a number field of degree $n$.  Let $\sigma_1, \ldots, \sigma_n$ denote the embeddings of $K$ in $\C$.  For any $\alpha \in K$, the \tbd{norm} of
$\alpha$ is given by
\[
    \mathcal{N}(\alpha) = \prod_{i=1}^n \sigma_i(\alpha).
\]
\end{definition}

We now define ideals, prime ideals, cosets, and quotient rings.
\begin{definition}\label{def:Okamoto:Ideal}
Let $R$ be a ring.  An \tbd{ideal} of $R$ is a subset $I \subseteq R$ with the following properties:
\begin{enumerate}
    \item $I$ is a subgroup of $(R,+)$, and
    \item if $a \in I$ and $r \in R$ then $ra \in I$.
\end{enumerate}
\end{definition}

\begin{definition}
An ideal $I$ of a ring $R$ is called a prime ideal if $I \neq R$ and $ab \in I$ implies $a \in I$ or $b \in I$.
\end{definition}

\begin{definition}
Let $I$ be an ideal of a ring $R$.  The set $a + I = \brace{ a + x \st a \in R, x \in I }$ is
called the coset of $I$ corresponding to $a$.  Addition and multiplication of cosets are defined as
follows:
\begin{itemize}
\item $\paren{a_1+I} + \paren{a_2+I} = \paren{a_1+a_2} + I$
\item $\paren{a_1+I} \cdot \paren{a_2+I} = \paren{a_1 \cdot a_2} + I$
\end{itemize}
\end{definition}

\begin{proposition}
Let $I$ be an ideal of a ring $R$.  The set of cosets of $I$ is a ring under the operations of
addition and multiplication defined above.  This new ring is called a quotient ring and is denoted
$R/I$.
\end{proposition}

Finally we mention three more well-known results that we will use later.  The second of these
results is a rewording of Proposition~1 from~\cite{OTU2000:QuantumPKC}.  The third is a
generalisation of Fermat's Little Theorem.

\begin{proposition}\label{prop:Okamoto:PrimeIdealQuotient}
Let $K$ be a number field and let $\mathfrak{p}$ be a non-zero prime ideal of $\integers{K}$.  Then
$\integers{K}/\mathfrak{p}$ is a finite field.  The cardinality of $\integers{K}/\mathfrak{p}$ is
called the \tbd{norm} of $\mathfrak{p}$ and denoted $\mathcal{N}(\mathfrak{p})$.
\end{proposition}

\begin{proposition}\label{prop:Okamoto:UniqueRep}
Let $K$ be a number field of degree $n$ and let $\mathfrak{p}$ be a prime ideal of $\integers{K}$.
Then there exist elements $\omega_1, \ldots, \omega_n \in \integers{K}$ and $e_1, \ldots, e_n \in
\Z$ such that the elements of $\integers{K}/\mathfrak{p}$ are uniquely represented by the elements
of
\[
    R = \left\{ \sum_{i=1}^{n} a_i \omega_i \st 0 \leq a_i < e_i, i = 1, \ldots, n \right\}.
\]
\end{proposition}

\begin{proposition}
Let $\mathfrak{p}$ be a prime ideal of $\integers{K}$, and let $g$ be a non-zero element from
$\integers{K}/\mathfrak{p}$.  Then $g^{N(\mathfrak{p})-1} \equiv 1 \pmod{\mathfrak{p}}$.
\end{proposition}

We now present the cryptosystem.  The steps basically correspond to the steps in~\cite{OTU2000:QuantumPKC}, although some minor variations have been made for
clarity.

To generate a key in this quantum public key cryptosystem, Alice performs the following steps:
\begin{algorithm}[QPKC Key Generation]\label{alg:Okamoto:KeyGen}
\begin{algone}
  \item Alice selects a set $\mathcal{K}$ of number fields, and integers $n$ and $k$.  (These parameters may be chosen to provide the desired level of security for the cryptosystem.)
  \item She randomly selects an algebraic number field $K$ from $\mathcal{K}$.
  \item She selects a prime ideal $\mathfrak{p}$ of $\integers{K}$, and a generator $g$ of the multiplicative group of the finite field $\integers{K}/\mathfrak{p}$.
  \item She chooses $n$ elements $p_1, \ldots , p_n$ from $\integers{K}/\mathfrak{p}$ such that the following two conditions are satisfied:
  \begin{algtwo}
    \item $\mathcal{N}(p_1), \ldots, \mathcal{N}(p_n)$ are coprime, and\label{step:Okamoto:KeyGen:CoprimeNormCondition}
    \item For any subset $\{ i_1, \ldots, i_k \} \subset \{ 1, \ldots, n \}$, the product $\prod_{j=1}^{k} p_{i_j}$
    is in the set $R$ defined in \propref{prop:Okamoto:UniqueRep}.\label{step:Okamoto:KeyGen:UniqueRepCondition}
  \end{algtwo}
  \item Alice uses the quantum algorithm for finding discrete logarithms described in \secref{sec:IntroQuantumAlgs:DLP}
  to find $q_1, \ldots, q_n$ such that $p_i \equiv
  g^{q_i} \pmod{\mathfrak{p}}$, where $q_i \in \Z_{\mathcal{N}(\mathfrak{p})-1}$ for $i = 1, \ldots, n$.
  \item She randomly selects a rational integer $d$ in $\Z_{\mathcal{N}(\mathfrak{p}) - 1}$, and computes the values $b_i = (q_i + d) \bmod (\mathcal{N}(\mathfrak{p})-1)$ for $i =
  1, \ldots, n$.
  \item Alice's public key is $(\mathcal{K}, n, k, b_1, \ldots, b_n)$ and her private key is $( K, \mathfrak{p}, g, d, p_1, \ldots, p_n )$.
\end{algone}
\end{algorithm}

Note that the condition in \stepref{step:Okamoto:KeyGen:UniqueRepCondition} seems complicated to
check, but based on the set of number fields selected, it may be possible to simplify it.  For
example, in~\cite{OTU2000:QuantumPKC} the authors present a version of this scheme that sets
$\mathcal{K}$ to be the set of all imaginary quadratic number fields.  In this particular case, by
using some further results from number theory, it can be shown that checking the condition amounts
to verifying that some bounds are met on the size of the norms of the $p_i$.  Similar
simplifications may be possible for other choices of $\mathcal{K}$, and one general method is
presented in~\cite{OTU2000:QuantumPKC} (although this method results in an encryption scheme with a
low information rate).

To encrypt a message for Alice, Bob performs the following steps:
\begin{algorithm}[QPKC Encryption]
\begin{algone}
  \item He starts with a message $m$ of length $\lfloor \log \binom{n}{k} \rfloor$ bits.
  \item He uses the following procedure to encode $m$ into a binary string $s = s_1 s_2 \cdots s_n$ of length $n$ bits
  and of Hamming weight $k$:
  \begin{algtwo}
    \item He sets $l \leftarrow k$.
    \item For $i$ from 1 to $n$:
    \begin{algthree}
      \item If $m \geq \binom{n-i}{l}$ then Bob sets $s_i \leftarrow 1$, $m \leftarrow m - \binom{n-i}{l}$, and $l \leftarrow l - 1$.
      \item Otherwise, he sets $s_i \leftarrow 0$.
    \end{algthree}
  \end{algtwo}
  \item Bob computes the encrypted message $c = \sum_{i=1}^{n} s_i b_i$.
\end{algone}
\end{algorithm}

To decrypt the ciphertext and recover the original message, Alice does the following:
\begin{algorithm}[QPKC Decryption]
\begin{algone}
  \item She computes $r = ( c - kd ) \bmod{(N(\mathfrak{p})-1)}$.
  \item She computes $u \in \integers{K}$ such that $u = g^r \bmod{\mathfrak{p}}$.
  \item She finds an element $v$ such that $u$ and $v$ are in the same coset of $\mathfrak{p}$, and
  $v$ is in the set $R$ defined in \propref{prop:Okamoto:UniqueRep}.
  \item Alice recovers $s$ from $v$ as follows:
  \begin{algtwo}
    \item For $i$ from 1 to $n$:
    \begin{algthree}
        \item If $p_i \vert v$ then she sets $s_i \leftarrow 1$.
        \item Otherwise she sets $s_i \leftarrow 0$.
    \end{algthree}
  \end{algtwo}
  \item Alice recovers $m$ from $s$ as follows:
  \begin{algtwo}
    \item She sets $m \leftarrow 0$, and $l \leftarrow k$.
    \item For $i$ from 1 to $n$:
    \begin{algthree}
      \item If $s_i = 1$, then set $m \leftarrow m + \binom{n-i}{l}$ and $l \leftarrow l - 1$.
    \end{algthree}
  \end{algtwo}
\end{algone}
\end{algorithm}

\begin{theorem}
The decryption procedure works properly.
\end{theorem}

\begin{proof}
First consider the value $u$ computed by Alice in the decryption procedure.  Note that
\begin{align*}
  u &= g^r \bmod{\mathfrak{p}} \\
    &= g^{c-kd} \bmod{\mathfrak{p}} \\
    &= g^{(\sum_{i=1}^{n} s_i b_i) - kd} \bmod{\mathfrak{p}} \\
    &= g^{(\sum_{i=1}^{n} s_i ( q_i + d ) ) -kd} \bmod{\mathfrak{p}} \\
    &= g^{(\sum_{i=1}^{n} s_i q_i ) + kd - kd } \bmod{\mathfrak{p}} \\
    &= \prod_{i=1}^n (g^{q_i})^{s_i} \bmod{\mathfrak{p}} \\
    &= \prod_{i=1}^n p_i^{s_i} \bmod{\mathfrak{p}}.
\end{align*}

Next consider the element $v \in R$ such that $v \equiv u \pmod{\mathfrak{p}}$.  We claim that in
fact $v = \prod_{i=1}^n p_i^{s_i}$.  Suppose that the claim is not true.  By the condition in
\stepref{step:Okamoto:KeyGen:UniqueRepCondition}, since exactly $k$ of the $s_i$ are 1 and the rest
are 0, $\prod_{i=1}^n p_i^{s_i}$ is an element of $R$.  Since the elements of $R$ are in distinct
cosets of $\mathfrak{p}$ it must be true that $v \not\equiv \prod_{i=1}^n p_i^{s_i}
\pmod{\mathfrak{p}}$. Finally, since $v \equiv u \pmod{\mathfrak{p}}$ we must have $u \not\equiv
\prod_{i=1}^n p_i^{s_i} \pmod{\mathfrak{p}}$ which is a contradiction. Thus $v = \prod_{i=1}^n
p_i^{s_i}$.

As pointed out in~\cite{OTU2000:QuantumPKC}, it is not always true that $\integers{K}$ is a unique
factorisation domain.  However, note that
\[
    \mathcal{N}(v) = \mathcal{N}\left(\prod_{i=1}^n p_i^{s_i}\right) = \prod_{i=1}^n \mathcal{N}(p_i)^{s_i}
\]
by the definition of the norm.  By the condition in
\stepref{step:Okamoto:KeyGen:CoprimeNormCondition}, $\mathcal{N}(p_1), \ldots, \mathcal{N}(p_n)$
were all chosen to be coprime.  As a result there is a unique decomposition of $\mathcal{N}(v)$
into a product of the $\mathcal{N}(p_i)$, and hence a unique decomposition of $v$ into a product of
the $p_i$.

The remainder of the decryption algorithm finds this unique decomposition of $v$ into a product of the $p_i$, recovering the correct values for the $s_i$, and
then correctly decodes the $s_i$ back to the message $m$.
\end{proof}

\end{section}

\begin{section}{Security Of The Scheme}\label{sec:Okamoto:Security}

Consider the task faced by an passive attacker Eve who wishes to determine the private key from the
public key. It is hard for Eve to determine the correct number field $K$ from the set
$\mathcal{K}$, since $\mathcal{K}$ could be exponentially large.  If the field $K$ were revealed in
some way, there could be exponentially many generators $g$ for the field.  Since only a small
number of elements from $\integers{K}/\mathfrak{p}$ are chosen as the $p_i$, it is unlikely that an
attacker could correctly determine even a small subset of the $p_i$, and in order for some known
attacks on similar schemes to succeed, a large subset is required. Further, even if a large subset
were determined, the attacker would still have to find the one-to-one correspondence between the
known $p_i$ and the $b_i$. This task should be difficult without knowledge of both $g$ and $d$
since the relationship between an element and its discrete logarithm tends to appear random.  These
observations from~\cite{OTU2000:QuantumPKC} are all heuristic, but they do seem to indicate that it
should be difficult for an attacker to determine the private key from the public key.

To determine the plaintext of a message from a ciphertext, the attacker must solve an instance of the following problem:
\begin{problem}[Subset-Sum Problem (SSP)]
Given the positive integers $c$ and $b_1, \ldots, b_n$, find $m_1, \ldots m_n \in \{ 0, 1 \}$ such
that $c = \sum_{i=1}^n m_i b_i$.
\end{problem}

SSP is known to be \NP-complete, and thus it is unlikely that there is a polynomial-time algorithm that solves general instances of the problem.  However,
there are algorithms that have been successful in solving instances that satisfy certain conditions.  The density of an instance of SSP is defined as
\[
    \frac{n}{\log\left( \underset{1 \leq i \leq n}{\max} \{b_i\} \right)}.
\]
There are algorithms based on the LLL algorithm that are generally successful at solving SSP
instances with a density less than 0.9408~\cite{CLOS1991:LowDensitySubsetSum}.  This and similar
attacks have been used to successfully cryptanalyse other schemes based on the hardness of SSP, and
so to avoid these attacks, we wish to ensure that we can choose parameters for this quantum
cryptosystem that result in a sufficiently high density. Indeed, as shown
in~\cite{OTU2000:QuantumPKC}, the implementation of the scheme with $\mathcal{K}$ chosen to be the
set of imaginary quadratic number fields results in a density that is at least $1$, and this
provides some evidence that the scheme could resist such an attack.

\end{section}
\end{chapter}

\begin{chapter}{Diffie-Hellman Key Establishment}\label{chap:DiffieHellman}

So far we have described a number of public key encryption schemes, which allow Bob to send a secret message to Alice even if they have never met before to
agree on a secret key.  In the following chapters, we will discuss key establishment protocols, in which Alice and Bob (who still may never have met before)
send a series of messages over a public channel, they each perform some mathematical operations, and they each obtain a copy of a secret key.  If Eve is
listening on the public channel and intercepts all of the messages sent between Alice and Bob, she should not be able to determine this secret key.  Once a
secret key has been established, Alice and Bob can use it to encrypt messages for one another using a symmetric key encryption scheme, for example.  The first
proposed key establishment protocol was the Diffie-Hellman protocol.

\begin{section}{The Protocol}\label{sec:DiffieHellman:Protocol}

The Diffie-Hellman key establishment protocol works as follows:
\begin{algorithm}[Diffie-Hellman Protocol]
\begin{algone}
  \item Alice and Bob agree on a group $G$ of prime order $p$ and a generator $g$ of $G$.  (These choices can be made public.)
  \item Alice selects an integer $a$ uniformly at random from $\{0,\ldots,p-1\}$.  She computes the value $g^a$ and sends it to Bob.
  \item Bob selects an integer $b$ uniformly at random from $\{0,\ldots,p-1\}$.  He computes the value $g^b$ and sends it to Alice.
  \item Bob uses $b$ and the value he receives from Alice to compute $(g^a)^b = g^{ab}$.
  \item Alice uses $a$ and the value she receives from Bob to compute $(g^b)^a = g^{ab}$.
\end{algone}
\end{algorithm}

At the end of the protocol, Alice and Bob share the secret value $g^{ab}$, which they can use to derive a secret key.

\end{section}

\begin{section}{Security Of The Protocol}

The only values that are sent on the public channel are $g^a$ and $g^b$.  This means that in order to determine the secret key, Eve must solve the following
problem:
\begin{problem}[Diffie-Hellman Problem (DHP)]
Let $G$ be a group of prime order $p$, and let $g$ be a generator of $G$.  Given $g$, $g^a$, and
$g^b$ where $a$ and $b$ are selected uniformly at random from $\{0,\ldots,p-1\}$, find $g^{ab}$.
\end{problem}

Note that if Eve can solve the Discrete Logarithm Problem (DLP) she can solve DHP: she can simply
compute $a$ from $g^a$, and then compute $(g^b)^a = g^{ab}$.  In other words, DHP is
polynomial-time reducible to DLP.  There are some groups in which it is also true that DLP is
reducible to DHP, but this is not known to be true in general: the equivalence of DLP and DHP in
general remains an open problem.  It is clear, however, that the group $G$ must be chosen carefully
so that DLP is computationally infeasible in $G$, such as the multiplicative group $\Z_p^*$ where
$p$ is prime, or the group of points on an elliptic curve over a finite field. For more examples of
such groups, see~\cite{MvOV1996:HAC}.

The most common attack on the Diffie-Hellman protocol is not to solve DHP directly but rather to
solve DLP.  Thus the algorithms discussed in \chapref{chap:ElGamal} are the best classical
algorithms currently known to break the scheme.  Recall that these algorithms all require
superpolynomial time and so the Diffie-Hellman protocol is widely thought to be secure against a
passive adversary with a classical computer.  However, because of the existence of a polynomial
time quantum algorithm for DLP as discussed in \secref{sec:IntroQuantumAlgs:DLP}, the protocol is
not secure against an adversary with a quantum computer.

\end{section}

\end{chapter}

\begin{chapter}{Buchmann-Williams Key Establishment}
\label{chap:BW}

The Buchmann-Williams key establishment protocols are protocols whose security is based on the
hardness of problems in algebraic number theory.  There are two versions of the protocol, one which
takes place in an imaginary quadratic number field~\cite{BW1988:BuchmannWilliamsImaginary} and
another which takes place in a real quadratic number field~\cite{BW1990:BuchmannWilliamsReal}.  The
imaginary version of the protocol is the Diffie-Hellman protocol set in a particular finite Abelian
group, whereas the real version of the protocol is a variation on the Diffie-Hellman protocol set
in a finite set that is ``group-like''.

\begin{section}{The Protocol}\label{sec:BW:Protocol}

Before presenting the protocol, we mention some important definitions and results.  For more details, see for example~\cite{C1993:CourseInComputationalANT}
or~\cite{J1999:PhDThesis}.

\begin{definition}
Let $\Delta$ be a non-square integer congruent to $0$ or $1\bmod{4}$.  The \tbd{quadratic field of discriminant $\Delta$} is
\[
  \Q[\sqrt{\Delta}\,] = \Q + \sqrt{\Delta}\Q.
\]
The \tbd{quadratic order of discriminant $\Delta$} is given by
\[
  \qo=\Z+\frac{\Delta + \sqrt{\Delta}}{2}\Z.
\]
If $\Delta < 0$ we call $\qo$ an imaginary quadratic order, and if $\Delta > 0$ we call $\qo$ a real quadratic order.  In either case, $\qo$ is a subring of
$\Q[\sqrt{\Delta}]$.
\end{definition}

\begin{definition}
A \tbd{fractional ideal} of $\qo$ is a subset of $\Q[\sqrt{\Delta}\,]$ of the form
\[
    \mathfrak{a} = q \left( a\Z + \frac{b+\sqrt{\Delta}}{2}\Z \right)
\]
where $q \in \Q$, $a, b \in \Z$, $a, q > 0$ and $b^2 \equiv \Delta \pmod{4a}$.  We denote $\mathfrak{a}$ by the triple $(q,a,b)$.  If $q=1$ the ideal is called
a \tbd{primitive ideal}.
\end{definition}

Like the ideals introduced in \defref{def:Okamoto:Ideal}, a fractional ideal of $\qo$ is invariant
under multiplication by elements of $\qo$.  However, unlike those ideals, a fractional ideal of
$\qo$ is not necessarily a subset of $\qo$.

We can define a multiplication operation on ideals as follows:
\begin{definition}
Let $\mathfrak{a}$ and $\mathfrak{b}$ be ideals of $\qo$.  The \tbd{product} of $\mathfrak{a}$ and $\mathfrak{b}$ is
\[
    \mathfrak{ab} = \left\{ \sum_{(a,b) \in U} ab \st U \subset \mathfrak{a} \times \mathfrak{b},\; \abs{U} < \infty \right\}.
\]
\end{definition}

The product $\mathfrak{ab}$ is also an ideal of $\qo$; that is, the set of ideals is closed under multiplication.  The order $\qo$ itself acts as a
multiplicative identity since $\mathfrak{a}\,\qo = \qo\mathfrak{a} = \mathfrak{a}$.

\begin{definition}
An ideal $\mathfrak{a}$ is said to be \tbd{invertible} if there exists an ideal $\mathfrak{a}^{-1}$ such that $\mathfrak{a}\mathfrak{a}^{-1} = \qo$.
\end{definition}

\begin{definition}
An ideal $\mathfrak{a}$ is said to be \tbd{principal} if there exists an element $\alpha \in \Q[\sqrt{\Delta}\,]$ such that $\mathfrak{a} = \alpha\qo$.
\end{definition}

The set of invertible ideals forms a group under multiplication; this group is denoted $\iideals$.
Every principal ideal is invertible, since $\paren{\alpha\qo}^{-1} = \alpha^{-1}\qo$, and in fact
the set of principal ideals forms a subgroup of $\iideals$; this subgroup is denoted $\pideals$. We
now come to a very important definition:

\begin{definition}
The \tbd{class group} of $\qo$ is the factor group $\iideals / \pideals$, and is denoted by $\cg$.  The \tbd{class number} of $\qo$ is the order of $\cg$, and
is denoted by $\cn$.
\end{definition}

Thus the class group $\cg$ is a set of equivalence classes, where two invertible ideals
$\mathfrak{a}$ and $\mathfrak{b}$ are in the same equivalence class if and only if there is some
principal ideal $\alpha \qo$ such that $\alpha \qo \mathfrak{a} = \mathfrak{b}$.  It turns out that
these equivalence classes form a finite Abelian group under the multiplication operation defined
above.

The above definitions are the same for both imaginary and real quadratic orders, but many of the
properties of these two types of quadratic orders are quite different.  We will first describe some
further properties of imaginary quadratic orders and their class groups and present the imaginary
Buchmann-Williams key establishment protocol from~\cite{BW1988:BuchmannWilliamsImaginary}.  We will
then describe some further properties of real quadratic orders and their class groups and present
the real Buchmann-Williams key establishment protocol.  The real version of the protocol was first
published in~\cite{BW1990:BuchmannWilliamsReal}, although the presentation
in~\cite{SBW1994:BuchmannWilliamsReal} is considerably more detailed and complete.

\begin{subsection}{The Imaginary Case}

In this section we assume that $\Delta < 0$ so that $\qo$ is an imaginary quadratic order.

\begin{definition}
Let $\mathfrak{a}$ be a primitive ideal of $\qo$ with the representation
\[
    \mathfrak{a} = a\Z + \frac{b + \sqrt{\Delta}}{2}\Z
\]
where $a, b \in \Z$, $a > 0$ and $b^2 \equiv \Delta \pmod{4a}$.  Let $c = \frac{b^2-\Delta}{4a}$.
Then $\mathfrak{a}$ is called a \tbd{reduced ideal} if $0 \leq b \leq a \leq c$, or if $ 0 < -b < a
< c$.
\end{definition}

When dealing with the equivalence classes that are the elements of a factor group, ideally we would
like to have a canonical representative of each equivalence class so that we can use those
representatives for computation.  (For example, in the factor group $\Z_p = \Z / p\Z$ we use the
representatives $0, 1, \ldots, p-1$ for the equivalence classes.)  In the case of imaginary
quadratic orders, each equivalence class in $\cg$ contains exactly one reduced ideal.  Thus we can
choose the set of reduced ideals to be the set of canonical representatives for the elements of
$\cg$.  There are algorithms that, given any ideal in $\iideals$, can efficiently compute the
equivalent reduced ideal; we can use these algorithms and modifications of them to compute reduced
products, powers, and inverses of all invertible ideals~\cite{J1999:PhDThesis}.

It is also important to note that the class number $\cn$ is typically close to $\sqrt{\Delta}$, so
there are approximately $\sqrt{\Delta}$ equivalence classes in $\cg$. Further, a well-supported
conjecture by Cohen and Lenstra predicts that $\cg$ is typically cyclic or ``nearly cyclic'' (for
example, the direct product of a large cyclic group and a much smaller
one)~\cite{C1993:CourseInComputationalANT}.

We can now describe the imaginary case of the Buchmann-Williams key establishment protocol.  The
idea is that, Alice and Bob agree on some element $\mathfrak{g} \in \cg$ and perform the standard
Diffie-Hellman protocol in the subgroup generated by $\mathfrak{g}$,
\[
  \left<\mathfrak{g}\right> = \left\{ \qo, \mathfrak{g}, \mathfrak{g}^2, \ldots, \mathfrak{g}^{r-1} \right\}
\]
where $r$ is the order of $\mathfrak{g}$.

The protocol works as follows:
\begin{algorithm}[Buchmann-Williams Protocol (Imaginary Case)]
\begin{algone}
  \item Alice and Bob agree on a discriminant $\Delta < 0$, $\Delta \equiv 0,1 \pmod 4$ and a reduced ideal $\mathfrak{g}$ of $\qo$.  (These choices can be made public.)
  \item Alice chooses an integer $a$ uniformly at random from $\{ 1, \ldots, \lfloor\sqrt{\Delta}\,\rfloor\}$.  She computes the value $\mathfrak{g}^a$ and sends it to Bob.
  \item Bob chooses an integer $b$ uniformly at random from $\{ 1, \ldots, \lfloor\sqrt{\Delta}\,\rfloor\}$.  He computes the value $\mathfrak{g}^b$ and sends it to Alice.
  \item Bob uses $b$ and the value he receives from Alice to compute $(\mathfrak{g}^a)^b = \mathfrak{g}^{ab}$.
  \item Alice uses $a$ and the value she receives from Bob to compute $(\mathfrak{g}^b)^a = \mathfrak{g}^{ab}$.
\end{algone}
\end{algorithm}

At the end of the protocol, Alice and Bob share the secret ideal $\mathfrak{g}^{ab}$, which they
can use to derive a secret key.

\end{subsection}

\begin{subsection}{The Real Case}\label{sec:BW:Protocol:Real}

In this section we assume that $\Delta > 0$ so that $\qo$ is a real quadratic order.  Let $n = \log
\Delta$.  We can still define a reduced ideal, but the definition changes slightly:
\begin{definition}
Let $\mathfrak{a}$ be a primitive ideal of $\qo$ with the representation
\[
    \mathfrak{a} = a\Z + \frac{b + \sqrt{\Delta}}{2}\Z
\]
where $a, b \in \Z$, $a > 0$ and $b^2 \equiv \Delta \pmod{4a}$.  Then $\mathfrak{a}$ is a
\tbd{reduced ideal} if
\[
  \abs{\sqrt{\Delta}-2a} < b < \sqrt{\Delta}.
\]
\end{definition}

We also define the units and the regulator of a real quadratic order:

\begin{definition}
An element $\varepsilon \in \qo$ is called a \tbd{unit} if there exists an element $\varepsilon' \in \qo$ such that $\varepsilon\varepsilon' = 1$.  The
\tbd{fundamental unit} of $\qo$ is the smallest positive unit greater than $1$ in $\qo$, and denoted $\varepsilon_\Delta$.
\end{definition}

\begin{definition}
The \tbd{regulator} of the real quadratic order $\qo$ is $\log \varepsilon_\Delta$, and denoted
$\reg$.
\end{definition}

Unlike the imaginary case, it is not true in the real case that each equivalence class of $\cg$ contains exactly one reduced ideal; we can say only that each
equivalence class contains a finite number of reduced ideals.  The class number is typically very small, often $\cn = 1, 2,$ etc., meaning that there are very
few equivalence classes in $\cg$ and each one contains many reduced ideals.  In fact, $\cn \reg \approx \sqrt{\Delta}$ and in this way, the regulator of the
order is in some way a measure for how many reduced ideals occur in each equivalence class.

Because $\cn$ is generally very small in a real quadratic order, $\cg$ is a poor choice for a group for the typical Diffie-Hellman key establishment. However,
in~\cite{S1972:RealQuadraticOrderInfrastructure} Shanks proposed a method that could be used to organise the set of reduced ideals in any equivalence class
into a structure that is not a group structure, but ``group-like'' in some respects, which he called the ``infrastructure'' of the class.

Shanks proposed a real-valued ``distance'' function that defines the distance between any two reduced ideals in the same equivalence class.  This function
implies an ordering of the reduced ideals: they can be arranged in order of increasing distance from the unit ideal $\qo$.

\begin{definition}
The distance between two reduced ideals $\mathfrak{a}$ and $\mathfrak{b}$ is denoted
$\idealdistancetwo{\mathfrak{a}}{\mathfrak{b}}$.  We will use $\idealdistance{\mathfrak{a}}$ as a
shorthand for $\idealdistancetwo{\mathfrak{a}}{\qo}$.
\end{definition}

Shanks also defined a function $\rho$ that given any reduced ideal would determine the next reduced
ideal in the ordering. By repeatedly applying the $\rho$ operator to $\qo$, we eventually obtain
all of the reduced ideals in the equivalence class, and then again obtain $\qo$.  In other words,
the ordering is a cyclical ordering of the reduced ideals.  The total distance around the cycle of
reduced ideals (using Shanks's distance function) is $\reg$ (the regulator of $\qo$).

We can perform several operations with the reduced ideals in this infrastructure, as described for
example in~\cite{J1999:PhDThesis}.  One of the most important operations is the following: given a
reduced ideal $\mathfrak{a}$ and a real number $x$, we can compute the last reduced ideal whose
distance from $\mathfrak{a}$ is no more than $x$ (modulo $\reg$).  If we think of the reduced
ideals as being arranged on a circle of circumference $\reg$, this operation corresponds to
starting at the point on the circle corresponding to $\mathfrak{a}$, proceeding around the
circumference a distance of $x$, and selecting the last reduced ideal we encounter.  As a result,
this ideal is sometimes called the ideal to the left of $x$ (relative to $\mathfrak{a}$).

\begin{definition}
We will denote the ideal to the left of $x$ (relative to $\mathfrak{a}$) by
$\ideallefttwo{x}{\mathfrak{a}}$. We will use $\idealleft{x}$ as a shorthand for
$\ideallefttwo{x}{\qo}$.
\end{definition}

We can also define the error of the ideal to the left of $x$ (relative to $\mathfrak{a}$) which
quantifies how well the true distance between $\ideallefttwo{x}{\mathfrak{a}}$ and $\mathfrak{a}$
approximates $x$:
\begin{definition}
The error of $\ideallefttwo{x}{\mathfrak{a}}$ is denoted $\idealerrortwo{x}{\mathfrak{a}}$ and is
defined by
\[
  \idealerrortwo{x}{\mathfrak{a}} = \blb x - \idealdistancetwo{\idealleft{x}}{\mathfrak{a}} \brb
\bmod{\reg}.
\]
We will use $\idealerror{x}$ as shorthand for $\idealerrortwo{x}{\qo}$.
\end{definition}

The concepts of the ideal to the left of $x$ and the error of this ideal are illustrated in
\figref{fig:BW:CycleOfReducedIdeals}.  In the figure, we are working relative to $\qo$.

\begin{figure}[ht]
\begin{center}
\fbox{\includegraphics{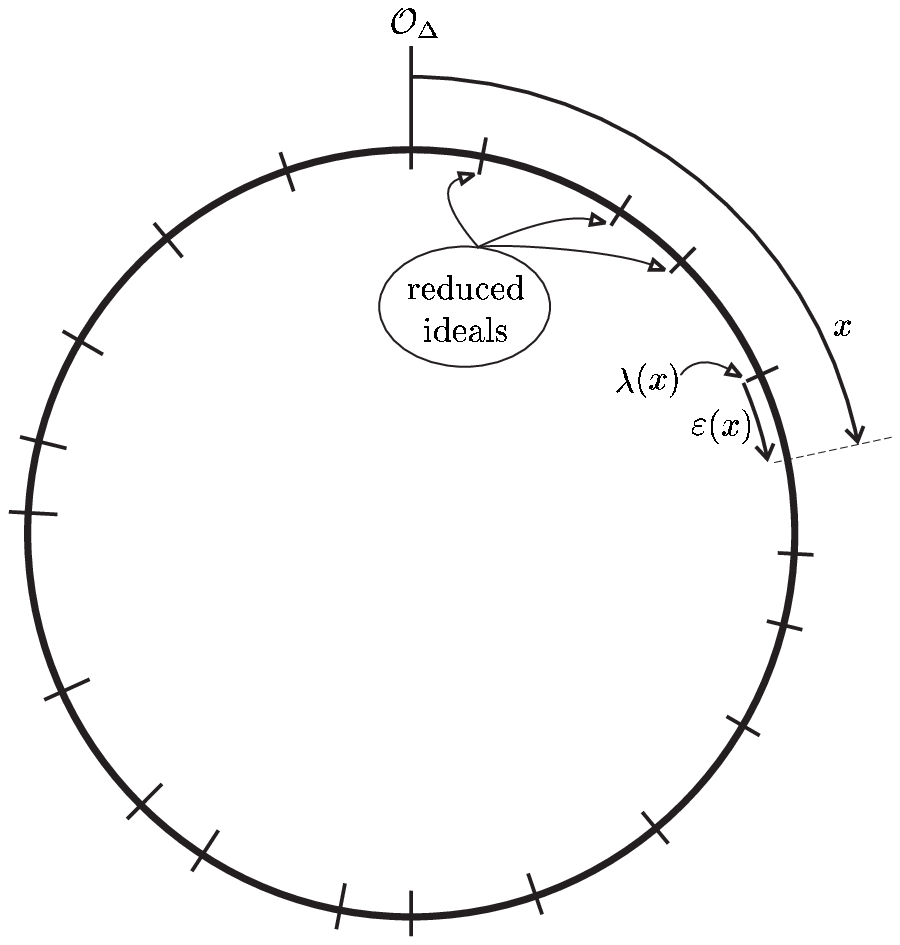}}
\caption{The Cycle Of Reduced Principal Ideals}
\label{fig:BW:CycleOfReducedIdeals}
\end{center}
\end{figure}

It should be noted that the distances with which the participants in the protocol must work are all
real numbers, and so to perform the required calculations exactly would require infinite precision.
The participants must therefore choose some finite precision within which to perform all of the
calculations, and as a result there may be round-off errors that propagate through the protocol.
These potential round-off errors force the participants to perform an extra ``clean-up'' round of
communication after the one usual round of a Diffie-Hellman-like exchange in order to make sure
that they share the same value.

We can now sketch the protocol.  There are many details of the implementation that are omitted in
the presentation below; for a more complete description of the protocol, refer
to~\cite{SBW1994:BuchmannWilliamsReal}.
\begin{algorithm}[Buchmann-Williams Protocol (Real Case)]
\begin{algone}
\item Alice and Bob agree on a discriminant $\Delta > 0$, $\Delta \equiv 0,1 \pmod 4$ and an equivalence class of $\cg$.  (These choices can be made public.)
\item Alice chooses an integer $a$ uniformly at random from $\{ 1, \ldots, \lfloor\sqrt{\Delta}\,\rfloor\}$.  She computes $\mathfrak{a}$,
the ideal to the left of $a$; that is, $\mathfrak{a} = \idealleft{a}$.  She also computes
$\idealerror{a}$ and sends $\mathfrak{a}$ and $\idealerror{a}$ to Bob.
\item Bob chooses an integer $b$ uniformly at random from $\{ 1, \ldots, \lfloor\sqrt{\Delta}\,\rfloor\}$.  He computes $\mathfrak{b}$,
the ideal to the left of $b$; that is, $\mathfrak{b} = \idealleft{b}$.  He also computes
$\idealerror{b}$ and sends $\mathfrak{b}$ and $\idealerror{b}$ to Alice.
\item Alice computes $\mathfrak{c}_A$, the ideal to the left of $a + \idealerror{b}$ (relative to
$\mathfrak{b}$); that is, \linebreak $\mathfrak{c}_A =
\ideallefttwo{a+\idealerror{b}}{\mathfrak{b}}$.
\item Bob computes $\mathfrak{c}_B$, the ideal to the left of $b + \idealerror{a}$ (relative to
$\mathfrak{a}$); that is, \linebreak $\mathfrak{c}_B =
\ideallefttwo{b+\idealerror{a}}{\mathfrak{a}}$.
\item Alice and Bob send each other one classical bit which allows them to determine whether $\mathfrak{c}_A = \mathfrak{c}_B$.  If this is not true, Alice and Bob make small adjustments
(see~\cite{SBW1994:BuchmannWilliamsReal}) after which they are certain that they have computed the
same ideal.
\end{algone}
\end{algorithm}

At the end of the protocol, Alice and Bob share a secret ideal which they can use to derive a secret key.

\end{subsection}

\end{section}

\begin{section}{Security Of The Protocol}

In this section we discuss the security of each of the two cases of the Buchmann-Williams protocol.  As we will see, both cases are susceptible to attacks with
a quantum computer.  In order to break the real case of the protocol, however, we will need two recently discovered quantum algorithms.

\begin{subsection}{The Imaginary Case}

Since the imaginary case of the Buchmann-Williams protocol is equivalent to the Diffie-Hellman
protocol, as mentioned in \chapref{chap:DiffieHellman} the scheme would be broken if the Discrete
Logarithm Problem (DLP) could be solved efficiently in the group $\cg$. There are no known
efficient classical algorithms to solve DLP in this group, however: the best known algorithms still
require superpolynomial time, like that in~\cite{J1999:PhDThesis}.  Furthermore,
in~\cite{BW1988:BuchmannWilliamsImaginary} it is mentioned that if an efficient algorithm to solve
DLP in $\cg$ did exist, it could likely be used to factor~$\Delta$.

The group could possibly admit attacks that did not depend on solving DLP but on solving the
Diffie-Hellman Problem (DHP) directly, but again, there is some evidence described
in~\cite{BW1988:BuchmannWilliamsImaginary} that these attacks could also lead to algorithms to
factor~$\Delta$.  These facts suggest that breaking the scheme with a classical computer is at
least as hard as the factoring problem, which we believe to be hard. We therefore believe the
protocol to be secure against a passive adversary with a classical computer.

However, as mentioned in \secref{sec:IntroQuantumAlgs:DLP}, there is an efficient quantum algorithm
to solve DLP.  In other words, this protocol is not secure against a quantum adversary.

\end{subsection}

\begin{subsection}{The Real Case}

Consider the following variation of DLP as proposed in~\cite{BW1990:BuchmannWilliamsReal}:
\begin{problem}[Principal Ideal Distance Problem (PIDP)]
Given a principal\linebreak ideal $\mathfrak{a}$ of a real quadratic order $\qo$, compute
$\idealdistance{\mathfrak{a}}$, its distance from $\qo$.
\end{problem}

Suppose an adversary can solve PIDP.  When Alice sends~$\mathfrak{a}$ and~$\idealerror{a}$ to Bob,
the adversary can compute $\idealdistance{\mathfrak{a}}$, and hence determine
$\idealdistance{\mathfrak{a}} + \idealerror{a} = a$.  The adversary then has knowledge of Alice's
private value $a$ (to some finite precision). With this knowledge, with good probability the
adversary can construct the shared secret in the same way Alice does, and the protocol is broken.
That is, an algorithm to solve PIDP would allow an adversary to break the real version of the
Buchmann-Williams key establishment protocol.

There is evidence that PIDP is hard to solve with a classical computer.  It is shown
in~\cite{BW1990:BuchmannWilliamsReal} that an efficient solution to PIDP would result in an
efficient algorithm to compute the regulator~$\reg$.  Further, it is shown
in~\cite{S1982:QuadraticFields} that an efficient algorithm to compute~$\reg$ would result in an
efficient algorithm to factor~$\Delta$. Thus, PIDP is at least as hard as the factoring problem,
which we believe to be hard with a classical computer.  We therefore believe the protocol to be
secure against a passive adversary with a classical computer.

However, as recently discovered by Hallgren~\cite{H2002:PellsEquation}, we can efficiently solve
PIDP with a quantum computer.  Therefore the real version of the Buchmann-Williams key
establishment protocol can be broken by a passive adversary with a quantum computer.

The remainder of this chapter introduces the new quantum algorithms that efficiently solve PIDP.
Suppose we are given a quadratic order~$\qo$.  We will describe two algorithms: one that
computes~$\reg$, and another that given $\reg$ solves PIDP.  The description of these algorithms
in~\cite{H2002:PellsEquation} is quite terse, and the presentation below attempts to provide more
details and to correct some of the minor errors in~\cite{H2002:PellsEquation}.  A similar but
independently constructed clarification of the algorithm to compute the regulator can be found
in~\cite{J2003:NotesOnPellsEquation}, along with much of the background material already presented
in this chapter.

\end{subsection}

\begin{subsection}{Computing The Regulator}\label{sec:BW:Regulator}

As mentioned above, much of the material in this section can also be found
in~\cite{J2003:NotesOnPellsEquation}. However, except where noted, the presentation here was
developed independently.

We work in the identity class of $\cg$.  Consider the function $g \colon \R \longrightarrow
\pideals\!\times \R$ defined by $g(x) =
\paren{ \idealleft{x}, \idealerror{x}}$ for all $x \in \R$.

\begin{proposition}\label{prop:BW:gPeriodic}
The function $g$ is one-to-one on the interval $[0, \reg)$ and periodic with period $\reg$.
\end{proposition}

\begin{proof}
If $g(x) = g(y)$ for any $x,y \in [0,\reg)$, then $\idealleft{x} = \idealleft{y}$ and
\begin{align*}
                               \idealerror{x} &= \idealerror{y} \\
            x - \idealdistance{\idealleft{x}} &\equiv y - \idealdistance{\idealleft{y}} \pmod{\reg}\\
            x - \idealdistance{\idealleft{x}} &\equiv y - \idealdistance{\idealleft{x}} \pmod{\reg}\\
                                            x &\equiv y \pmod{\reg}.
\end{align*}
Further, $x,y \in [ 0, \reg )$ so it follows that $x=y$, and $g$ is one-to-one on $[0,\reg)$.

Since the distance around the cycle of reduced ideals is $\reg$, it follows that for any $x \in \R$,
\begin{align*}
    g(x+\reg) &= \lb \idealleft{x+\reg}, \idealerror{x+\reg} \rb \\
              &= \lb \idealleft{x}, \idealerror{x} \rb \\
              &= g(x)
\end{align*}
so $g$ is periodic with period $\reg$.
\end{proof}

We now have a periodic function with a domain of $\R$.  We would like to use techniques similar to
those from \secref{sec:IntroQuantumAlgs:IHSP} to find the period of the function, but in order to
compute with the function we first have to modify it slightly so that its domain is some discrete
set. Such a modification will also have the effect of making the function no longer ``perfectly''
periodic, but we will still be able to recover a close approximation to the period.

More specifically, as in~\cite{H2002:PellsEquation} we can define what it means for a function with
an \emph{integer} domain to be ``periodic'' with a \emph{real} (not necessarily integer) period. We
use the definition presented in~\cite{J2003:NotesOnPellsEquation}:
\begin{definition}\label{def:BW:PeriodicIntegerDomain}
Let $X$ be any set.  A function $f \colon \Z \longrightarrow X$ is called \tbd{weakly periodic with
period $s \in \R$} if for all integers $k$, $0 \leq k < s$, and for all non-negative integers $j$,
either
\begin{enumerate}
\item $f(k) = f( k + \lfloor j s \rfloor )$, or
\item $f(k) = f( k + \lceil j s \rceil )$.
\end{enumerate}
For brevity we will write $f(k) = f(k + \bracket{js})$ to indicate that one of the above conditions
is satisfied.  The satisfied condition may vary with $k$ and $j$.
\end{definition}
(Here we use the notation $\floor{x}$ to denote the largest integer less than or equal to $x$, and
$\ceil{x}$ to denote the smallest integer greater than or equal to $x$.  We will also use the
notation $\round{x}$ to denote the closest integer to $x$.)

We now define a weakly periodic function by slightly modifying the function $g$.  Given a positive
integer $N$, we define $\hat{g} \colon \Z \longrightarrow \pideals\!\times\Z$ by $\hat{g}(j) =
\paren{ \idealleft{\tfrac{j}{N}}, \floor{N \idealerror{\tfrac{j}{N}}} }$ for all integers $j$.

As we will see, this function does not precisely satisfy the definition of a weakly periodic
function, but by choosing $N$ wisely we can ensure that it satisfies the definition for a large
fraction of the integers $k$, $0 \leq k < s$.  This statement is made more specific in the
following theorem:

\begin{theorem}\label{thm:BW:GHatPeriodic}
If $N \geq n\frac{32\Delta}{3}$ then the function $\hat{g}$ is one-to-one on the interval
$[0,N\reg)$ and $\hat{g}(k) = \hat{g}(k + \bracket{jN\reg})$ for at least a $\paren{1 -
\frac{1}{\bigO{n}}}$ fraction of the integers $k \in [0,N\reg)$.
\end{theorem}

Note that in the analogous theorem in~\cite{H2002:PellsEquation}, it is stated without proof that
we should require only $N \geq n\sqrt{\Delta}$.  However, precise analysis
in~\cite{J2003:NotesOnPellsEquation} proves the existence of the lower bound on $N$ given in
\thmref{thm:BW:GHatPeriodic}; so we use that lower bound here.

\begin{proof}

First we show that $\hat{g}$ is one-to-one on the interval $[0,N\reg)$.

If $\hat{g}(j) = \hat{g}(k)$ for any $j,k \in [0,N\reg)$ then $\idealleft{\frac{j}{N}} =
\idealleft{\frac{k}{N}}$ and
\begin{align*}
  \lfloor N \idealerror{\tfrac{j}{N}} \rfloor &= \floor{ N \idealerror{\tfrac{k}{N}} }\\
     N \idealerror{\tfrac{j}{N}} &= N \idealerror{\tfrac{k}{N}} + \sigma
\end{align*}
for some $\sigma$ with $\abs{\sigma} < 1$.  Dividing both sides by $N$, we get
\begin{align*}
    \idealerror{\tfrac{j}{N}} &= \idealerror{\tfrac{k}{N}} + \tfrac{\sigma}{N} \\
    \tfrac{j}{N} - \idealdistance{\idealleft{\tfrac{j}{N}}} &\equiv \tfrac{k}{N} - \idealdistance{\idealleft{\tfrac{k}{N}}} + \tfrac{\sigma}{N} \pmod{\reg} \\
    \tfrac{j}{N} - \idealdistance{\idealleft{\tfrac{j}{N}}} &\equiv \tfrac{k}{N} - \idealdistance{\idealleft{\tfrac{j}{N}}} + \tfrac{\sigma}{N} \pmod{\reg} \\
    \tfrac{j}{N} &\equiv \tfrac{k}{N} + \tfrac{\sigma}{N} \pmod{\reg}.
\end{align*}

Since $0 \leq j, k < N\reg$, we must have $0 \leq \frac{j}{N}, \frac{k}{N} < \reg$.  Therefore,
$\frac{j}{N} = \frac{k}{N} + \frac{\sigma}{N}$.  Further, since $j,k \in \Z$ and $\abs{\sigma} <
1$, we must in fact have $\sigma = 0$, so $j=k$ and thus $\hat{g}$ is one-to-one on $[0,N\reg)$.

Now we wish to show that $\hat{g}(k) = \hat{g}(k+\bracket{jN\reg})$ for a sufficiently large
fraction of the integers $k \in [0,N\reg)$. Choose any reduced ideal $\mathfrak{a}$, and consider
the interval between the distance of $\mathfrak{a}$ and the distance of the next ideal in the
cycle; in other words, the interval $I = \left[\idealdistance{\mathfrak{a}},
\idealdistance{\rho(\mathfrak{a})}\right)$. By a proven bound on the distance between consecutive
ideals developed in~\cite{J2003:NotesOnPellsEquation} we know that $I$ has length at least
$\frac{3}{32\Delta}$.

Since we are given $N \geq n\frac{32\Delta}{3}$, we know that $\frac{3}{32\Delta} \geq
\frac{n}{N}$, and hence there are at least $n$ integers $k$ in $[0,N\reg)$ for which $\frac{k}{N}
\in I$.  Let $S$ be the set of all such integers.

For any $k \in S$, $\idealleft{\frac{k}{N}} = \mathfrak{a}$. Further, if $k \neq \min S$ and $k
\neq \max S$, and $\sigma$ is chosen such that $0 \leq \sigma < 1$, then $\max S < k \pm \sigma <
\min S$, so
\begin{equation}\label{equ:BW:leftIdealRounding}
    \idealleft{\tfrac{k\pm\sigma}{N}} = \idealleft{\tfrac{k}{N}} = \mathfrak{a}.
\end{equation}

Fix some value of $j$, and define the quantity $\sigma_1 = jN\reg - \floor{jN\reg}$.  We assume
that $jN\reg \not\in \Z$, so $0 < \sigma_1 < 1$.

We consider two cases:
\begin{enumerate}
\item If we round down the value of $jN\reg$,
\begin{align*}
    \idealleft{\tfrac{k + \floor{jN\reg}}{N}} &= \idealleft{\tfrac{k+jN\reg - \sigma_1}{N}}\\
                                             &= \idealleft{\tfrac{k- \sigma_1}{N}+j\reg}\\
                                             &= \idealleft{\tfrac{k- \sigma_1}{N}}\\
                                             &= \idealleft{\tfrac{k}{N}}\quad\mbox{by~\equref{equ:BW:leftIdealRounding}.}
\end{align*}
\item Similarly, if we round up the value of $jN\reg$,
\begin{align*}
    \idealleft{\tfrac{k + \ceil{jN\reg}}{N}} &= \idealleft{\tfrac{k+jN\reg + (1-\sigma_1)}{N}}\\
                                             &= \idealleft{\tfrac{k+ (1- \sigma_1)}{N}}\\
                                             &= \idealleft{\tfrac{k}{N}}\quad\mbox{by~\equref{equ:BW:leftIdealRounding}.}
\end{align*}
\end{enumerate}

Thus for any $k \in S$, $k \neq \min S$, $k \neq \max S$,
\begin{equation}
    \idealleft{\tfrac{k + \floor{jN\reg}}{N}} = \idealleft{\tfrac{k}{N}} = \idealleft{\tfrac{k +
    \ceil{jN\reg}}{N}}.\label{equ:BW:RoundedIdealsEqual}
\end{equation}

Now define the quantity $\sigma_2 = N\idealerror{\frac{k}{N}} - \floor{N\idealerror{\frac{k}{N}}}$.
Then $0 \leq \sigma_2 < 1$.

Again, we consider two cases:
\begin{enumerate}
\item If $\sigma_1 \leq \sigma_2$, then
\begin{align*}
    \floor{N \idealerror{\tfrac{k + \floor{jN\reg}}{N}}} &= \floor{N \idealerror{\tfrac{k - \sigma_1}{N}+j\reg}}\\
                                                         &= \floor{N \lb \tfrac{k - \sigma_1}{N} - \idealdistance{\idealleft{\tfrac{k - \sigma_1}{N}}}\rb}\\
                                                         &= \floor{N \lb \tfrac{k}{N}  - \tfrac{\sigma_1}{N} - \idealdistance{\idealleft{\tfrac{k}{N}}\rb}}\\
                                                         &= \floor{N \idealerror{\tfrac{k}{N}} - \sigma_1} \\
                                                         &= \floor{N \idealerror{\tfrac{k}{N}}} \quad\mbox{since $\sigma_1 \leq \sigma_2$.}
\end{align*}
\item If $\sigma_1 > \sigma_2$, then similarly
\begin{align*}
    \floor{N \idealerror{\tfrac{k + \ceil{jN\reg}}{N}}}  &= \floor{N \idealerror{\tfrac{k + (1-\sigma_1)}{N}+j\reg}}\\
                                                        &= \floor{N \idealerror{\tfrac{k}{N}} + (1-\sigma_1)} \\
                                                        &= \floor{N \idealerror{\tfrac{k}{N}}} \quad\mbox{since $1-\sigma_1 < 1-\sigma_2$.}
\end{align*}
\end{enumerate}

Thus for any $k \in S$ with $k \neq \min S$, $k \neq \max S$,
\begin{equation}
    \floor{N \idealerror{\tfrac{k + \bracket{jN\reg}}{N}}} = \floor{N
    \idealerror{\tfrac{k}{N}}}.\label{equ:BW:RoundedDistancesEqual}
\end{equation}

Combining \equref{equ:BW:RoundedIdealsEqual} and \equref{equ:BW:RoundedDistancesEqual}, we see that
\begin{equation}\label{equ:BW:GHatPeriodic}
    \hat{g}(k + \bracket{jN\reg}) = \hat{g}(k).
\end{equation}

Since $\abs{S} \geq n$ and $k$ can take on all but two of the values in $S$,
\equref{equ:BW:GHatPeriodic} is satisfied for at least a $\paren{1 - \frac{2}{n}}$ fraction of the
integers $k \in S$.  The same argument can be made for the interval between any two reduced ideals.
Thus \equref{equ:BW:GHatPeriodic} is satisfied for at least a $\lb 1 - \frac{2}{n} \rb$ fraction of
the integers $k \in [0,N\reg)$. (This bound can in fact be improved to a $\paren{ 1 - \frac{1}{n}}$
fraction with a slightly different analysis like that of~\cite{J2003:NotesOnPellsEquation}, but the
bound presented here is still sufficient.)

Therefore as required, $\hat{g}(k + \bracket{jN\reg}) = \hat{g}(k)$ for at least a $\paren{1 -
\frac{1}{\bigO{n}}}$ fraction of the integers $k \in [0,N\reg)$.
\end{proof}

We can now describe the core of the quantum algorithm to compute the regulator.  The algorithm will
calculate an approximation to $s = N\reg$, from which we can recover $\reg$.  Like in
\algref{alg:IntroQuantumAlgs:IHSPBounded}, we assume that we have a unitary operator $\U_{\hat{g}}$
that maps $\ket{x}\ket{y} \longmapsto \ket{x}\ket{y \oplus \hat{g}(x)}$.  We also assume that we
are given an integer $m > 3s^3$, although if the approximate size of $s$ is unknown, we can use a
technique similar to that of \secref{sec:IntroQuantumAlgs:IHSP}, where we repeatedly double $m$
until the algorithm succeeds.

\begin{algorithm}[Core Of Computing The Regulator]\label{alg:BW:Regulator}
\begin{algone}
\item Start in the state $\ket{0}\ket{0} \in \Hilbert{m} \tensor \Hilbert{l}$, where $l$ is chosen such that $l$ bits
are sufficient to encode any point in the range of $\hat{g}$.
\item Apply $\QFT{m}$ to the first register.\label{step:BW:Regulator:QFT}
\item Apply $\U_{\hat{g}}$ to the system.\label{step:BW:Regulator:Ug}
\item Measure the second register.\label{step:BW:Regulator:Measurement1}
\item Apply $\QFT{m}$ to the first register.\label{step:BW:Regulator:QFT2}
\item Measure the first register to obtain the integer $y$.  If $y > \frac{m}{n}$, begin the
procedure again.\label{step:BW:Regulator:Measurement2}
\item Otherwise, return $y$.
\end{algone}
\end{algorithm}

\begin{theorem}\label{thm:BW:RegulatorResult}
With probability in $\bigO{\frac{1}{\log{s}}}$, the output of \algref{alg:BW:Regulator} satisfies
$y = \round{k\frac{m}{s}}$ for some integer $k$.
\end{theorem}

\begin{proof}[Proof Sketch]
After \stepref{step:BW:Regulator:Ug} our system is in the state $\ket{\phi} =
\frac{1}{\sqrt{m}}\sum_{x=0}^{m-1} \ket{x}\ket{\hat{g}(x)}$. When we measure the second register in
\stepref{step:BW:Regulator:Measurement1}, we see a value $z$, and we leave the first register in a
superposition of all states in which $z$ appears in the second register.

We now determine this superposition.  Let $m = \floor{ps} + r$ where $p, r \in \Z$ and $0 < r \leq
s$; in other words, $m \approx ps$. By \thmref{thm:BW:GHatPeriodic} we know that for an inverse
polynomial fraction of the integers $k$ with $0 \leq k < s$, $\hat{g}(k) = \hat{g}(k +
\bracket{js})$.  So with high probability, if $z = \hat{g}(k)$, then $z = \hat{g}(k +
\bracket{js})$ for all $j$, $0 \leq j < p$. Thus we can say that after
\stepref{step:BW:Regulator:Measurement1} we leave the first register in a state that is ``close
to''
\[
    \ket{\psi} = \frac{1}{\sqrt{p}} \sum_{j=0}^{p-1} \ket{k + \bracket{js}}.
\]

This is not exactly the state of the first register, since the function $\hat{g}$ is not exactly
weakly periodic.  The consequences of using this approximate data are not explicitly analysed
in~\cite{H2002:PellsEquation} or~\cite{J2003:NotesOnPellsEquation}.  However, both claim that
because of the large fraction of integers $k$ for which $\hat{g}(k) = \hat{g}(k+\bracket{js})$ (see
\thmref{thm:BW:GHatPeriodic}) the approximation is close enough for the algorithm to succeed.
Similarly, the remainder of the analysis here assumes that the first register is in the exact state
$\ket{\psi}$.

After applying $\QFT{m}$ to $\ket{\psi}$ in \stepref{step:BW:Regulator:QFT2}, we obtain the state
\begin{align*}
    &\frac{1}{\sqrt{pm}}\sum_{j=0}^{p-1} \sum_{x=0}^{m-1} e^{2\pi i x \frac{k + \bracket{js}}{m}} \ket{x} \\
    &= e^{2\pi i \frac{k}{m}}\frac{1}{\sqrt{pm}}\sum_{x=0}^{m-1} \sum_{j=0}^{p-1} e^{2\pi i x \frac{\bracket{js}}{m}}
    \ket{x}.
\end{align*}

It is interesting to note that the global phase coefficient $e^{2\pi i \frac{k}{m}}$ does not
affect the probability distribution of the results of measuring this state, since $\abs{e^{2\pi i
\frac{k}{m}}}^2 = 1$. (In other words, we can assume without loss of generality that $k = 0$.  It
is a general property of the Fourier sampling method used by this algorithm that given a group $G$
and a subset $X$ of $G$, the distributions induced by applying the method to the superpositions
$\sum_{x\in X}\ket{x}$ and $\sum_{x\in X}\ket{g+x}$ are identical for every $g \in
G$~\cite{H2002:PellsEquation}.)

The probability of obtaining a particular measurement result $y$ in
\stepref{step:BW:Regulator:Measurement2} is therefore given by
\begin{equation}
    P_y = \abs{\frac{1}{\sqrt{pm}}\sum_{j=0}^{p-1} e^{2\pi i y
    \frac{\bracket{js}}{m}}}^2.\label{equ:BW:ProbabilityDistY}
\end{equation}
As in~\cite{H2002:PellsEquation} (but in more detail) we now analyse this distribution.

Fix a value $y = \round{k \frac{m}{s}}$ for some integer $k$, and let $y = k \frac{m}{s} +
\varepsilon$ where $-\frac{1}{2} \leq \varepsilon < \frac{1}{2}$.  For each $j$, $0 \leq j < p$,
let $\bracket{js} = js + \delta_j$, where $-1 \leq \delta_j < 1$.

Then note that
\begin{align}
    y \frac{\bracket{js}}{m} &= \paren{\frac{k}{s} + \frac{\varepsilon}{m}}\paren{js + \delta_j}\notag\\
                           &= kj + \frac{\varepsilon j s}{m} + \frac{k\delta_j}{s} +
                           \frac{\varepsilon\delta_j}{m}.\label{equ:BW:pyExponent}
\end{align}

Recall that in \stepref{step:BW:Regulator:Measurement2} we accepted only values of $y$ that
satisfied $y \leq \frac{m}{n}$.  Thus
\begin{align}
    \frac{km}{s} + \varepsilon &\leq \frac{m}{n} \notag\\
    \frac{k}{s} + \frac{\varepsilon}{m} &\leq \frac{1}{n} \notag\\
    \abs{\frac{k\delta_j}{s} + \frac{\varepsilon\delta_j}{m}} &\leq \abs{\frac{\delta_j}{n}} \notag\\
    \abs{\frac{k\delta_j}{s} + \frac{\varepsilon\delta_j}{m}} &< \frac{1}{n}.
    \label{equ:BW:InequalityBeta}
\end{align}

Further note that $\frac{\varepsilon j s}{m} = \parfrac{j}{p}\parfrac{\varepsilon s p}{m}$ and
\begin{equation}
    \abs{\frac{\varepsilon sp}{m}} \approx \abs{\frac{\varepsilon sp}{sp}}
                                   \leq \frac{1}{2}. \label{equ:BW:IneqaulityAlpha}
\end{equation}

We now appeal to Claim 3.2 of~\cite{H2002:PellsEquation}, which we re-state more precisely but do
not prove.  (See~\cite{J2003:NotesOnPellsEquation} for a proof of the claim's correctness.) The
claim is the following:
\begin{proposition}\label{prop:BW:HallgrenClaim32}
Let $n$ and $q$ be positive integers, let $\alpha$ be a constant, $\abs{\alpha} \leq \frac{3}{4}$,
and let $\beta \colon \Z \rightarrow \R$ be a function such that $\abs{\beta(j)} \leq \frac{1}{n}$
for all $j$, $0 \leq j < q-1$.  Then there exists a constant $c$ such that if $n \in \bigO{\log
q}$,
\[
    \abs{\sum_{j=0}^{q-1} e^{2 \pi i \paren{\frac{j}{q} \alpha + \beta(j)}}}^2 \geq c q^2.
\]
\end{proposition}

If we let $\beta(j) = \frac{k\delta_j}{s} + \frac{\varepsilon\delta_j}{m}$ and $\alpha =
\frac{\varepsilon sp}{m}$, then by \equref{equ:BW:InequalityBeta} and
\equref{equ:BW:IneqaulityAlpha}, respectively, the hypotheses of \propref{prop:BW:HallgrenClaim32}
are satisfied.

Further, combining \equref{equ:BW:ProbabilityDistY} and \equref{equ:BW:pyExponent}, we see that
\begin{align*}
    P_y &= \abs{ \frac{1}{\sqrt{pm}} \sum_{j=0}^{p-1} \exp\paren{2\pi i \paren{kj + \frac{\varepsilon j s}{m} + \frac{k\delta_j}{s} +
                           \frac{\varepsilon\delta_j}{m}}} }^2 \\
        &= \frac{1}{pm} \abs{ \sum_{j=0}^{p-1} \exp\paren{2\pi i \paren{\frac{\varepsilon j s}{m} + \frac{k\delta_j}{s} +
                           \frac{\varepsilon\delta_j}{m}}} }^2 \\
        &= \frac{1}{pm} \abs{\sum_{j=0}^{p-1} e^{2\pi i \paren{\paren{\frac{j}{p}}\alpha + \beta(j)}}
        }^2.
\end{align*}

We can therefore apply \propref{prop:BW:HallgrenClaim32} to the above sum, and deduce that
\begin{align*}
    P_y \geq \frac{1}{pm} cp^2 \approx \frac{c}{s}.
\end{align*}

Finally, we calculate the number of $y$ that satisfy the conditions of the theorem. By the
condition in \stepref{step:BW:Regulator:Measurement2} we know that $0 \leq y \leq \frac{m}{n}$.
Using the fact that $\log{s} > n$, we can obtain a lower bound on the number of such $y$ by
counting only those that satisfy
\begin{equation}
0 \leq y \leq \frac{m}{\log{s}}.\label{equ:BW:Regulator:RangeOfY}
\end{equation}
If $0 \leq \round{k\frac{m}{s}} \leq \frac{m}{\log{s}}$, then $0 \leq k \leq \frac{s}{\log{s}}$
(approximately); so there are $\frac{s}{\log{s}}$ values of $y$ that satisfy the conditions of the
theorem and \equref{equ:BW:Regulator:RangeOfY}.

The probability that the output of the algorithm satisfies the conditions of the theorem is
therefore at least
\begin{align*}
    \frac{s}{\log{s}}\cdot P_y &= \frac{s}{\log{s}}\cdot\frac{c}{s} \\
                          &= \frac{c}{\log{s}}
\end{align*}
which is in $\bigO{\frac{1}{\log{s}}}$ as required.
\end{proof}

The basic statement of \thmref{thm:BW:RegulatorResult} is that the probability of measuring such an
integer $y$ is considerably higher than selecting the integer $y$ uniformly at random from $0,
\ldots, m-1$.

The remainder of the algorithm to compute the regulator is purely classical and similar to
\algref{alg:IntroQuantumAlgs:IHSPBounded} to solve the bounded case of the Integer Hidden Subgroup
Problem (IHSP); we do not prove its correctness here (see~\cite{H2002:PellsEquation}
or~\cite{J2003:NotesOnPellsEquation}).  The algorithm works as follows: we run
\algref{alg:BW:Regulator} twice, obtaining integers $y_1$ and $y_2$ which by
\thmref{thm:BW:RegulatorResult} with high probability are equal to $\round{k_1\frac{m}{s}}$ and
$\round{k_2\frac{m}{s}}$ for some integers $k_1$ and $k_2$.  Also, with high probability,
$\gcd(k_1,k_2) = 1$.  Applying the continued fraction algorithm from
\secref{sec:IntroQuantumAlgs:IHSP} to $\frac{y_1}{y_2}$, and using an algorithm
from~\cite{H2002:PellsEquation} to test whether a given integer is ``close to'' a multiple of $s$,
we can recover the integer $k_1$.

Once we have recovered $k_1$, we can compute
\begin{align*}
    a &= \round{\frac{ k_1 m }{y_1}} \\
      &= \round{\frac{ k_1 m }{k_1 \frac{m}{s}}} \\
      &= \round{s}.
\end{align*}
The final step of the algorithm is to compute $\frac{a}{N}$; note that
\begin{align*}
    \abs{a - s} &< 1 \\
    \abs{a - N\reg} &< 1 \\
    \abs{\frac{a}{N} - \reg} &< \frac{1}{N}
\end{align*}

In other words, this polynomial time quantum algorithm allows us to determine the regulator to
arbitrary precision depending on our choice of $N$.

\end{subsection}

\begin{subsection}{Solving The Principal Ideal Distance Problem}\label{sec:BW:PIDP}

After computing the regulator, we can use another new quantum algorithm
from~\cite{H2002:PellsEquation} to solve PIDP.  Given an ideal $\mathfrak{a}$, let the (unknown)
distance of $\mathfrak{a}$ be $a$.  The goal of the algorithm is to find $a$. We will use an
algorithm similar to the algorithm in \secref{sec:IntroQuantumAlgs:DLP} for solving the Discrete
Logarithm Problem (DLP).

We begin by defining a new periodic function, as suggested
in~\cite{Z2003:ZalkaPrivateCommunication}, that is based on the function $g$ from the previous
section, although this new function has a two-dimensional domain.  Consider the function $h \colon
\Z \times \R \longrightarrow \pideals \times \R$ defined by
\begin{align*}
    h(j,x) &= g(aj+x)\\
           &= \paren{\idealleft{aj+x}, \idealerror{aj+x}}.
\end{align*}

We briefly justify that this function is periodic with a two-dimensional period given by $p_1 =
(0,\reg)$ and $p_2 = (-1,a)$:
\begin{enumerate}
\item Note that $h( (j,x) + p_1 ) = h( j, x+\reg ) = g( aj + x + \reg ) = g(aj+x)$ since $g$ is
periodic with period $\reg$ by \propref{prop:BW:gPeriodic}.
\item Note that $h( (j,x) + p_2 ) = h( j-1, x+a ) = g( a(j-1) + x + a ) = g(aj+x)$.
\end{enumerate}
Thus $h$ is indeed a periodic function with the given two dimensional period.  Note that the
unknown value $a$ appears in $p_2$, so if we could find the period of this function, we could solve
PIDP.

As in the previous section, we cannot easily compute with $h$, since the domain is not a discrete
set.  We therefore modify $h$ slightly, and define a new function $\hat{h} \colon \Z \times \Z
\rightarrow \pideals \times \Z$ by
\[
    \hat{h}(j_1,j_2) = \paren{ \idealleft{a j_1 +\frac{j_2}{N}}, \floor{N\idealerror{a j_1+\frac{j_2}{N}}}
    }.
\]
(Note that $\hat{h}(j_1,j_2) = \hat{g}\paren{aj_1 + \frac{j_2}{N}}$ where the domain of $\hat{g}$
has been extended to $\R$ in the natural way.)  This function $\hat{h}$ is the function proposed
in~\cite{H2002:PellsEquation} and it is the function on which the solution to PIDP is based.

At first it seems as though we cannot evaluate $\hat{h}$ since we do not know $a$. However, using
an algorithm described in detail in~\cite{SBW1994:BuchmannWilliamsReal}, given $\mathfrak{a}$ we
can compute the ideal to the left of $\idealdistance{\mathfrak{a}}j_1$ and its error; that is, we
can compute $\idealleft{aj_1}$ and $\idealerror{aj_1}$.  Since the value $\frac{j_2}{N}$ is known,
we can use the methods described in \secref{sec:BW:Protocol:Real} (in fact the same methods that
Alice and Bob use to carry out the real Buchmann-Williams protocol) to compute $\idealleft{aj_1 +
\frac{j_2}{N}}$ and $\idealerror{aj_1 + \frac{j_2}{N}}$.  (Note that at the end of these
computations we still do not know the value of $a$.)

We will solve a problem similar to the Hidden Subgroup Problem (HSP), although in this case we do
not have a hidden subgroup, but instead a hidden ``group-like set''.  Consider the set $T = \brace{
(s,t) \in \Z \times \Z \st \paren{as + \tfrac{t}{N}} \bmod{\reg} < \tfrac{1}{N} }$. The function
$\hat{h}$ is constant on $T$ because the interval $\left[0,\frac{1}{N}\right)$ is short enough that
it contains only the ideal $\qo$, and thus for any $x$ in the interval $\idealleft{x} = \qo$ and
$\floor{N\idealerror{x}} = 0$.  So for any $(s,t) \in T$, $\hat{h}(s,t) = ( \qo, 0 )$.

Next consider a coset of $T$, say $T + (u,v)$.  We can write this coset as follows:
\begin{align*}
    T + (u,v)
    &= \brace{ (s,t) \in \Z \times \Z \st \paren{a(s-u) + \tfrac{t-v}{N}} \bmod{\reg} < \tfrac{1}{N}}\\
    &= \brace{ (s,t) \in \Z \times \Z \st \paren{as+\tfrac{t}{N}-\paren{us+\tfrac{v}{N}}} \bmod{\reg} <
    \tfrac{1}{N}}.
\end{align*}
In other words, this coset of $T$ is the set of points $(s,t)$ such that $as + \frac{t}{N}$ is in
the interval of length $\frac{1}{N}$ starting from $us+\tfrac{v}{N}$.  We will denote this interval
by $I_{(u,v)}$.

Unlike in an instance of the typical HSP, the function $h$ is not necessarily constant on the
cosets of $T$. For example, suppose there is an ideal $\mathfrak{b}$ in the interval $I_{(u,v)}$.
Then for the points $x \in I_{(u,v)}$ after $\mathfrak{b}$, $\idealleft{x} = \mathfrak{b}$, but for
the rest of the points $x$, $\idealleft{x} = \rho^{-1}(\mathfrak{b})$ (the previous ideal in the
cyclical ordering). Similarly, if there is a point $y \in I_{(u,v)}$ such that the distance from
$y$ to $\idealleft{y}$ is a multiple of $\frac{1}{N}$ then the value of $\floor{N\idealerror{x}}$
will change depending on whether $x$ occurs before or after $y$.

It is in fact true that $\hat{h}$ could take on at most $3$ values on any coset of $T$, in the case
where the corresponding interval contains both an ideal $\mathfrak{b}$ and a value $y$ such that
the distance from $y$ to $\idealleft{y}$ is a multiple of $\frac{1}{N}$.  Consequently, $\hat{h}$
must be constant on at least $\frac{1}{3}$ of the elements in the coset.  Although the implications
of $\hat{h}$ being only ``approximately constant'' on the cosets of $T$ are not explicitly analysed
in~\cite{H2002:PellsEquation}, the fraction of the elements on which $\hat{h}$ is constant is
sufficient to allow the algorithm to succeed.

It should also be noted that two cosets of $T$ may overlap without being exactly equal, since we
could have $au + \frac{v}{N} \approx au' + \frac{v'}{N}$ with $v \neq v'$ and $u \neq u'$. However,
for fixed $u$ and $0 \leq v < N\reg$ it is true that the cosets $\brace{ T + (u,v) }$ are disjoint,
and $h$ is distinct and approximately constant on these disjoint cosets.

First we select the parameters for the algorithm using the following algorithm:
\begin{algorithm}[Parameter Selection For PIDP]\label{alg:BW:PIDPParamSel}
\begin{algone}
\item Compute the regulator $\reg$ using the algorithm from \secref{sec:BW:Regulator}.
\item Choose an integer $m > 2\reg$.
\item Choose an integer $b > n\frac{32\Delta}{3}$ and compute the continued fraction expansion of $b\reg$
to find $p,q \in \Z$ such that $\abs{b\reg - \frac{p}{q}} \leq \frac{1}{4qm}$.
\item Let $N = qb$.
\item Output $( \reg, m, N )$.
\end{algone}
\end{algorithm}

\begin{proposition}\label{prop:BW:PIDPParamCond}
The output of \algref{alg:BW:PIDPParamSel} satisfies $\abs{ N\reg - \round{N\reg} } \leq
\frac{1}{4m}$.
\end{proposition}

\begin{proof}
We know that
\begin{align*}
    \abs{ b\reg - \frac{p}{q} } &\leq \frac{1}{4qm} \\
    \abs{ \frac{N}{q}\reg - \frac{p}{q}} &\leq \frac{1}{4qm} \\
    \abs{ N\reg - p } &\leq \frac{1}{4m}.
\end{align*}

Now since $p$ is an integer and its distance from $N\reg$ is less than $\tfrac{1}{2}$, we must have
$p = \round{N\reg}$.  Thus
\begin{align*}
    \abs{ N\reg - \round{N\reg} } &\leq \frac{1}{4m} \\
\end{align*}
as required.
\end{proof}

Once the parameters have been selected, we can run the following algorithm to solve PIDP.

\begin{algorithm}[Core Of Solving PIDP]\label{alg:BW:PIDP}
\begin{algone}
\item Start in the state $\ket{0}\ket{0}\ket{0} \in \Hilbert{mp} \tensor \Hilbert{p} \tensor \Hilbert{l}$, where $p = \round{N\reg}$
and $l$ is chosen such that $l$ bits are sufficient to encode any point in the range of $\hat{h}$.
\item Apply $\QFT{mp} \otimes \QFT{p}$ to the first two registers.\label{step:BW:PIDP:QFT}
\item Apply $\U_{\hat{h}}$ to the system.\label{step:BW:PIDP:Uh}
\item Measure the third register.\label{step:BW:PIDP:Measurement1}
\item Apply $\QFT{mp} \otimes \QFT{p}$ to the first two registers.\label{step:BW:PIDP:QFT2}
\item Measure the first two registers to obtain the integers $(s,t)$.  If $t > \frac{p}{n}$, begin the
procedure again.\label{step:BW:PIDP:Measurement2}
\item Otherwise, return $(s,t)$.
\end{algone}
\end{algorithm}

\begin{theorem}\label{thm:BW:PIDPResult}
With probability in $\bigO{\frac{1}{\log{\paren{N\reg}}}}$ the output $(s,t)$ of
\algref{alg:BW:PIDP} satisfies
\[
    \frac{s - \gamma_t}{mN} \equiv at \bmod{\reg}
\]
for some $\gamma_t$ with $\abs{\gamma_t} \leq \frac{1}{2}$.
\end{theorem}

\begin{proof}[Proof Sketch]
First we claim that for any integer $s$, there is exactly one integer $t$ with $0 \leq t < N\reg$
such that $(s,t) \in T$.  To see this, suppose that $(s,t_1), (s,t_2) \in T$, with $0 \leq t_1, t_2
< N\reg$.  Let $\varepsilon_i = \paren{as + \frac{t_i}{N}} \bmod{\reg}$ for $i=1,2$, and define the
integers $k_1$ and $k_2$ such that $as + \frac{t_i}{N} = \varepsilon_i + k_i\reg$. Then $0 \leq
\varepsilon_1, \varepsilon_2 < \frac{1}{N}$, so
\begin{align*}
    \abs{ \varepsilon_1 - \varepsilon_2 } &< \tfrac{1}{N} \\
    \abs{ as + \tfrac{t_1}{N} - k_1\reg - as - \tfrac{t_2}{N} + k_2\reg } &< \tfrac{1}{N} \\
    \abs{ \tfrac{t_1 - t_2}{N} + (k_2 - k_1) \reg } &< \tfrac{1}{N} \\
    \abs{ \tfrac{t_1 - t_2}{N} } \bmod{\reg} &< \tfrac{1}{N} \\
    \abs{ t_1 - t_2 } \bmod{N\reg} &< 1
\end{align*}
Thus since $0 \leq t_1, t_2 < N\reg$ we must have $t_1 = t_2$.

We now define
\[
    \hat{T} = \brace{ (s,t) \in \Z \times \Z \st 0 \leq t < N\reg, \paren{as +
\tfrac{t}{N}} \bmod{\reg} < \frac{1}{N}}
\]
and it follows that for each $s \in \Z$, there is a unique element $(s,t) \in \hat{T}$.

For each $s \in \Z$, we can use this unique element $(s,t)$ to define $\sigma_s$ such that
\linebreak $\paren{as-\frac{t}{N}} \bmod{\reg} = \frac{\sigma_s}{N}.$ By the definition of
$\hat{T}$, $0 \leq \sigma_s < 1$.

Then note that for each $(s,t) \in \hat{T}$,
\begin{equation}
    as + \frac{t}{N} - \frac{\sigma_s}{N} = k\reg \label{equ:BW:ElementOfT}
\end{equation}
for some integer $k$.

Now note the following:
\begin{enumerate}
\item $t < N\reg$, so $k\reg < as + \reg$, and
\item $\sigma_s < 1$, so $as - \frac{1}{N} < k\reg$.
\end{enumerate}
Combining these two inequalities we obtain
\[
    \frac{as}{\reg} - \frac{1}{N\reg} < k < \frac{as}{\reg} + 1.
\]
Therefore, with high probability, $k = \ceil{\frac{as}{\reg}}$.

Rewriting \equref{equ:BW:ElementOfT} we see that
\begin{align}
    as + \frac{t}{N} - \frac{\sigma_s}{N} &= \ceil{\frac{as}{\reg}}\reg \notag\\
    t &= \ceil{\frac{as}{\reg}}N\reg - asN + \sigma_s \label{equ:BW:RearrangedT}
\end{align}

After \stepref{step:BW:PIDP:Uh} of \algref{alg:BW:PIDP} our system is in the state
\[
    \ket{\phi} = \frac{1}{p\sqrt{m}}\sum_{x=0}^{mp-1}\sum_{y=0}^{p-1}\ket{x}\ket{y}\ket{h(x,y)}.
\]
When we measure the third register in \stepref{step:BW:PIDP:Measurement1}, we see a value $z$, and
we leave the first two registers in a superposition of all states in which $z$ appears in the third
register.

This value of $z$ effectively specifies an interval $I$ of length $\frac{1}{N}$, since its first
coordinate is an ideal, and its second coordinate specifies a distance past that ideal (rounded
down to a multiple of $\frac{1}{N}$).  The interval $I$ will be approximately equal to the interval
$I_{(0,v)}$ for some value of $v$, $0 \leq v < N\reg$; this approximation is also sufficient for
our purposes. In other words, the measurement in \stepref{step:BW:PIDP:Measurement1} fixes a value
$v$ such that the first two registers of our system are in the state
\[
    \ket{\psi} = \frac{1}{p\sqrt{m}}\doublesum_{\substack{0 \leq x < mp \\
                                                0 \leq y < p \\
                                                (x,y) \in T + (0,v)}} \ket{x}\ket{y}.
\]

Since we are about to apply the same Fourier sampling technique discussed in
\secref{sec:BW:Regulator}, we can again make use of the fact that the distributions induced by
applying the technique to superpositions of the elements of a set $X$ and of some coset of $X$ are
identical.  In this case, we can therefore assume without loss of generality that $v = 0$, or in
other words, that the superposition in $\ket{\psi}$ is over the elements of $T$.  Further, since $0
\leq y < p$, we can more precisely assume that the superposition is over the elements of $\hat{T}$.
Thus without loss of generality, we can assume that
\begin{align*}
    \ket{\psi} &= \frac{1}{p\sqrt{m}}\doublesum_{\substack{0 \leq x < mp \\
                                                 (x,y) \in \hat{T}}} \ket{x}\ket{y} \\
               &= \frac{1}{\sqrt{mp}} \sum_{x=0}^{mp-1} \ket{x}\ket{\ceil{\frac{ax}{\reg}}N\reg - axN +
    \sigma_x} \quad\mbox{by \equref{equ:BW:RearrangedT}.}
\end{align*}

Again we mention that this is not exactly the state of the first two registers because of the
numerous approximations we have made along the way.  However, these approximations have all been
small enough so as to allow the remainder of the algorithm to succeed.

Temporarily let $y = \ceil{\frac{ax}{\reg}}N\reg - axN + \sigma_x$.  Then after applying $\QFT{mp}
\tensor \QFT{p}$ in \stepref{step:BW:PIDP:QFT2}, we obtain the state
\begin{align*}
    &\frac{1}{mp\sqrt{p}} \sum_{x=0}^{mp-1}
    \sum_{u=0}^{mp-1} \sum_{v=0}^{p-1} \exp\paren{2\pi i u
    \frac{x}{mp}} \exp\paren{2\pi i v \frac{y}{p}} \ket{u}\ket{v} \\
    = &\frac{1}{mp\sqrt{p}} \sum_{x=0}^{mp-1}
    \sum_{u=0}^{mp-1} \sum_{v=0}^{p-1} \exp\paren{2\pi i \frac{xu+yvm}{mp}} \ket{u}\ket{v}.
\end{align*}

The probability of obtaining a particular measurement result $(s,t)$ is therefore given by
\begin{equation}
    P_{(s,t)} = \abs{ \frac{1}{mp\sqrt{p}} \sum_{x=0}^{mp-1} \exp\paren{2\pi i \frac{xs+ytm}{mp}}
    }^2 \label{equ:BW:ProbabilityDistST}.
\end{equation}
As in~\cite{H2002:PellsEquation} (but in more detail) we now analyse this distribution.

The condition on $(s,t)$ given in the statement of the theorem is equivalent to
\begin{equation}
    \frac{s}{mN} - \frac{\gamma_t}{mN} - at = k\reg\label{equ:BW:PIDPResultEquality}
\end{equation}
for some integer $k$.  Note the following:
\begin{enumerate}
\item $\gamma_t \leq \frac{1}{2}$, so $-\frac{1}{2mN}-at \leq k\reg$, and
\item $\gamma_t \geq -\frac{1}{2}$ and $s \leq mN\reg$, so $\reg + \frac{1}{2mN} - at \geq k\reg$.
\end{enumerate}
Combining these two inequalities, we obtain
\[
    -\frac{at}{\reg}-\frac{1}{2mN\reg}      \leq k \leq - \frac{at}{\reg} + 1 + \frac{1}{2mN\reg} .
\]
Therefore, with high probability, $k = \ceil{-\frac{at}{\reg}} = -\floor{\frac{at}{\reg}}$.

Rewriting \equref{equ:BW:PIDPResultEquality}, we see that
\begin{align}
    \frac{s}{mN} - \frac{\gamma_t}{mN} - at &= - \floor{\frac{at}{\reg}}\reg \notag\\
    s &= atmN - \floor{\frac{at}{\reg}}mN\reg + \gamma_t \label{equ:BW:RearrangedS}
\end{align}

Now note that
\begin{align*}
    xs+ytm
    &= xatmN - x\floor{\frac{at}{\reg}}mN\reg + x\gamma_t + \ceil{\frac{ax}{\reg}}N\reg tm - axNtm +
    \sigma_x tm \\
    &= mN\reg\paren{ t \ceil{\frac{ax}{\reg}} - x \floor{\frac{at}{\reg}} } + x\gamma_t + tm\sigma_x.
\end{align*}

Now define $\lambda$ such that $p = N\reg + \lambda$.  Since $p = \round{N\reg}$, by
\propref{prop:BW:PIDPParamCond} $\abs{\lambda} \leq \frac{1}{4m}$.  Then $mN\reg = mp - m\lambda$,
and taking the above equation modulo $mp$, we obtain
\begin{align*}
    xs + ytm &\equiv -\lambda m \paren{ t \ceil{\frac{ax}{\reg}} - x \floor{\frac{at}{\reg}} } + x\gamma_t + tm\sigma_x \pmod{mp}.
\end{align*}

Finally, define $\delta_t$ and $\delta_x$ such that $\floor{\frac{at}{\reg}} = \frac{at}{\reg} -
\delta_t$ and $\ceil{\frac{ax}{\reg}} = \frac{ax}{\reg} + \delta_x$.  Then $0 \leq \delta_t,
\delta_x < 1$.  Thus the above equation becomes
\begin{align}
    xs + ytm &\equiv -\lambda m \paren{ t \frac{ax}{\reg} + t\delta_x - x \frac{at}{\reg} + x\delta_t } + x\gamma_t + tm\sigma_x
    \pmod{mp}\notag\\
    &\equiv -\lambda mt\delta_x - \lambda mx\delta_t + x\gamma_t + tm\sigma_x \pmod{mp}\notag\\
    &\equiv x\paren{ \gamma_t - \lambda m \delta_t } + tm\paren{\sigma_x - \lambda\delta_x} \pmod{mp}
    \label{equ:BW:pstExponent}
\end{align}

Note that
\begin{align}
    \abs{ \gamma_t - \lambda m \delta_t } &\leq \abs{\gamma_t} + \abs{ - \lambda m \delta_t } \notag\\
                                          &< \frac{1}{2} + \frac{1}{4m}\cdot m \cdot 1 \notag\\
                                          &= \frac{3}{4}.
                                          \label{equ:BW:IneqaulityAlpha2}
\end{align}

Also
\begin{align}
    \abs{ \frac{tm\paren{\sigma_x - \lambda\delta_x}}{mp} } &\leq \abs{ \frac{ tm \paren{ 1 - 0
    }}{mp}} \notag\\
    &= \frac{t}{p} \notag\\
    &\leq \frac{1}{n}\label{equ:BW:IneqaulityBeta2}
\end{align}
since in \stepref{step:BW:PIDP:Measurement2} we accepted only values of $t$ with $t \leq
\frac{p}{n}$.

If we let $\alpha = \abs{ \gamma_t - \lambda m \delta_t }$ and $\beta(x) = \frac{tm\paren{\sigma_x
- \lambda\delta_x}}{mp}$ then by \equref{equ:BW:IneqaulityAlpha2} and
\equref{equ:BW:IneqaulityBeta2}, respectively, the hypotheses of \propref{prop:BW:HallgrenClaim32}
are again satisfied.

Further, combining \equref{equ:BW:ProbabilityDistST} and \equref{equ:BW:pstExponent}, we see that
\begin{align*}
    P_{(s,t)} &= \abs{ \frac{1}{mp\sqrt{p}} \sum_{x=0}^{mp-1} \exp\paren{2\pi i \frac{x\paren{\gamma_t - \lambda m \delta_t }
    + tm\paren{\sigma_x - \lambda\delta_x}}{mp}}}^2 \\
    &= \frac{1}{m^2p^3} \abs{ \sum_{x=0}^{mp-1} e^{2\pi i \paren{\paren{\frac{x}{mp}} \alpha +
    \beta(x)}}}^2
\end{align*}

We can therefore apply \propref{prop:BW:HallgrenClaim32} to the above sum, and deduce that
\begin{align*}
    P_{(s,t)} \geq \frac{1}{m^2p^3} c (mp)^2 \approx \frac{c}{N\reg}.
\end{align*}

Finally, we calculate the number of $(s,t)$ that satisfy the conditions of the theorem. By the
condition in \stepref{step:BW:PIDP:Measurement2} we know that $0 \leq t \leq \frac{p}{n}$. Using
the fact that $\log{\paren{N\reg}} > n$, we can obtain a lower bound on the number of such $(s,t)$
by counting only those that satisfy
\begin{equation}
0 \leq t \leq \frac{p}{\log{\paren{N\reg}}}.\label{equ:BW:PIDP:RangeOfT}
\end{equation}
Since $p = \round{N\reg}$, then $0 \leq t \leq \frac{N\reg}{\log{\paren{N\reg}}}$ (approximately),
and for each value of $t$, there must be at least one value of $s$ such that $(s,t) \in \hat{T}$.
Thus there are at least $\frac{N\reg}{\log{\paren{N\reg}}}$ values $(s,t)$ that satisfy the
conditions of the theorem and \equref{equ:BW:PIDP:RangeOfT}.

The probability that the output of the algorithm satisfies the conditions of the theorem is
therefore at least
\begin{align*}
    \frac{N\reg}{\log{\paren{N\reg}}}\cdot P_{(s,t)} &= \frac{N\reg}{\log{\paren{N\reg}}}\cdot\frac{c}{N\reg} \\
                          &= \frac{c}{\log{\paren{N\reg}}}
\end{align*}
which is in $\bigO{\frac{1}{\log{\paren{N\reg}}}}$ as required.
\end{proof}

The remainder of the algorithm to solve PIDP is purely classical, and we do not prove its
correctness here (see~\cite{H2002:PellsEquation}).  The algorithm works as follows: we run
\algref{alg:BW:PIDP} until we obtain ordered pairs $(s_1, t_1)$ and $(s_2, t_2)$ with $\gcd( t_1,
t_2 ) = 1$.  By \thmref{thm:BW:PIDPResult} with high probability the ordered pairs satisfy
$\frac{s_i - \gamma_{t_i}}{mN} \equiv a t_i \bmod{\reg}$ for $i=1,2$.  We then use the extended
Euclidean algorithm to find integers $x,y$ such that $x t_1 + y t_2 = 1$, and compute $\tilde{a} =
\frac{x s_1 + y s_2}{mN} \bmod{\reg}$. As proven in~\cite{H2002:PellsEquation}, $\abs{ a -
\tilde{a} } \leq 1$.

It is acknowledged in~\cite{H2002:PellsEquation} that we would like to compute $a$ to a higher
accuracy, but no specific method to do so is described.  One such method would be to consider the
interval $\bracket{ \tilde{a}-1, \tilde{a}+1 }$, select $N$ equally-spaced points in the interval,
and perform a binary search among those $N$ points, finally selecting the smallest point
$\tilde{\tilde{a}}$ for which $\idealleft{\tilde{\tilde{a}}} = \mathfrak{a}$.  Then we know that
\[
    \abs{ \tilde{\tilde{a}} - a } < \frac{1}{N}
\]

Therefore there exists a polynomial-time quantum algorithm to solve PIDP (that is, to determine the
distance of a principal ideal to arbitrary precision, depending on our choice of $N$).

\end{subsection}

\end{section}

\end{chapter}

\begin{chapter}{Conclusions And Future Work}\label{chap:Conclusions}
\markright{CHAPTER \thechapter. \ CONCLUSIONS AND FUTURE WORK}

While large-scale quantum computers are not currently technologically feasible, this thesis has
demonstrated that if they ever become realistic, they will pose a serious threat to much of our
secret communication.  Today's most widely-used public key cryptosystems, such as the RSA
cryptosystem studied in \chapref{chap:RSA} and the ElGamal cryptosystem studied in
\chapref{chap:ElGamal}, as well as the popular key establishment protocols like the Diffie-Hellman
protocol from \chapref{chap:DiffieHellman}, are open to polynomial-time attacks with a quantum
computer.

Other less popular cryptosystems have been proposed that rely on the hardness of other problems,
and these schemes may be candidates for systems that resist quantum cryptanalysis.  For example,
the McEliece cryptosystem described in \chapref{chap:McEliece} does not seem to fit into a
framework in which it could be attacked with today's known set of quantum algorithms.  However,
such less popular schemes may suffer from a lack of efficiency compared to the more commonly used
algorithms, and they may not have received the same degree of academic scrutiny as their more
established counterparts.  However, not all of these alternative classical schemes are
quantum-resistant: for example, the real version of the Buchmann-Williams key establishment
protocol from \chapref{chap:BW} does not seem to suffer from classical vulnerabilities, but recent
developments in quantum algorithm theory have exposed some quantum weaknesses.

This thesis has also touched on cryptosystems of historical importance, even though such
cryptosystems may no longer be feasible choices given the existence of known classical attacks. The
lattice-based schemes presented in \chapref{chap:AjtaiDwork} and \chapref{chap:GGH} are based on
hard problems that are fundamentally different from the standard cryptosystems in use today,
although unlike the McEliece cryptosystem, they have been shown to have serious classical
weaknesses.  Nonetheless, they could provide a starting point for further investigation of
lattice-based cryptography.  The NTRU scheme studied in \chapref{chap:NTRU} is an example of
another scheme that could resist a quantum attack if its recently discovered classical
vulnerabilities can be overcome.

Another new class of cryptosystems is made up of schemes that use a quantum computer to aid the
parties who wish to communicate securely; the quantum scheme described in \chapref{chap:Okamoto} is
one concrete example of a scheme from this class.  Since these cryptosystems are necessarily quite
new, and since they are currently only of theoretical interest, they have not received much
attention in the academic community.  However, as attackers begin to include quantum computers in
their arsenals, the legitimate parties to secure communication may be able to stay one step ahead
by also using quantum computers.

Further, while this thesis has touched on many of today's important cryptosystems, there are still
many more that could bear further investigation in a quantum setting.  Examples of such
cryptosystems would be newly-proposed schemes where operations are carried out in a braid
group~\cite{AAG1999:AlgebraicMethodForPKC,AAFG2001:BraidGroupCrypto}.  This thesis could also be
extended by performing quantum security analyses of public key signature schemes, which attempt to
provide authentic (as opposed to confidential) communication.  Many of the cryptosystems discussed
in this thesis, such as the RSA, ElGamal, and NTRU schemes, have associated signature schemes that
are based on the hardness of similar problems.  Another extension would be to analyse some of the
popular symmetric key cryptosystems in use today to see whether they might be susceptible to
quantum attacks.

This thesis has gathered together many results from several areas of mathematics and has presented
them in a practical way.  It has attempted to provide a clear presentation of the basics of public
key cryptography (for the cryptographic beginner) and a concise introduction to many of the basics
of quantum computation (for the quantum beginner).  Each cryptosystem has been presented along with
enough background material to make its basic concepts easy to understand.  In some cases, such as
that of the quantum scheme in \chapref{chap:Okamoto}, the presentation has involved making some
minor corrections and clarifications to ambiguities in the original papers. In other cases, such as
that of the new quantum algorithms presented in \secref{sec:BW:Regulator} and \secref{sec:BW:PIDP},
the presentation has also been expanded considerably to provide a more detailed and precise
analysis geared to be more accessible to those without expertise in the field.  Wherever possible,
parallels have been drawn between similar cryptosystems, or between cryptosystems built on similar
ideas.  For many of the cryptosystems mentioned in this thesis, this may be the first time that the
schemes have been considered in a quantum setting.

The main goal of this thesis, however, has been to make it clear that the encryption schemes in
current use will not provide a high level of security in a quantum setting.  While it would be
unwise to claim that cryptosystems that currently resist quantum attacks will necessarily continue
to resist them, it would also be unwise to ignore the possibility that large-scale quantum attacks
will one day be feasible.  We need to start investigating alternative quantum-resistant
cryptosystems now, in the event that we one day need to make use of them.

\end{chapter}

\bibliographystyle{amsalpha}
\bibliography{thesis_bibliography}

\end{document}